\documentclass[12pt]{article}
\usepackage[utf8]{inputenc}
\usepackage[margin=25truemm]{geometry}
\usepackage{amsmath,amssymb}
\usepackage{physics}
\usepackage{tensor}
\usepackage{mathtools}
\usepackage{bm}
\usepackage{cite}
\usepackage{url}

\numberwithin{equation}{section}

\begin{document}

\begin{titlepage}
\renewcommand{\thefootnote}{\fnsymbol{footnote}}
\begin{normalsize}
\begin{flushright}
\begin{tabular}{l}
\end{tabular}
\end{flushright}
\end{normalsize}

~~\\

\vspace*{0cm}
    \begin{Large}
       \begin{center}
         {Description of curved spacetimes by finite-size matrices \\
         in the type IIB matrix model }
       \end{center}
    \end{Large}
\vspace{1cm}

\begin{center}
           Keiichiro H{\sc attori}$^{1)}$\footnote
            {
e-mail address:
hattori.keiichiro.17@shizuoka.ac.jp},
           Tatsuya S{\sc eko}$^{1)}$\footnote
            {
e-mail address:
seko.tatsuya.18@shizuoka.ac.jp}
           {and}
           Asato T{\sc suchiya}$^{1),2)}$\footnote
           {
e-mail address: tsuchiya.asato@shizuoka.ac.jp}
\\
      \vspace{1cm}

$^{1)}$
 {\it Graduate School of Science and Technology, Shizuoka University}\\
               {\it 836 Ohya, Suruga-ku, Shizuoka 422-8529, Japan}\\
        \vspace{0.3cm}

 $^{2)}$
{\it Department of Physics, Shizuoka University}\\
                {\it 836 Ohya, Suruga-ku, Shizuoka 422-8529, Japan}\\

\end{center}

\vspace{1cm}

\begin{abstract}
\noindent
The type IIB matrix model is expected to give a nonperturbative formulation of superstring theory.
Its covariant derivative interpretation provides a method to describe
curved spacetimes in the model. There, matrices are identified with certain covariant derivatives which can be
viewed as infinite-size matrices.
Here, by using the Berezin-Toeplitz quantization, we develop a 
method 
to regularize these matrices as finite-size ones, which is needed 
to calculate quantum effects in the interpretation or in particular
to apply the interpretation to the results of numerical simulations.
As examples, we examine the cases of $T^{2n}$ and $S^2$ in detail.

\end{abstract}
\end{titlepage}
\vfil\eject

\setcounter{footnote}{0}

\tableofcontents

\section{Introduction} \label{sec1}
Matrix models are expected to provide 
a nonperturbative formulation of superstring theory \cite{Banks:1996vh,Ishibashi:1996xs}.
The type IIB matrix model \cite{Ishibashi:1996xs} is one of such models (for recent studies of the model, see \cite{Hirasawa:2024dht,Anagnostopoulos:2026utg,Anagnostopoulos:2026qvz,Asano:2024def,Asano:2024edo,Steinacker:2024unq,Steinacker:2024huv,Brandenberger:2024ddi,Laliberte:2024iof,Manta:2024vol,Manta:2025inq,Manta:2025tcl,Hattori:2024btt,Gohara:2025zfh,Ho:2025htr,Gass:2025bqr,Liao:2025yfb,Steinacker:2026qzk,Chou:2025rwy,Chou:2025moy}\footnote{In particular, see \cite{Hartnoll:2024csr,Komatsu:2024bop,Komatsu:2024ydh,Hartnoll:2025ecj} for the polarized model\cite{Bonelli:2002mb}. }). The action is given by
    \begin{equation}
      S = -\frac{1}{g^2} \Tr(\frac{1}{4} [A_a, A_b][A^a, A^b] - \frac{1}{2} \bar{\psi} \Gamma^a [A_a, \psi]) \ ,  \label{action}
    \end{equation}
where $A_a \ (a=1, \ldots, 10)$ 
are ten $N\times N$ Hermitian matrices, and $\psi$ are sixteen $N\times N$
Hermitian matrices carrying ten-dimensional Majorana-Weyl spinor indices.
The action is obtained by dimensional reduction of ten-dimensional 
$U(N)$ $\mathcal{N}=1$ super Yang-Mills theory  to zero dimensions.
Hence, space-time does not exist a priori in the model but is expected to emerge dynamically from the degrees of freedom of the matrices. In this paper, we focus on the Euclidean version of the model.

If the type IIB matrix model indeed provides a nonperturbative formulation of 
superstring theory, it should be capable of describing
gravity, and thus curved spacetimes.
A framework for such a description is known as the covariant derivative interpretation of matrix models \cite{Hanada:2005vr} 
(see also \cite{Hanada:2006gg,Hanada:2006ei,Furuta:2006kk,Saitou:2007sf,Matsuo:2008yd,Asano:2012mn,Sakai:2017dxi,Sakai:2019cmj,Hattori:2024btt}, and see \cite{Steinacker:2024unq,Ho:2025htr} for other approaches to this issue.).
In this interpretation, Einstein's equations can be derived from the equations of motion of the type IIB matrix model, and general coordinate as well as local Lorentz transformations are embedded in the U($N$) symmetry of the model.
The matrices can be regarded as covariant derivatives $\nabla_{(a)}$
on a $d$-dimensional manifold $M$\footnote{The covariant derivatives 
carrying indices with parenthesis 
$\nabla_{(a)}$ can be viewed as infinite-size matrices, while ordinary covariant
derivatives
$\nabla_a$ are not. The construction of $\nabla_{(a)}$ will be reviewed 
in \ref{sec2.1}.}, 
which contain the geometric information of the underlying space-time.
However, since the covariant derivatives are infinite-dimensional operators, 
they must be regularized to perform quantum 
analyses.
In particular, for applications to the results of numerical simulations\cite{Anagnostopoulos:2022dak,Hirasawa:2024dht,Anagnostopoulos:2026utg,Anagnostopoulos:2026qvz}, 
it is necessary to regularize them as finite-size matrices.

In this paper, we develop a finite-matrix regularization of the 
covariant derivatives $\nabla_{(a)}$
when the manifold $M$ is a closed connected Kähler manifold of real dimension $2n$.
This provides a finite-matrix description of the manifold $M$.
Our method generalizes the approach developed previously for 
closed Riemann surfaces ($n=1$ case) \cite{Hattori:2024btt} to higher-dimensional Kähler 
manifolds.
As in that case, we employ the Berezin-Toeplitz (BT) quantization \cite{Bordemann:1993zv,Hawkins_2000} for the regularization.

The BT quantization provides a systematic method to regularize 
fields on manifolds as finite-size matrices.
The finite-size matrices obtained through the BT quantization are 
called Toeplitz operators.
A Toeplitz operator associated with a field $\chi$ is constructed 
by projecting 
$\chi$ onto the space of zero modes of the Dirac operator.
When the topological charge $p$ of the background gauge field is 
sufficiently large, the vanishing theorem and the Atiyah-Singer 
index theorem ensure that the number of normalizable, linearly 
independent zero modes is finite.
Therefore, the Toeplitz operator of the field $\chi$ is a finite-size
matrix.
As the topological charge $p$ increases, 
the number of independent zero modes and hence the matrix 
size also increases.
In the large-$p$ (large-matrix) limit, the Toeplitz operators 
exhibit desirable regularization properties: the product of 
Toeplitz operators corresponds to the Toeplitz operator of the 
product of fields, and their commutator corresponds to the 
Toeplitz operator of the Poisson bracket of the fields.
Thus, projecting a field onto the Dirac zero-mode space yields a 
finite-matrix representation that provides a well-behaved 
regularization in the large-$p$ limit.

The main result of this paper is a construction of finite-size matrices $\mathcal{P}_{(a)}$ corresponding to the covariant derivative $-i\nabla_{(a)}$.
The explicit expression of $\mathcal{P}_{(a)}$ is given in the 
following.
Let $\varphi$ denote an element on which the covariant derivative 
acts, and 
$T(\varphi)$ its Toeplitz operator obtained by the BT quantization.
Then, $\mathcal{P}_{(a)}$ acts on $T(\varphi)$ as
    \begin{equation}
      \mathcal{P}_{(a)} T(\varphi)
      = \hbar_p^{-1} T(\partial_{(a)}X^A) [T(X^A), T(\varphi)] - \frac{1}{2} \hbar_p^{-1} [T(\partial_{(a)}X^A), T(X^A)] T(\varphi) \ .
      \label{eq1.2}
    \end{equation}
Here, $\{X^A\}_{A=1, \ldots, D}$
are the embedding coordinate functions of the 
$2n$-dimensional Kähler manifold $M$
into a $D$-dimensional Euclidean space $\mathbb{R}^D$, and $T(X^A)$
and $T(\partial_{(a)}X^A)$ denote Toeplitz operators for $X^A$ and 
$\partial_{(a)}X^A$, respectively. $\hbar_p$ is a quantity proportional to $1/p$.
The operator $\mathcal{P}_{(a)}$ maps 
a finite-size matrix $T(\varphi)$
to another matrix of the same size, and therefore can itself be regarded as a finite-size square matrix.
In the large-$p$ limit, corresponding to the large-matrix limit, 
$\mathcal{P}_{(a)}$
reproduces the covariant derivative:
\begin{equation}
      \mathcal{P}_{(a)} T(\varphi)
      \sim T(-i \nabla_{(a)} \varphi) \ . 
      \label{eq1.3}
    \end{equation}
This demonstrates that 
$\mathcal{P}_{(a)}$ provides a proper finite-matrix 
regularization of the covariant derivative $-i\nabla_{(a)}$.

The remainder of this paper is organized as follows.
In Section 2, we review the covariant derivative interpretation 
of matrix models and the BT quantization.
In Section 3, we construct the finite-matrix 
regularization of covariant derivatives on closed 
connected Kähler manifolds of real dimension $2n$.
Section 4 presents explicit examples for $T^{2n}$ and $S^{2}$.
Finally, Section 5 is devoted to conclusion and discussion.
Appendix A contains the computation of the commutator $[\mathcal{P}_{(a)},\mathcal{P}_{(b)}]$
of the matrix-regularized covariant derivatives.

\section{Reviews} \label{sec2}

\subsection{Covariant derivative interpretation of matrix models} \label{sec2.1}

In this subsection, we review the covariant derivative 
interpretation of matrix models\cite{Hanada:2005vr}.  
One possible way to describe curved spacetime in matrix models is to interpret the matrices simply as covariant derivatives:
$A_a = -i\nabla_a$.
Here the index $a$ refers to that in the local Lorentz frame.  
However, each component of the covariant derivative cannot be regarded as a matrix as it stands, for the following two reasons.  
First, $\nabla_a$ is not globally defined on a curved manifold; it transforms between different coordinate patches.  
Second, the product of covariant derivatives does not correspond to matrix multiplication.  
For instance, for $A_1 = -i\nabla_1$ and $A_2 = -i\nabla_2$,
\begin{align}
A_1 A_2 = -i \partial_1 A_2 - i \Omega_1^{2c} A_c \ ,
\end{align}
where $\Omega^{bc}_a$ denotes the spin connection.  
Since $A_1 = -i\nabla_1$ acts on the second index of $A_2$, the space on which $A_1$ acts differs from that for $A_2$.  
To overcome these two issues, we consider a fiber bundle whose fiber is the representation space of the regular representation of $G = \mathrm{Spin}(d)$ or $\mathrm{Spin}^c(d)$.

Let $M$ be a $d$-dimensional manifold with local coordinate patches $\{U_i\}$, and let $G = \mathrm{Spin}(d)$ if $M$ admits a spin structure, otherwise $G = \mathrm{Spin}^c(d)$\footnote{$\mathrm{Spin}^c(d)$ is defined by $\mathrm{Spin}^c(d)= (\mathrm{Spin}(d) \times U(1)) / Z_2$.
Here, we assume that $M$ possesses at least a $\mathrm{Spin}^c(d)$ structure.}.  
The covariant derivative on each patch $U_i$ is given by
    \begin{equation}
      \nabla_a^{[i]}
      = e_a^{\mu [i]}(x) \left(\partial_\mu + \frac{1}{2} \Omega^{bc [i]}_\mu(x) \mathcal{O}_{bc} + A_\mu^{[i]}(x)\right). 
      \label{covariant derivative}
    \end{equation}
where  $e_a^{\mu [i]}(x)$ is the vielbein, $\Omega^{bc[i]}_\mu(x)$ is the spin connection, $\mathcal{O}_{bc}$ is the Lorentz generators\footnote{Note that $\mathcal{O}_{bc}$ here is twice $\mathcal{O}_{bc}$ in \cite{Hanada:2005vr}.}, and $A^{[i]}_\mu(x)$ is a $U(1)$ gauge field belonging to the $U(1)$ part of $\mathrm{Spin}^c(d)$.  
(When $G = \mathrm{Spin}(d)$, one may set $A_\mu = 0$.)  
The index $[i]$ labels quantities defined on the patch $U_i$.

To regard the covariant derivative as a matrix, we need to find a vector space on which it acts as an \textit{endomorphism}.  
For this purpose, we begin with the regular representation of the group $G$. It is a space of smooth functions from $G$ to $\mathbb{C}$:
$V_{\text{reg}} = \{ f : G \to \mathbb{C} \}$,
The group action on it is defined by
\begin{align}
(\hat{h}f)(g) = f(h^{-1}g),
\qquad f \in V_{\text{reg}}, \;\; h,g \in G \ .
\end{align}

The regular representation has two important properties.  
First, it is reducible and decomposes into irreducible representations as
   \begin{equation}
      V_{\text{reg}}
      = \bigoplus_r (\underbrace{V_r \oplus \cdots \oplus V_r}_{d_r})\ , \label{irreducible decomposition}
    \end{equation}
where $V_r$ is the representation space of an irreducible representation $r$ of $G$, and $d_r$ is its dimension.  
For $G = \mathrm{Spin}(2) \cong U(1)$, this corresponds to the Fourier expansion.  
Indeed, by the Peter-Weyl theorem, the representation matrices of the irreducible representations of $G$ form an orthonormal basis of smooth functions on $G$, and any $f \in V_{\text{reg}}$ can be expanded as
\begin{align}
f(g) = \sum_{r:\mathrm{irr.}} f^{\langle r\rangle}_{ij} \sqrt{d_r}\, R^{\langle r^\ast\rangle}_{ij}(g)\ ,
\label{Peter-Weyl}
\end{align}
where the sum runs over all irreducible representations $r$ of $G$, and $R^{\langle r^\ast\rangle}_{ij}(g)$ denotes the representation matrix of the complex-conjugate representation.  
The orthonormality of these matrices is expressed as
\begin{align}
\int dg\, R^{\langle r\rangle}_{ij}(g)^\ast R^{\langle r'\rangle}_{kl}(g)
= \frac{1}{d_r}\,\delta^{\langle r\rangle\langle r'\rangle}\delta_{ik}\delta_{jl}\ ,
\label{orthonormality}
\end{align}
where $dg$ is the Haar measure normalized by $\int dg = 1$.  
From \eqref{Peter-Weyl}, we see that under the group action $h \in G$,
\begin{align}
f^{\langle r\rangle}_{ij} \;\to\; R^{\langle r\rangle}_{ik}(h)\, f^{\langle r\rangle}_{kj}\ ,
\end{align}
which means that the first index $i$ transforms under representation $r$ while the second index $j$ remains invariant.  
Thus, $f^{\langle r\rangle}_{ij}$ corresponds to $d_r$ copies of representation $r$, establishing the decomposition 
\eqref{irreducible decomposition}.

The second important property of $V_{\text{reg}}$ is the following isomorphism, valid for any representation $r$ of $G$:
\begin{align}
V_r \otimes V_{\text{reg}} \cong 
\underbrace{V_{\text{reg}} \oplus \cdots \oplus V_{\text{reg}}}_{d_r}.
\label{isomorphism}
\end{align}
This can be seen as follows.  
Let $\Phi^i(g) \in V_r \otimes V_{\text{reg}}$, where the index $i$ transforms under representation $r$:
\begin{align}
(\hat{h}\Phi^i)(g) = R\indices{^i_j}(h)\,\Phi^j(h^{-1}g)\ .
\end{align}
Then, the isomorphism \eqref{isomorphism} is realized by
\begin{align}
\Phi^{(i)}(g) = R\indices{^{(i)}_j}(g^{-1})\,\Phi^j(g)\ ,
\end{align}
where $i=1,\ldots,d_r$.  
Although $R\indices{^i_j}$ and $R\indices{^{(i)}_j}$ represent the same object, we distinguish the index $(i)$, which labels the $d_r$ copies of $V_{\text{reg}}$ that remain inert under the $G$-action.  
Indeed, under $h \in G$,
\begin{align}
(\hat{h}\Phi^{(i)})(g)
&= R\indices{^{(i)}_j}(g^{-1})(\hat{h}\Phi^j)(g)
 = R\indices{^{(i)}_j}(g^{-1})R\indices{^j_k}(h)\,\Phi^k(h^{-1}g)
 = \Phi^{(i)}(h^{-1}g) \ ,
\end{align}
showing that the label $(i)$ is unchanged by the action of $h$ and each element of $\Phi^{(i)}$ belongs to $V_{\mathrm{reg}}$.

Let $E_{\text{reg}}$ be a fiber bundle over $M=\bigcup_i U_i$ with fiber $V_{\text{reg}}$ and structure group $G$.  
Denote by $\Gamma(E_{\text{reg}})$ the set of smooth global sections of $E_{\text{reg}}$.  
Since an element of $V_{\text{reg}}$ is a smooth function on $G$, each local section $f^{[i]}(x,g)$ defined on $U_i \times G$ satisfies the following transition condition on an overlap $U_i \cap U_j$:
\begin{align}
f^{[j]}(x,g) = f^{[i]}(x, t_{ij}(x)g)\ ,
\end{align}
where $t_{ij}(x)$ is the transition function of the bundle.

A similar isomorphism to \eqref{isomorphism} holds for the corresponding bundles:
\begin{align}
\Gamma(E_r \otimes E_{\text{reg}}) \cong 
\underbrace{\Gamma(E_{\text{reg}}) \oplus \cdots \oplus \Gamma(E_{\text{reg}})}_{d_r} \ ,
\end{align}
where $E_r$ is a bundle associated with $E_{\text{reg}}$ whose fiber is $V_r$.  
The isomorphism is realized by
\begin{align}
f^{[i]}_{(k)}(x,g)
= R\indices{_{(k)}^l}(g^{-1})\, f^{[i]}_l(x,g) \ .
\end{align}

We now show that $\Gamma(E_{\text{reg}})$ is the space on which the covariant derivative acts as an endomorphism.  
Since the covariant derivative acts as
\begin{align}
\nabla_a : \Gamma(E_{\text{reg}}) \to \Gamma(T M \otimes E_{\text{reg}}) \ ,
\end{align}
using the isomorphism
\begin{align}
\Gamma(T M \otimes E_{\text{reg}}) \cong 
\underbrace{\Gamma(E_{\text{reg}}) \oplus \cdots \oplus \Gamma(E_{\text{reg}})}_{d},
\end{align}
we can define
\begin{align}
\nabla_{(a)} : \Gamma(E_{\text{reg}}) \to
\underbrace{\Gamma(E_{\text{reg}}) \oplus \cdots \oplus \Gamma(E_{\text{reg}})}_{d} \ ,
\end{align}
where the label $(a)$ enumerates the $d$ copies of $\Gamma(E_{\text{reg}})$.  
Each component $\nabla_{(a)}$ thus acts as an endomorphism on $\Gamma(E_{\text{reg}})$.

On each patch $U_i$, $\nabla_{(a)}$ and $\nabla_a$ are related by
\begin{align}
\nabla^{[i]}_{(a)} = R\indices{_{(a)}^b}(g^{-1}_{[i]})\, \nabla^{[i]}_b,
\end{align}
where $R\indices{_{(a)}^b}(g^{-1}_{[i]})$ is the representation matrix for the vector representation of $g^{-1}_{[i]}\in G$.  
On the overlap $U_i \cap U_j$, $\nabla^{[i]}_a$ and $\nabla^{[j]}_a$ satisfy
\begin{align}
\nabla^{[i]}_a = R\indices{_a^b}(t_{ij}(x))\, \nabla^{[j]}_b \ ,
\end{align}
and hence
\begin{align}
\nabla^{[i]}_{(a)} &= R\indices{_{(a)}^b}(g^{-1}_{[i]}) R\indices{_b^c}(t_{ij}(x))\, \nabla^{[j]}_c \notag\\
&= R\indices{_{(a)}^c}\!\left((t_{ij}^{-1}(x)g_{[i]})^{-1}\right)\! 
\nabla^{[j]}_c
= \nabla^{[j]}_{(a)}.
\end{align}
Thus, each component $\nabla_{(a)}$ is globally defined on $M$.  
In summary, $\nabla_{(a)} = R\indices{_{(a)}^b}(g^{-1})\,\nabla_b$ acts as an endomorphism
\begin{align}
\nabla_{(a)}: \Gamma(E_{\text{reg}}) \to \Gamma(E_{\text{reg}}),
\end{align}
and can therefore be regarded as an infinite-size matrix.

The action of $\nabla_{(a)}$ on $f^{[i]}(x,g) \in \Gamma(E_{\text{reg}})$ is given by
\begin{align}
\nabla^{[i]}_{(a)} f^{[i]}(x,g)
= R\indices{_{(a)}^b}(g^{-1}) e^{\mu[i]}_{b}(x)
\left(
\partial_\mu + \frac{1}{2}\Omega^{cd[i]}_{\mu}(x)\mathcal{O}_{cd}
+ A^{[i]}_\mu(x)
\right)
f^{[i]}(x,g),
\end{align}
and the Lorentz generators act as
\begin{align}
i\epsilon_{ab}(\mathcal{O}^{ab}f^{[i]})(x,g)
= f^{[i]}\!\left(x,(1+i\epsilon_{ab}M^{ab})^{-1}g\right) - f^{[i]}(x,g) \label{Lorenz}
\end{align}
where $M^{ab}$ are matrices of the fundamental representation of $G$.

We define the inner product on $\Gamma(E_{\text{reg}})$ as
\begin{align}
(u,v) = \int e\, d^d x\, dg\; u^{[i]}(x,g)^\ast v^{[i]}(x,g),
\end{align}
where $e = \det e^\mu_a$ and $dg$ is the Haar measure.
Then, it can be shown that each component of $-i\nabla_{(a)}$ is Hermitian.

Hence, each component $-i\nabla_{(a)}$ with the parenthesized index acts as an endomorphism on $\Gamma(E_{\text{reg}})$ and can be interpreted as an infinite-size Hermitian matrix.  
Under this interpretation, the type IIB matrix model action \eqref{action} can be rewritten as
\begin{align}
S &= -\frac{1}{4g^2}
\mathrm{Tr}\!\left(
[A_{(a)}, A_{(b)}][A_{(c)}, A_{(d)}]\delta^{(a)(c)}\delta^{(b)(d)}
\right)
+ \frac{1}{2g^2}
\mathrm{Tr}\!\left(
\bar{\psi}^{(\alpha)}(\Gamma^{(a)}){}\indices{_{(\alpha)}^{(\beta)}}[A_{(a)},\psi_{(\beta)}]
\right),
\label{action2}
\end{align}
where
\begin{align}
A_{(a)},\, \psi_{(\alpha)} : \Gamma(E_{\text{reg}}) \to \Gamma(E_{\text{reg}}) \ .
\end{align}

In the covariant derivative interpretation, one finds that Einstein's equations can be derived from the equations of motion of the IIB matrix model.  
Let us consider only the bosonic part of the action \eqref{action2} and add a mass term:
\begin{align}
S &= -\frac{1}{4g^2}
  \mathrm{Tr}\!\left([A_{(a)},A_{(b)}][A^{(a)},A^{(b)}]\right)
  + \frac{m^2}{g^2}\,\mathrm{Tr}\!\left(A_{(a)}A^{(a)}\right).
\end{align}
The equation of motion then reads
\begin{align}
[A^{(a)},[A_{(b)},A_{(a)}]] - 2m^2 A_{(b)} = 0 \ .
\label{equation of motion}
\end{align}

We now consider a classical solution of the form
\begin{align}
A_{(a)} =
\begin{cases}
-i\nabla_{(a)}, & a = 1,\ldots,d, \\
0, & a = d+1,\ldots,10\ ,
\end{cases}
\label{ansatz}
\end{align}
where $\nabla_{(a)} = R\indices{_{(a)}^b}(g^{-1})\,\nabla_b$ and $\nabla_b$ is defined in \eqref{covariant derivative}.  
Substituting \eqref{ansatz} into \eqref{equation of motion} gives
\begin{align}
-[\nabla^{(a)}, [\nabla_{(b)}, \nabla_{(a)}]] - 2m^2 \nabla_{(b)} = 0 \ ,
\end{align}
for $a,b=1,\ldots,d$, which is equivalent to
\begin{align}
[\nabla^a, [\nabla_a, \nabla_b]] - 2m^2 \nabla_b = 0 \ .
\label{eom}
\end{align}

Assuming vanishing torsion and setting $A_\mu = 0$ in 
\eqref{covariant derivative}, the left-hand side of \eqref{eom} 
can be computed as
\begin{align}
[\nabla^a, [\nabla_a, \nabla_b]] - 2m^2 \nabla_b
 = \frac{1}{2}(\nabla^a R_{ab}{}^{cf})\,\mathcal{O}_{cf}
   - (R^c{}_b + 2m^2 \delta^c_b)\,\nabla_c \ .
   \label{eom2}
\end{align}
Thus, Eq. \eqref{eom2} is satisfied if the following conditions hold:
\begin{align}
\nabla^a R_{ab}{}^{cf} = 0, \qquad
R_{bc} = -2m^2 \delta_{bc}\ .
\end{align}
The first equation follows from the Bianchi identity
$\nabla_{[a}R_{bc]fg}=0$ and the second equation.  
The second equation represents the $d$-dimensional Einstein equation with a cosmological constant
\begin{align}
\Lambda = -(d-2)m^2 \ .
\end{align}

\subsection{Examples: $T^2$ and $S^2$} \label{sec2.22}

In this subsection, we explicitly present the covariant derivatives $\nabla_{(a)}$ on $T^2$ and $S^2$ as concrete examples \cite{Hanada:2005vr}. Note that these manifolds possess the spin structure.   
The irreducible representations of $\mathrm{Spin}(2) \cong U(1)$ can be classified by an integer or half-integer charge $s$.  
The representation matrix (a $1\times 1$ matrix) of charge $s$ is given by
\begin{align}
R^{\langle s\rangle}(\theta) = e^{2is\theta} \ , \qquad (\theta \in [0,2\pi)) \ .
\end{align}
From \eqref{Lorenz}, the Lorentz generator $\mathcal{O}_{+-}$ is represented as a derivative with respect to $\theta$:
\begin{align}
\mathcal{O}_{+-} = \frac{i}{4}\frac{\partial}{\partial\theta} \ .
\end{align}
Here, the superscripts $\pm$ are defined by $V^{\pm} = V^1 \pm iV^2$ for a contravariant vector $V^a$,  
so that the charges of $V^{\pm}$ are $s = \pm 1$, respectively.


We first consider the two-dimensional torus $T^2$.  
To cover the entire $T^2$, four coordinate patches are needed, which we denote by $(z^{[i]},\bar{z}^{[i]})$ ($i=1,\ldots,4$).  
The metric is given by
\begin{align}
g = \frac{1}{2}\,\mathrm{d}z^{[i]}\!\otimes\! \mathrm{d}\bar{z}^{[i]} + \frac{1}{2}\,\mathrm{d}\bar{z}^{[i]}\!\otimes\! \mathrm{d}z^{[i]} \ .
\end{align}
Since the spin connection on $T^2$ vanishes, the corresponding covariant derivatives are given by
\begin{align}
\nabla^{[i]}_{(+)} &= \frac{1}{2}\, e^{2i\theta^{[i]}}\partial_{z^{[i]}} \ , \\
\nabla^{[i]}_{(-)} &= \frac{1}{2}\, e^{-2i\theta^{[i]}}\partial_{\bar{z}^{[i]}} \ .
\end{align}


Next, we consider the two-sphere $S^2$ with a uniform and isotropic metric.  
To cover the entire $S^2$, two coordinate patches are sufficient.  
Let $(z,\bar{z})$ be the stereographic coordinates projected from the north pole.  
The metric is given by
\begin{align}
g = \frac{2}{(1+|z|^2)^2}\,\mathrm{d}z\otimes \mathrm{d}\bar{z} + \frac{2}{(1+|z|^2)^2}\,\mathrm{d}\bar{z}\otimes \mathrm{d}z \ .
\label{S^2 metric}
\end{align}
We adopt the following zweibein:
\begin{align}
e^1 = \frac{\mathrm{d}z + \mathrm{d}\bar{z}}{1 + |z|^2} \ , 
\qquad
e^2 = -\,\frac{i(\mathrm{d}z - \mathrm{d}\bar{z})}{1 + |z|^2} \ .
\label{S^2 zweibein}
\end{align}
Its inverse is given by
\begin{align}
e_1 = \frac{1}{2}(1+|z|^2)(\partial_z + \partial_{\bar{z}}) \ ,
\qquad
e_2 = \frac{i}{2}(1+|z|^2)(\partial_z - \partial_{\bar{z}}) \ .
\label{S^2 inverse of zweibein}
\end{align}
From the torsion-free condition
\begin{align}
\mathrm{d}e^a + \Omega^a_{\;b} \wedge e^b = 0 \ ,
\end{align}
the spin connection is determined as
\begin{align}
\Omega^{12} = -\,i\,\frac{\bar{z}\,\mathrm{d}z - z\,\mathrm{d}\bar{z}}{1 + |z|^2} \ .
\label{S^2 spin connection}
\end{align}

Then, the covariant derivatives $\nabla_{(a)}$ are obtained as
\begin{align}
\nabla^{[z]}_{(+)} &= 
\frac{1}{2} e^{2i\theta}
\left[
(1+|z|^2)\,\partial_z - \frac{i}{2}\bar{z}\,\partial_\theta
\right] \ ,
\notag\\[4pt]
\nabla^{[z]}_{(-)} &= 
\frac{1}{2} e^{-2i\theta}
\left[
(1+|z|^2)\,\partial_{\bar{z}} + \frac{i}{2}z\,\partial_\theta
\right] \ .
\end{align}
It can be shown that the following commutation relations hold:
\begin{align}
[\nabla_{(+)}, \nabla_{(-)}] &= \frac{i}{4}\frac{\partial}{\partial\theta} = \mathcal{O}_{+-} \ \label{eq2.37},\\
[\mathcal{O}_{+-}, \nabla_{(\pm)}] &= \mp \frac{1}{2}\nabla_{(\pm)} \ \label{eq2.38}.
\end{align}
Therefore, $2i\nabla_{(\pm)}$ and $-2O_{+-}$ form the $\mathfrak{su}(2)$ Lie algebra.

\subsection{Berezin-Toeplitz quantization} \label{sec2.2}

A field can be described as a section of a vector bundle.  
In this subsection, we review the BT quantization \cite{Adachi:2021ljw,Adachi:2021aux,Adachi:2022mln}, which provides a method for regularizing a section of a vector bundle as a finite-size matrix.  
The finite-size matrices obtained through the 
BT quantization are called Toeplitz operators.  
In the large-matrix limit, Toeplitz operators exhibit desirable regularization properties.  
Here, we review the BT quantization of a section of a vector bundle on a closed connected Kähler $2n$-dimensional manifold \cite{Adachi:2022mln}.

Let $(g, J, \omega)$ be the Kähler structure on a closed
connected Kähler $2n$-dimensional manifold $M$,  
where $g$ is the Riemannian metric, $J$ is the complex structure, and $\omega$ is the symplectic form satisfying
\begin{align}
\omega(\cdot, \cdot) = g(J\cdot, \cdot) \ .
\end{align}

The Kähler potential $K$ is a locally defined function satisfying $\omega = i\partial\bar{\partial}K$,  
where $\partial$ and $\bar{\partial}$ denote the Dolbeault differential operators.  
The volume form is given by $\mu = \omega^n / n!$.  
In local real coordinates $\{x^\mu\}_{\mu=1,\ldots,2n}$, this can be expressed as $\mu = \sqrt{g}\, d^{2n}x = \sqrt{g}\, \mathrm{d}x^1 \wedge \cdots \wedge \mathrm{d}x^{2n}$.  
For a general vector bundle $F$, its connection and curvature are denoted respectively by $A^F$ and $R^F = \mathrm{d}A^F + A^F \wedge A^F$,  
where $A^F$ is the connection one-form of $F$.
We denote the space of smooth sections of $F$ by $\Gamma(F)$.

Let $E$ be an Hermitian vector bundle with finite rank over $M$.
Using the BT quantization, a section of $E$ is regularized as a finite-size matrix
in the following.
$E$ is expressed in terms of 
two finite-rank Hermitian 
vector bundles $E_1$ and $E_2$ over $M$
as the Hom-bundle 
$\mathrm{Hom}(E_2, E_1)$, 
whose fiber at a point $x \in M$ consists of linear maps from the fiber of $E_2$ at $x$ to that of $E_1$.  
A section of $E_i$ is given locally by $\nabla^{E_i}=\mathrm{d}+A^{E_i}$ where $A^{E_i}$ is 
the connection 1-form of $E_i$.

In order to quantize a section $\chi \in \Gamma(\mathrm{Hom}(E_2, E_1))$ by the BT quantization,  
we consider the twisted spinor bundles $\Gamma(S_c \otimes L^{\otimes p} \otimes E_i)$ ($i=1,2$).  
Here, $p$ is an integer and $L$ is a holomorphic line bundle equipped with a connection whose curvature is proportional to the symplectic form.  
$S_c$ denotes the spin-$c$ bundle, and its local expression for the connection is given by
\begin{align}
\nabla^{S_c} = \mathrm{d} + \frac{1}{4}\,\Omega_{(2n)}^{ab}\gamma_{(2n)a}\gamma_{(2n)b}
 - \frac{1}{2}\sum_{m=1}^n \Omega_{m \bar{m}}\ .
\end{align}
Here, $\{\gamma_{(2n)a}\}_{a=1,\ldots,2n}$ are the $2n$-dimensional gamma matrices satisfying  
$\{\gamma_{(2n)a}, \gamma_{(2n)b}\} = 2\delta_{ab} I_{2^n}$, where $I_{2^n}$ is the $2^n\times 2^n$ identity matrix.  
$\Omega_{ab}$ is the spin connection, and for a local orthonormal frame $\{V_m, \bar{V}_m\}$ ($m=1,\ldots,n$) defined by
\begin{align}
V^m = \frac{1}{\sqrt{2}}\left(V^{2m-1} + iV^{2m}\right), 
\qquad 
V^{\bar{m}} = \frac{1}{\sqrt{2}}\left(V^{2m-1} - iV^{2m}\right),
\end{align}
we introduce the local complex structure.  
The line bundle $L$ is a holomorphic line bundle whose curvature $R^L$ is proportional to the symplectic form:
\begin{align}
R^L = -i k \omega.
\label{eq2.41}
\end{align}

The integer $k$ is chosen so that, for any two-cycle $\Sigma \subset M$,
\begin{align}
\frac{i}{2\pi}\int_{\Sigma} R^L \in \mathbb{Z}.
\end{align}
This condition is equivalent to requiring that the symplectic form $\frac{k}{2\pi}\omega$ belongs to the second integral cohomology class $H^2(M,\mathbb{Z})$.  
Manifolds satisfying this condition and admitting such a holomorphic line bundle $L$ are called quantizable manifolds.  
For the two-dimensional case $M=\Sigma$, this quantization condition reads $k=2\pi/\!\int_M\omega$.  

If the Kähler potential $K$ is given locally, the connection on $L$ can be written as
\begin{align}
\nabla^L = \mathrm{d} + A^L \ , 
\qquad 
A^L = -\frac{k}{2}(\partial - \bar{\partial})K \ .
\end{align}
A section $\chi \in \Gamma(\mathrm{Hom}(E_2,E_1))$ can be viewed as linear map 
$\Gamma(S_c \otimes L^{\otimes p} \otimes E_2)
   \longrightarrow 
   \Gamma(S_c \otimes L^{\otimes p} \otimes E_1)$.
Here, at each point $x\in M$, the Dirac operator acts on the fibers of $S_c \otimes L^{\otimes p}$ as the standard spin-$c$ Dirac operator.  
The inner product on $\Gamma(S_c \otimes L^{\otimes p} \otimes E_i)$ ($i=1,2$) is defined by
\begin{align}
(\psi',\psi)
 = \int_M \mu\,(\psi')^{\dagger} \cdot \psi \ ,
\end{align}
where $(\psi')^{\dagger} \cdot \psi$ denotes the fiberwise Hermitian inner product,  
and the norm is defined as $\|\psi\| = \sqrt{(\psi,\psi)}$.

We now define the Dirac operator $D^{(E_i)}$ acting on $\Gamma(S_c \otimes L^{\otimes p} \otimes E_i)$ ($i=1,2$) as
\begin{align}
D^{(E_i)} 
&= i \gamma_{(2n)}^a \nabla_a^{S_c \otimes L^{\otimes p} \otimes E_i}
\notag\\
&= i \gamma_{(2n)}^a e_a^{\;\;\mu}
\left(
\partial_\mu + \frac{1}{4}\,\Omega_{bc \mu}\gamma_{(2n)}^b \gamma_{(2n)}^c
 - \frac{1}{2}\sum_{m=1}^n \Omega_{m \bar{m} \mu}
 + p A^L_\mu + A^{E_i}_\mu
\right) \ .
\end{align}
To make this expression concrete, we specify the representation of the gamma matrices.  
We define the chirality matrix by
\begin{align}
\gamma_{(2n+1)} = (-i)^n \gamma_{(2n)}^1 \gamma_{(2n)}^2 \cdots \gamma_{(2n)}^{2n}
\ .
\end{align}
This matrix is Hermitian and satisfies $\{\gamma_{(2n)}^a, \gamma_{(2n+1)}\} = 0$ ($a = 1,\ldots,2n$) and $(\gamma_{(2n+1)})^2 = I_{2^n}$.  
Let $\sigma^a$ ($a=1,2,3$) denote the Pauli matrices and $\otimes$ the Kronecker product.  
Then, in the Weyl representation, the gamma matrices can be written as
    \begin{align}
      & \gamma_{(2)}^1 = \sigma^1, \quad \gamma_{(2)}^2 = \sigma^2 \ , \\
      & \gamma_{(2n + 2)}^i = \sigma^2 \otimes \gamma_{(2n)}^i \quad (i = 1 , \ldots, 2n) \ , \\
      & \gamma_{(2n + 2)}^{2n + 1} = \sigma^2 \otimes \gamma_{(2n)} \ , \\
      & \gamma_{(2n + 2)}^{2n + 2} = -\sigma^1 \otimes I_{2^n} \ . 
    \end{align}
The chirality matrix $\gamma_{(2n+1)} = \sigma^3 \otimes I_{2^{n-1}}$ anticommutes with the Dirac operator $D^{(E_i)}$.  
Hence, $D^{(E_i)}$ can be decomposed as
\begin{align}
D^{(E_i)} =
\begin{pmatrix}
0 & D^{(E_i)}_- \\
D^{(E_i)}_+ & 0
\end{pmatrix} \ .
\end{align}
Here, $D^{(E_i)}_\pm$ act on the subspaces of positive and negative chirality, respectively.  
Let $|+\rangle$ denote the eigenstate of $\sigma^3$ with eigenvalue $+1$,  
and consider the positive-chirality subspace spanned by $|+\rangle \otimes \mathbb{C}^{2^{n-1}}$.  
As shown in \cite{Adachi:2022mln}, for sufficiently large $p$, the normalizable 
zero modes of $D^{(E_i)}$ are proportional 
to $|+\rangle^{\otimes n}$.  
This means that they have positive chirality so that $\mathrm{Ker}\,D^{(E_i)}_-$ becomes trivial, $\mathrm{Ker}\,D^{(E_i)}_- = \{0\}$, which is known as the vanishing theorem.   
By the Atiyah-Singer index theorem, for large $p$, the number of normalizable zero modes of $D^{(E_i)}$ is given by
\begin{align}
\dim \mathrm{Ker}\, D^{(E_i)} 
= \int_M \mathrm{Td}(T^{(1,0)}M) \wedge \mathrm{ch}(L^{\otimes p} \otimes E_i) \ ,
\label{dimKer}
\end{align}
where $\mathrm{Td}$ and $\mathrm{ch}$ denote the Todd class and the Chern 
character, respectively,  
and $T^{(1,0)}M$ is the holomorphic tangent bundle of $M$.  
Expanding Eq. \eqref{dimKer}, we obtain
    \begin{equation}
      \mathrm{dimKer}D^{(E_i)}
      = \frac{\mathrm{rank}(E_i)}{(2\pi \hbar_p)^n} \int_M \mu + O(p^{n-1})\ , 
    \end{equation}
where 
    \begin{equation}
      \hbar_p = \frac{1}{k p} \ .
    \end{equation}
Thus, the dimension of $\mathrm{Ker}\, D^{(E_i)}$ grows as $p^n$ for large $p$.

Let $\Pi^{(E_i)}$ denote the projection from $\Gamma(S_c \otimes L^{\otimes p} \otimes E_i)$ onto $\mathrm{Ker} D^{(E_i)}$.  
For a section $\chi \in \Gamma(\mathrm{Hom}(E_2,E_1))$, the Toeplitz operator associated with $\chi$ is defined as the map on the zero-mode space of the Dirac operator:
\begin{align}
T^{(E_1,E_2)}(\chi) = \Pi^{(E_1)} \chi\, \Pi^{(E_2)} \ .
\end{align}
That is, $T^{(E_1,E_2)}(\chi)$ acts as a linear operator from the finite-dimensional vector space $\mathrm{Ker} D^{(E_2)}$ to $\mathrm{Ker} D^{(E_1)}$.  
Let $\{\psi_I^{(E_i)}\}_{I=1,\ldots,\dim\mathrm{Ker} D^{(E_i)}}$ be an orthonormal basis of $\mathrm{Ker} D^{(E_i)}$.  
Then, the matrix elements of the Toeplitz operator are given by
\begin{align}
T^{(E_1,E_2)}(\chi)_{IJ} 
= (\psi_I^{(E_1)},\, \chi \psi_J^{(E_2)}) 
= \int_M \mu\, (\psi_I^{(E_1)})^\dagger \cdot \chi\, \psi_J^{(E_2)} \ .
\end{align}
Since $\dim\mathrm{Ker} D^{(E_i)}$ is finite, $T^{(E_1,E_2)}(\chi)$ is represented by a finite-size matrix.

The Toeplitz operator has an asymptotic expansion in the large-$p$ (semiclassical) limit.   
Then the product of two Toeplitz operators admits the following expansion:
\begin{align}
T^{(E_1,E_2)}(\chi)\, T^{(E_2,E_3)}(\chi')
 = \sum_{i=0}^{\infty} \hbar_p^i\, T^{(E_1,E_3)}\!\left(C_i(\chi,\chi')\right) \ , \label{eq2.9}
\end{align}
where $C_i$ are bidifferential operators mapping
\begin{align}
C_i : \Gamma(\mathrm{Hom}(E_2,E_1)) \times \Gamma(\mathrm{Hom}(E_3,E_2))
 \longrightarrow \Gamma(\mathrm{Hom}(E_3,E_1)) \ .
\end{align}
The first few coefficients take the explicit forms
\begin{align}
C_0(\chi,\chi') &= \chi\,\chi'\ , 
\notag\\
C_1(\chi,\chi') &= -\frac{1}{2}\, G^{\alpha\beta}(\nabla_\alpha \chi)(\nabla_\beta \chi')\ ,
\notag\\
C_2(\chi,\chi') &= \frac{1}{8} G^{\alpha\beta}G^{\gamma\delta}
\left[
(\nabla_\alpha \chi)\!\left(i R_{\beta\gamma\mu\nu}W^{\mu\nu} - 2R^{E_2}_{\beta\gamma}\right)
(\nabla_\delta \chi') 
+ (\nabla_\alpha \nabla_\gamma \chi)(\nabla_\beta \nabla_\delta \chi')
\right] \ . \label{eq2.10}
\end{align}
Here, $G^{\alpha\beta} = g^{\alpha\beta} + iW^{\alpha\beta}$, where $g^{\alpha\beta}$ is the inverse metric and $W^{\alpha\beta}$ is defined by $\omega_{\mu\nu}W^{\mu\rho} = \delta^\rho_\nu$,  
so that $W^{\mu\nu}$ represents the Poisson tensor.  
$R_{\alpha\beta\gamma\delta}$ is the Riemann curvature tensor associated with the Levi-Civita connection,  
and $R^{E_i}_{\alpha\beta}$ is the curvature of the bundle $E_i$.  
$\nabla_\alpha$ denotes the covariant derivative acting on $\chi$.
For instance, for $\chi \in \Gamma(\mathrm{Hom}(E_2,E_1))$, 
\begin{align}
\nabla_\alpha \chi = \partial_\alpha \chi + A^{E_1}_\alpha \chi - \chi A^{E_2}_\alpha \ .
\end{align}
From \eqref{eq2.9} and \eqref{eq2.10}, we obtain the semiclassical correspondence:
\begin{align}
\lim_{p\to\infty} 
\big\| T^{(E_1,E_2)}(\chi)\,T^{(E_2,E_3)}(\chi') - T^{(E_1,E_3)}(\chi\chi') \big\| = 0 \ , \label{eq2.11}\\
\lim_{p\to\infty} 
\big\| i\hbar_p^{-1}\!\left[ T(f\bm{1}), T^{(E_1,E_2)}(\chi) \right]
- T^{(E_1,E_2)}(\{f,\chi\}) \big\| = 0 \ \label{eq2.12},
\end{align}
where $f \in C^{\infty}(M)$ and $\left[ T(f\bm{1}), T^{(E_1,E_2)}(\chi) \right]$ is the commutator-like operation defined by
\begin{equation}
  \left[ T(f\bm{1}), T^{(E_1,E_2)}(\chi) \right]
  = T^{(E_1,E_1)}(f\bm{1}_{E_1}) T^{(E_1,E_2)}(\chi)-T^{(E_1,E_2)}(\chi) T^{(E_2,E_2)}(f\bm{1}_{E_2})\ ,\label{eq2.36}
\end{equation}
and $\{f,\chi\}$ is the 
covariantized Poisson bracket defined by
\begin{align}
\{f,\chi\} = W^{\alpha\beta} (\partial_\alpha f)(\nabla_\beta \chi) \ .
\end{align}

Finally, we show that the Dirac zero-mode equation reduces to a simpler differential equation for a holomorphic section.  
As mentioned earlier, for sufficiently large $p$, a normalizable zero mode $\psi$ of the Dirac operator $D$ acting on $\Gamma(S_c \otimes L^{\otimes p} \otimes E)$ is expressed as
\begin{align}
\psi = f\, |+\rangle^{\otimes n}, 
\qquad f \in \Gamma(L^{\otimes p} \otimes E) \ . 
\label{zero mode}
\end{align}
The spin-connection term $\Omega_{ab}\gamma_{(2n)}^a \gamma_{(2n)}^b$ is computed as
\begin{align}
\Omega_{ab}\gamma_{(2n)}^a \gamma_{(2n)}^b
= \Omega_{m\bar{l}}\gamma_{(2n)}^m \gamma_{(2n)}^{\bar{l}}
 + \Omega_{\bar{l}m}\gamma_{(2n)}^{\bar{l}} \gamma_{(2n)}^m
= 2\Omega_{m\bar{l}}\gamma_{(2n)}^m \gamma_{(2n)}^{\bar{l}}
 - 2\sum_{m=1}^{n}\Omega_{m\bar{m}} \ .
\label{spin connection}
\end{align}
In the first equality we used $\Omega_{ml}=\Omega_{\bar{m}\bar{l}}=0$\footnote{
The inverse of the vielbein, $e_m$ and $e_{\bar{m}}$ $(m=1,\ldots,n)$,
satisfy $Je_m=ie_m$ and $Je_{\bar{m}}=-ie_{\bar{m}}$.},
and in the second equality, $\Omega_{m\bar{l}}=-\Omega_{\bar{l}m}$ and 
$\{\gamma_{(2n)}^m, \gamma_{(2n)}^{\bar{l}}\}=2\delta_{m\bar{l}}I_{2^n}$.  
Using \eqref{zero mode} and \eqref{spin connection} together with $\gamma_{(2n)}^m |+\rangle^{\otimes n}=0$,\\
$\gamma_{(2n)}^m\gamma_{(2n)}^{\bar{l}} |+\rangle^{\otimes n} = 2\delta_{m\bar{l}} |+\rangle^{\otimes n}$, and $e^\mu_{\;\bar{m}}=0$, we obtain
\begin{align}
D\psi 
= i\, e^{\bar{\mu}}_{\;\bar{m}} \gamma_{(2n)}^{\bar{m}} |+\rangle^{\otimes n}
\left(\partial_{\bar{\mu}} + p A^L_{\bar{\mu}} + A^E_{\bar{\mu}}\right) f \ .
\end{align}
Since a complex number $c_m$ satisfying $c_m\gamma_{(2n)}^{\bar{m}} |+\rangle^{\otimes n}=0$ must vanish, $c_m=0$,  
the condition $D\psi=0$ implies
\begin{align}
&\forall m\in\{1,\ldots,n\}:\quad
e^{\bar{\mu}}_{\;\bar{m}}
\left(\partial_{\bar{\mu}} + p A^L_{\bar{\mu}} + A^E_{\bar{\mu}}\right) f = 0 \nonumber \\
&\Rightarrow\quad
\left(\partial_{\bar{\mu}} + p A^L_{\bar{\mu}} + A^E_{\bar{\mu}}\right) f = 0 \ .
\end{align}
Thus, the Dirac zero-mode equation $D\psi=0$ reduces to the following differential equation for $f \in \Gamma(L^{\otimes p}\otimes E)$:
\begin{align}
\left(\partial_{\bar{\mu}} + p A^L_{\bar{\mu}} + A^E_{\bar{\mu}}\right) f = 0 \ .
\label{eq2.48}
\end{align}
Hence, the zero mode of the Dirac operator is given by $\psi = f\,|+\rangle^{\otimes n}$, where $f$ satisfies \eqref{eq2.48}.

\section{Regularization of covariant derivatives as finite-size matrices}

In section~2.1, within the covariant derivative interpretation of matrix models,  
we have seen that the matrices $A_{(a)}$ correspond to the covariant derivatives $-i\nabla_{(a)}$.
Since $\nabla_{(a)}$ are differential operators, the matrices $A_{(a)}$ are infinite-dimensional.  
To investigate quantum corrections, however,  
we must regularize them as finite-size matrices.  
In particular, to interpret the results of numerical simulations of the matrix model,
a regularization of the covariant derivatives $\nabla_{(a)}$ by finite-size 
matrices is necessary.  

In this section, we construct such a 
finite-matrix regularization 
based on the Berezin-Toeplitz quantization reviewed in 
section~2.3.  
We assume that the underlying manifold $M$ is a closed connected 
Kähler $2n$-dimensional manifold.  
In sections~3.1 and 3.2, we provide finite-size matrices acting on rectangular matrices and square matrices as the finite-matrix regularization of $\nabla_{(a)}$, respectively.   
In section~3.3, we show that for $n=1$, our construction reproduces the results 
obtained in \cite{Hattori:2024btt}. In this section, we denote $\mathrm{Spin}(2n)$ or $\mathrm{Spin}^c(2n)$ by $G$ 
. 

\subsection{Case of rectangular Toeplitz operator 
for $\varphi$} \label{sec3.1}
In this subsection, we consider the case in which the Toeplitz operator for $\varphi$ is a rectangular matrix. 

Let $M$ be a closed connected K\"{a}hler $2n$-dimensional 
manifold.  
As in section~2.1, we consider the fiber bundle $E_{\mathrm{reg}}$ 
over $M$,  
whose fiber is the representation space of the regular 
representation of $G$. 
Let $\Gamma(E_{\text{reg}})$ denote the space of global smooth sections of $E_{\text{reg}}$.

The covariant derivative $\nabla_{(a)}$ ($a=1,\ldots,2n$) introduced in 
section 2.1 acts on the elements of 
$\Gamma(E_{\mathrm{reg}})$. Hence, the finite-size matrices obtained as regularization of $\nabla_{(a)}$
is considered to act on the finite-size matrices 
obtained as 
regularization of elements of $\Gamma(E_{\mathrm{reg}})$.
The latter matrices are given by Toeplitz operators
in the BT quantization.
We denote the Toeplitz operator for 
$\varphi \in \Gamma(E_{\mathrm{reg}})$ by $T(\varphi)$, and
denote the finite-size matrices
for $-i\nabla_{(a)}$ acting 
on $T(\varphi)$ by $\mathcal{P}_{(a)}$.
In the following, we will show that 
$\mathcal{P}_{(a)}$ is given by
\begin{align}
\mathcal{P}_{(a)}T(\varphi)
  = \hbar_p^{-1}\, T(\partial_{(a)}X^A)[
    T(X^A),T(\varphi)]
   - \frac{1}{2}\hbar_p^{-1}\,[T(\partial_{(a)}X^A), T(X^A)]\,T(\varphi) \ .
   \label{eq3.1}
\end{align}
Here, $\hbar_p = \frac{1}{kp}$ with $k$ a certain constant is the 
semiclassical parameter associated with the BT quantization,  
and $X^A$ ($A=1,\ldots,D$) are the embedding coordinates for $M$ 
embedded
isotopically in $D$-dimensional Euclidean space $\mathbb{R}^D$ so that 
$\partial_a X^A \partial_bX^A=\delta_{ab}$ where $\partial_a=e_a^{\mu}\partial_\mu$\footnote{Existence of such an embedding is 
guaranteed by Nash's embedding theorem.}.
$\partial_{(a)} X^A$ is defined by $R^{\langle v \rangle}_{(a)b}
(g^{-1}) e_b^{\mu} \partial_\mu X^A$ where
$R^{\langle v \rangle}_{(a)b}(g^{-1})$ is the representation 
matrix of the 
vector representation for $g^{-1} \in G$.
$T(X^A)$ and $T(\partial_{(a)} X^A)$ denote
the Toeplitz operators for
$X^A$ and $\partial_{(a)} X^A$, respectively.
As seen below, since $T(\varphi)$ is a rectangular matrix, 
$[T(X^A), T(\varphi)]$ is not an ordinary commutator but defined 
by 
\begin{align}
[T(X^A), T(\varphi)] = T(X^A) T(\varphi) - T(\varphi) T'(X^A) \ ,
\end{align}
where $T(X^A)$ and $T'(X^A)$ are both
Toeplitz operator for $X^A$ but square 
matrices with different matrix sizes.
$\mathcal{P}_{(a)}$ defined 
in \eqref{eq3.1}
maps a rectangular matrix
$T(\varphi)$ to another rectangular matrix 
$\mathcal{P}_{(a)} T(\varphi)$ with the same matrix size, so that
it can be viewed as a square matrix.

We will first construct 
$T(\varphi)$ and $T(-i \nabla_{(a)} \varphi)$ in section 3.1.1, and $T(X^A)$, $T'(X^A)$ and $T(\partial_{(a)} X^A)$ in section 3.1.2.
Next, in section 3.1.3, we will show 
\begin{align}
    \mathcal{P}_{(a)} T(\varphi)=T(-i \nabla_{(a)} \varphi) + O(1/p)\ . \label{pa}
\end{align}
This implies that $\mathcal{P}_{(a)} T(\varphi)$ agree with $T(-i \nabla_{(a)} \varphi)$ in the large-$p$ limit. 
We will  also show that $\mathcal{P}_{(a)}$ is Hermitian for finite $p$.
From these, we conclude that $\mathcal{P}_{(a)}$ correctly represents
the matrix regularization of 
$-i \nabla_{(a)}$. 
We will finally see how $[\mathcal{P}_{(a)}, \mathcal{P}_{(b)}]$ behaves
for large $p$.

\subsubsection{$T(\varphi)$ and $T(-i \nabla_{(a)} \varphi)$}

We first consider the Toeplitz operator $T(\varphi)$
for $\varphi  \in \Gamma(E_{\mathrm{reg}})$.
Since the regular representation
is infinite dimensional, we cannot directly 
apply the BT quantization to $\varphi$.
Thus, we expand $\varphi$ using the Peter-Weyl theorem.
The expansion coefficients 
are fields with finite-dimensional
irreducible representation
of $G$, to which
we can apply the BT quantization.
Then, we construct
$T(\varphi)$ such that its matrix elements are
given by 
the Toeplitz operators for the expansion coefficients.

From the Peter-Weyl theorem,  
$\varphi(x, g) \  (x \in M, \ g\in G)$ 
can be expanded in a similar way to \eqref{Peter-Weyl}:
    \begin{equation}
      \varphi(x, g)
      =\sum_{r: \mathrm{irr.}} \varphi^{\langle r \rangle}_{i(j)}(x) \sqrt{d_r} R^{\langle r^* \rangle}_{i(j)}(g) \ . 
      \label{eq3.2}
    \end{equation}
Here, the sums over repeated indices are implicitly taken.
The index $i$ in $\varphi^{\langle r \rangle}_{i(j)}(x)$
transforms under $r$ representation of
$G$, while
the index $(j)$ not\footnote{Unlike section \ref{sec2.1}, parentheses in $(j)$
are put on to emphasize that the index $(j)$ does not transform under $G$.}.
Thus, for fixed $r$ and $(j)$, 
$\varphi^{\langle r \rangle}_{i(j)}(x)$ behaves as a field with $r$ representation.

Let $E_r$ and $E_{\mathrm{trivial}}$ be fiber bundles whose fibers are
the spaces of $r$ representation and the trivial representation, respectively.
Then, $\varphi^{\langle r \rangle}_{i(j)}(x)$
can be viewed as a section of Hom-bundle $\mathrm{Hom}(E_{\mathrm{trivial}}, E_{r})$.
$\varphi^{\langle r \rangle}_{i(j)}(x)$ can therefore be regularized
to finite-size matrices by applying the BT quantization.
We denote
the Toeplitz operators for 
$\varphi^{\langle r \rangle}_{i(j)}(x)$ 
by $T^{(r, 1)} \left(\varphi^{\langle r \rangle}_{(j)}\right)$, where
$1$ stands for the trivial representation\footnote{We adopt this 
notation for simplicity, although we should use $T^{(E_r, E_{\mathrm{trivial}})} \left(\varphi^{\langle r \rangle}_{(j)}\right)$ rather than $T^{(r, 1)} \left(\varphi^{\langle r \rangle}_{(j)}\right)$
if we follow the notation in 
\ref{sec2.2} faithfully. }.
Namely, we have
    \begin{equation}
      T^{(r, 1)} \left(\varphi^{\langle r \rangle}_{(j)}\right)_{IJ}
      = \int_M \mu \  \left(\psi^{\langle r \rangle}_{I; i}(x) \right)^{\dagger} \varphi^{\langle r \rangle}_{i(j)}(x) \ \psi^{\langle 1 \rangle}_{J}(x) \ .
      \label{eq3.3}
    \end{equation}
Here, $\left\{\psi^{\langle r \rangle}_{I; i}(x)\right\}_{I=1, \ldots, \mathrm{dimKer}D^{(E_{r})}}$ is
an orthonormal basis in the space of normalizable zero modes for the Dirac operator $D^{(E_r)}$ denoted by $\mathrm{Ker} D^{(E_{r})}$. 
The index $I = 1, \ldots, \mathrm{dimKer}D^{(E_{r})}$ labels
the Dirac zero modes.
The index $i = 1, \ldots, d_r$ for
the fiber of $E_r$ transforms under $r$ representation
of $G$. 
Similarly, $\{\psi^{\langle 1 \rangle}_{J}(x)\}_{J=1, \ldots, \mathrm{dimKer}D^{(E_{\mathrm{trivial}})}}$
is an orthonormal basis in $\mathrm{Ker} D^{(E_{\mathrm{trivial}})}$.
The index $J=1, \ldots, \mathrm{dimKer}D^{(E_{\mathrm{trivial}})}$ labels the Dirac zero modes.
Since the fiber of $E_{\mathrm{trivial}}$ is the space of the trivial representation of
$G$, 
$\psi^{\langle 1 \rangle}_{J}(x)$ does not possess the index for a
fiber.
Table \ref{tab1} summarizes what each index stands for.

Note that in \eqref{eq3.3} the sum is taken over the 
repeated index $i$.
For this reason, the index $i$ is absent in $\varphi^{\langle r \rangle}_{i(j)}$ appearing in
the LHS of \eqref{eq3.3}.
The integrand in the RHS of \eqref{eq3.3} is invariant under $G$:
\begin{align}
\left(R^{\langle r \rangle}_{ii'}(h) \psi^{\langle r \rangle}_{I; i'}(x) \right)^{\dagger} R^{\langle r \rangle}_{ii''}(h) 
\varphi^{\langle r \rangle}_{i''(j)}(x) \psi^{\langle 1 \rangle}_{J}(x) = \left(\psi^{\langle r \rangle}_{I; i}(x) 
\right)^{\dagger} \varphi^{\langle r \rangle}_{i(j)}(x) \ \psi^{\langle 1 \rangle}_{J}(x) \ ,
\end{align}
where $h \in G$, and
the fact that the representation matrix $R$ for $h \in G$
is unitary is used. 
Hence, the Toeplitz operators \eqref{eq3.3} are 
independent of the choice of coordinate patches.

\begin{figure}[t]
      \centering
      \includegraphics[width=0.6\linewidth]{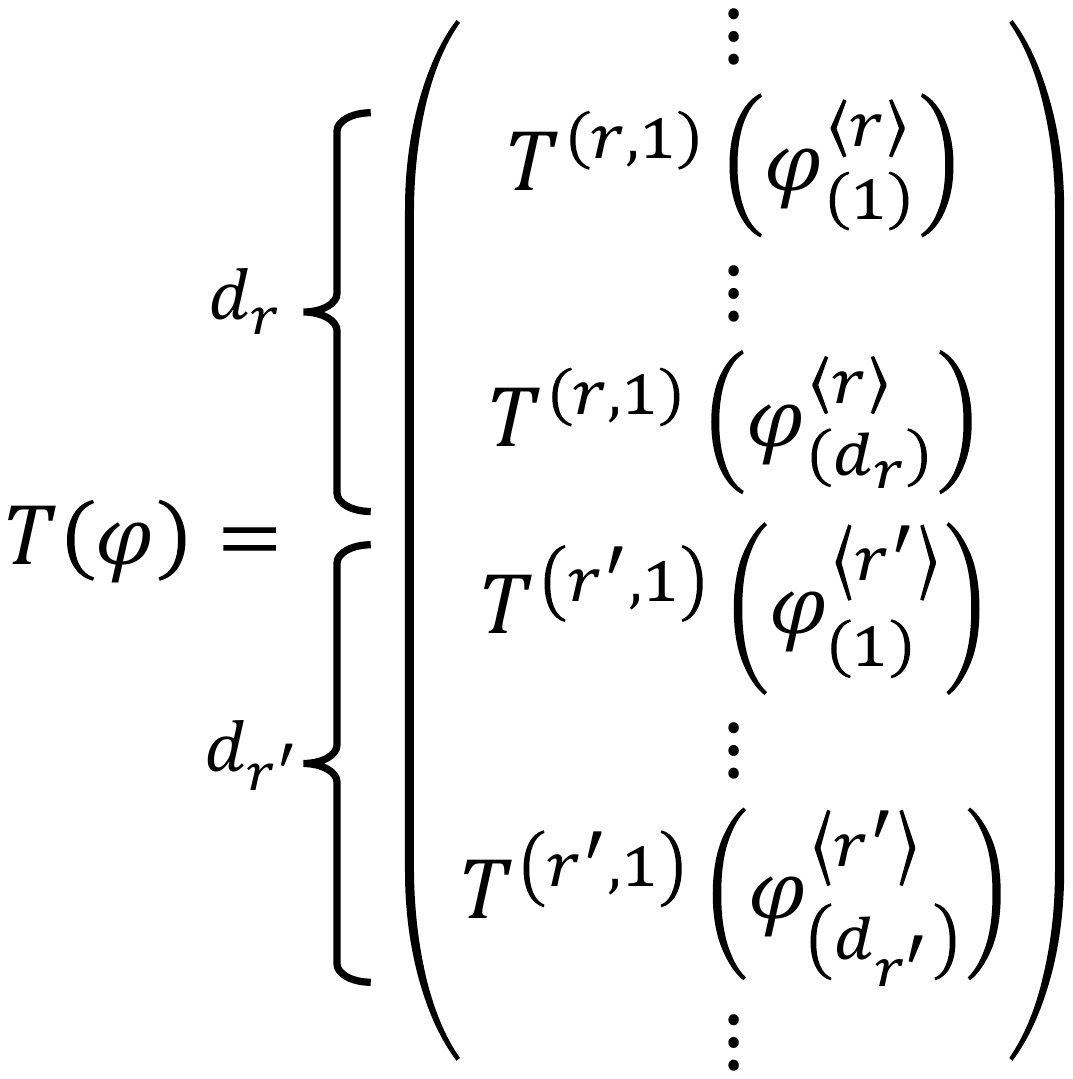}
      \caption{$T(\varphi)$ is constructed such that
its $(r,j)$ block is the Toeplitz operator 
$T^{(r, 1)} \left(\varphi^{\langle r \rangle}_{(j)}\right)$.
 $T^{(r, 1)} \left(\varphi^{\langle r \rangle}_{(j)}\right)$ is a retangular matrix
with the size $\mathrm{dimKer}D^{(E_{r})} \times \mathrm{dimKer}D^{(E_{\mathrm{trivial}})}$. Since $j = 1, \ldots,  d_r$, $d_r$ copies of these rectangular matrices
are arrayed. $T(\varphi)$ is a rectangular matrix with size 
$\left(\sum'_{r:\mbox{irr.}} d_r \mathrm{dimKer}D^{(E_{r})}\right) \times \mathrm{dimKer}D^{(E_{\mathrm{trivial}})}$.}
      \label{fig1}
    \end{figure}

By using the Toeplitz operators $T^{(r, 1)} \left(\varphi^{\langle r \rangle}_{(j)}\right)$,
we construct the Toeplitz operator $T(\varphi)$ for $\varphi$
as follows: 
    \begin{align}
      (\text{the $(I,J)$ component of the $(r,j)$ block in $T(\varphi)$})
      &= T^{(r, 1)} \left(\varphi^{\langle r \rangle}_{(j)}\right)_{IJ} \ ,
      \label{eq3.4}
    \end{align}
where $j=1, \ldots, d_r$, 
$I=1, \ldots, \mathrm{dimKer}D^{(E_{r})}$
and $J=1, \ldots, \mathrm{dimKer}D^{(E_{\mathrm{trivial}})}$. 
The shape of 
$T(\varphi)$ is shown in Fig. \ref{fig1}. 

We should note here that the sum in \eqref{eq3.2} is taken over all irreducible representations $r$
of $G$. 
This implies that
the matrix size of $T(\varphi)$ is infinite although the size of 
each $(r, j)$ block is finite.
We introduce a cutoff $\Xi$ for representations $r$ and restrict 
the sum 
$\sum_{r: \mathrm{irr.}}$ to the one over the irreducible 
representations with their Casimir less
than or equal to $\Xi$\footnote{From a point of view of 
the Kaluza-Klein expansion, 
discarding representations with large Casimir 
corresponds to discarding modes with large mass in the base space. } \footnote{In the case
of two dimensions, $\mathrm{Spin}(2) \cong U(1)$, and the irreducible representations
are specified by integer or half-integer charges $s$.
In this case, the sum $\sum_{s}$ is restricted to the sum over the charges $s$
satisfying
$|s|\leq \Xi$. }.  
We denote this restricted sum by $\sum'_{r:\mbox{irr.}}$.
We finally take the $p \rightarrow \infty$ and $\Xi \rightarrow \infty$ limit 
keeping  $\Xi \ll p$.
Thus, introducing the cutoff to $r$ makes the size of $T(\varphi)$ in
\eqref{eq3.4} finite. 
In particular, $T(\varphi)$ is a rectangular matrix with size 
$\left(\sum'_{r:\mbox{irr.}} d_r \mathrm{dimKer}D^{(E_{r})}\right) \times \mathrm{dimKer}D^{(E_{\mathrm{trivial}})}$. 
 
    \begin{table}[t]
      \centering
      \caption{Summary of indices}
      \begin{tabular}{cc} \hline
        $\langle r \rangle$ &  irreducible rep. of $G$ \\
        &   ($r=1$ corresponds to the trivial rep., while $r=v$ to the vector rep..) \\
         $(j), (l), \ldots$  & indices not transformed under $G$\\
        $i, k, \ldots$  & indices for representations of $G$ 
       (indices of fibers) \\ 
        $I, J, \ldots$ & indices for the Dirac zero mode\\ \hline
      \end{tabular}
      \label{tab1}
    \end{table}

We have seen that  
the Toeplitz operator $T(\varphi)$ is given by \eqref{eq3.4}.
We are now ready for constructing
the Toeplitz operators $T(-i \nabla_{(a)}\varphi)$
for $-i \nabla_{(a)}\varphi \in \Gamma(E_{\text{reg}}) \ (a=1, \ldots, 2n)$.
As seen in section \ref{sec2.1}, 
$\nabla_{(a)} \varphi(x, g) \ (x \in M, \ g \in G)$
is given by
    \begin{equation}
      \nabla_{(a)} \varphi(x, g)
      = R^{\langle v \rangle}_{(a)b}(g^{-1}) e_b^{\mu}(x) \left(\partial_\mu + \frac{1}{2} \Omega^{cd}_{\mu}(x) \mathcal{O}_{cd} + A_\mu(x)\right) \varphi(x, g) \ , \label{eq3.5}
    \end{equation}
where the Lorentz generator $\mathcal{O}_{cd}$ acts on $g$ 
in $\varphi(x, g)$.
By using the Peter-Weyl theorem, we rewrite (\ref{eq3.5}) as 
    \begin{equation}
      \nabla_{(a)} \varphi(x, g)
      = \sum_{r'} \left(\nabla_b \varphi^{\langle r' \rangle}_{k(l)}(x)\right) \sqrt{d_{r'}} R^{\langle v \rangle}_{b(a)}(g) R^{\langle r'^* \rangle}_{k(l)}(g) \ , 
    \end{equation}
where 
\begin{align}
\nabla_b = e_b^{\mu}(x) \left(\partial_\mu + \frac{1}{2} \Omega^{cd}_{\mu}(x) \mathcal{O}_{cd} + A_\mu(x)\right)
\end{align}
with
    \begin{equation}
      \mathcal{O}_{cd} \ \varphi^{\langle r' \rangle}_{k(l)}(x)
      = \left(\mathcal{O}_{cd}^{\langle r' \rangle}\right)_{kk'} \varphi^{\langle r' \rangle}_{k'(l)}(x) \ .
    \end{equation}
Here, $\mathcal{O}_{cd}^{\langle r' \rangle}$ are
the $r'$ representation matrices for the Lorentz generators.
By irreducibly decomposing a tensor representation $v \otimes r'^*$,
$R^{\langle v \rangle}_{b(a)}(g) R^{\langle r'^* \rangle}_{k(l)}(g)$
are represented in terms of the Clebsch-Gordan coefficients 
$C^{v, r'^*; r}_{b(a), k(l); i(j)}$ as  
    \begin{equation}
      R^{\langle v \rangle}_{b(a)}(g) R^{\langle r'^* \rangle}_{k(l)}(g)
      = \sum_{r} C^{v, r'^*; r}_{b(a), k(l); i(j)} R^{\langle r^* \rangle}_{i(j)}(g) \ .
      \label{eq3.9}
    \end{equation}

Thus, we obtain
    \begin{equation}
      \nabla_{(a)} \varphi(x, g)
      = \sum_{r} \left(\sum_{r'} \sqrt{\frac{d_{r'}}{d_r}} C^{v, r'^*; r}_{b(a), k(l); i(j)} \left(\nabla_b \varphi^{\langle r' \rangle}_{k(l)}(x)\right)\right) \sqrt{d_{r}} R^{\langle r^* \rangle}_{i(j)}(g) \ . \label{eq3.10}
    \end{equation}
Since the sums are taken over $b$ and $k$,
$\sum_{r'} \sqrt{\frac{d_{r'}}{d_r}} C^{v, r'^*; r}_{b(a), k(l); i(j)} \left(\nabla_b \varphi^{\langle r' \rangle}_{k(l)}(x)\right)$
is a field with $r$ representation of 
$G$. 

In a similar way as the case of 
$\varphi(x, g)$, we can construct the Toeplitz operators
for $-i \nabla_{(a)} \varphi(x, g)$ with 
$a=1, \ldots, 2n$ as follows:
    \begin{align}
      &(\text{the $(I, J)$ element of the $(r,j)$ block of $T(-i \nabla_{(a)} \varphi)$}) \nonumber\\
      &= T^{(r, 1)} \left(-i \sum_{r'} \sqrt{\frac{d_{r'}}{d_r}} C^{v, r'^*; r}_{b(a), k(l); (j)} \left(\nabla_b \varphi^{\langle r' \rangle}_{k(l)}\right)\right)_{IJ} \ .
      \label{eq3.11}
    \end{align}
Here the RHS represents the Toeplitz operators for $-i \sum_{r'} \sqrt{\frac{d_{r'}}{d_r}} C^{v, r'^*; r}_{b(a), k(l); i(j)} \left(\nabla_b \varphi^{\langle r' \rangle}_{k(l)}(x)\right)$.
Namely, 
    \begin{align}
      &T^{(r, 1)} \left(-i \sum_{r'} \sqrt{\frac{d_{r'}}{d_r}} C^{v, r'^*; r}_{b(a), k(l); (j)} \left(\nabla_b \varphi^{\langle r' \rangle}_{k(l)}(x)\right)\right)_{IJ} \nonumber\\
      &= \int_M \mu \  \left(\psi^{\langle r \rangle}_{I; i}(x) \right)^{\dagger} (-i) \sum_{r'} \sqrt{\frac{d_{r'}}{d_r}} C^{v, r'^*; r}_{b(a), k(l); i(j)} \left(\nabla_b \varphi^{\langle r' \rangle}_{k(l)}(x)\right) \psi^{\langle 1 \rangle}_{J}(x) \ .
      \label{eq3.12}
    \end{align}
Note that the Clebsch-Gordan coefficients in 
the LHS of \eqref{eq3.12} do not have the index $i$, since in the RHS
the index $i$ is contracted with that of the Dirac zero modes. 

\subsubsection{$T(X^A)$, $T'(X^A)$ and $T(\partial_{(a)} X^A)$}
 
Next, we construct the Toeplitz operators for the embedding functions
$X^A \ (A=1, \ldots, D)$ and their derivatives $\partial_{(a)} X^A \ (a=1, \ldots, 2n, \ A=1, \ldots, D)$.
They are defined such that 
$\mathcal{P}_{(a)} T(\varphi)$ defined in \eqref{eq3.1}
agree with $T(-i \nabla_{(a)} \varphi)$ in the large-$p$ limit.

\begin{figure}[t]
      \centering
      \includegraphics{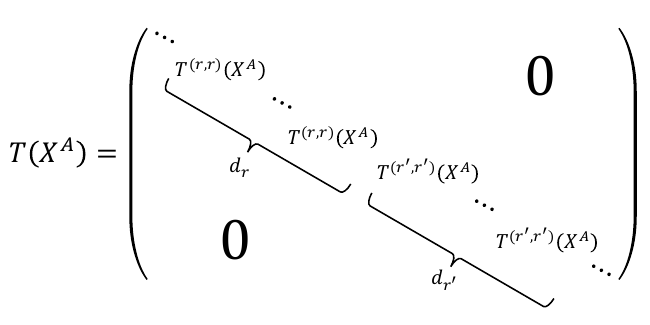}
      \caption{$T(X^A)$ are matrices whose $((r,j),(r,j))$ block ($j=1,\ldots,d_r$) is given by the Toeplitz operator 
      $T^{(r, r)} (X^A)$ and whose off-diagonal blocks are zero. $T^{(r, r)} (X^A)$ are square matrices with the size $\mathrm{dimKer}D^{(E_{r})} \times \mathrm{dimKer}D^{(E_{r})}$. 
      Thus, $T(X^A)$ are square matrices with the size $\left(\sum_{r}' d_r \mathrm{dimKer}D^{(E_{r})}\right) \times \left(\sum_{r}' d_r \mathrm{dimKer}D^{(E_{r})}\right)$. }
      \label{fig2}
    \end{figure}

Since $T(\varphi)$ is a rectangular matrix with the size
$\left(\sum_{r}' d_r \mathrm{dimKer}D^{(E_{r})}\right) \times \mathrm{dimKer}D^{(E_{\mathrm{trivial}})}$, 
the Toeplitz operators for $X^A$ should be different
depending on whether they act on $T(\varphi)$ from the left or the right.
$T(X^A)$ and $T'(X^A)$ denote the Toeplitz operators for $X^A$ that act on $T(\varphi)$ from the left and the right, respectively. $T(X^A)$ are square matrices with size $\left(\sum_{r}' d_r \mathrm{dimKer}D^{(E_{r})}\right) \times \left(\sum_{r}' d_{r} \mathrm{dimKer}D^{(E_{r})}\right)$, while $T'(X^A)$ are square matrices with size $\mathrm{dimKer}D^{(E_{\mathrm{trivial}})} \times \mathrm{dimKer}D^{(E_{\mathrm{trivial}})}$. 
 $T(X^A)$ and  $T'(X^A)$ are defined as follows:
    \begin{align}
      & (\text{the $(I, J)$ element of the $((r, j), (r', l))$
      block in $T(X^A)$})
      = \delta^{\langle r \rangle \langle r' \rangle} \delta_{jl} T^{(r, r)} (X^A \bm{1}_{E_r})_{IJ} \ , 
      \label{eq3.13} \\
      & (\text{the $(I, J)$ element of $T'(X^A)$})
      = T^{(1, 1)} (X^A)_{IJ} \ , 
      \label{eq3.13.2}
    \end{align}
where $T^{(r, r)} (X^A \bm{1}_{E_r})$ and $T^{(1, 1)} (X^A)$ are the Toeplitz operators for 
$X^A \bm{1}_{E_r}$ and $X^A$, respectively:
    \begin{align}
      & T^{(r, r)} (X^A \bm{1}_{E_r})_{IJ}
      = \int_M \mu \  \left(\psi^{\langle r \rangle}_{I; i}(x) \right)^{\dagger} X^A(x) \ \psi^{\langle r \rangle}_{J;i}(x) \ , \nonumber\\
      & T^{(1, 1)} (X^A)
      = \int_M \mu \  \left(\psi^{\langle 1 \rangle}_{I}(x) \right)^{\dagger} X^A(x) \ \psi^{\langle 1 \rangle}_{J}(x) \ . 
    \end{align}
$T(X^A)$ are block-diagonal matrices whose $((r,j),(r,j))$ block ($j=1,\ldots,d_r$) is given by the Toeplitz operator $T^{(r, r)} (X^A)$. 
There are $d_r$ copies of $T^{(r, r)} (X^A)$.
The shape of $T(X^A)$ are shown in Fig.\ref{fig2}. 
  $T'(X^A)$ are the Toeplitz operators for $X^A$ with the trivial representation.
  
We denote the Toeplitz operators for $\partial_{(a)} X^A$ by $T(\partial_{(a)} X^A)$. 
$T(\partial_{(a)} X^A)$ act on $T(\varphi)$ only from the left, and
they are square matrices with size $\left(\sum_{r} d_r \mathrm{dimKer}D^{(E_{r})}\right) \times \left(\sum_{r} d_{r} \mathrm{dimKer}D^{(E_{r})}\right)$ defined by
    \begin{align}
      &(\text{the $(I,J)$ element of the $((r, j), (r', l))$ block in $T(\partial_{(a)} X^A)$}) \nonumber\\
      &= T^{(r, r')} \left(\sqrt{\frac{d_{r'}}{d_r}} C^{v, r'^*; r}_{b(a), (l); (j)} \omega_{cb} \partial^c X^A\right)_{IJ}, 
      \label{eq3.14}
    \end{align}
where the RHS represents the Toeplitz operators for $\sqrt{\frac{d_{r'}}{d_r}} C^{v, r'^*; r}_{b(a), k(l); i(j)} \omega_{cb} \partial^c X^A$: 
    \begin{equation}
      T^{(r, r')} \left(\sqrt{\frac{d_{r'}}{d_r}} C^{v, r'^*; r}_{b(a), (l); (j)} \omega_{cb} \partial^c X^A\right)_{IJ}
      = \int_M \mu \  \left(\psi^{\langle r \rangle}_{I; i}(x) \right)^{\dagger} \sqrt{\frac{d_{r'}}{d_r}} C^{v, r'^*; r}_{b(a), k(l); i(j)} \omega_{cb} \partial^c X^A(x) \ \psi^{\langle r' \rangle}_{J;k}(x) \ . \label{eq3.1717}
    \end{equation}
Here, $\omega_{cb} = e_c^{\mu} e_b^{\nu} \omega_{\mu \nu}$ with 
$\omega_{\mu \nu}$ being the symplectic form.
If a representation $r$ does not appear 
in the irreducible decomposition of a tensor representation $v \otimes r'^*$, the $((r, j), (r', l))$ blocks in
$T(\partial_{(a)} X^A)$ vanish. 
Note that
the Clebsch-Gordan coefficients do not have the indices $i$ and $k$ in the LHS of \eqref{eq3.1717} because
they are contracted with those of zero modes in the RHS.

\subsubsection{$\mathcal{P}_{(a)}$}

We are now ready for showing \eqref{pa}.
As stated in the beginning of this subsection, 
we note that $[T(X^A), T(\varphi)]$ is not an ordinary commutator, but defined as
$[T(X^A), T(\varphi)] = T(X^A) T(\varphi) - T(\varphi) T'(X^A)$\footnote{$[T(\partial_{(a)}X^A), T(X^A)]$ in \eqref{eq3.1} is 
an ordinary commutator.}.
The first term in the RHS of \eqref{eq3.1} is calculated as
    \begin{align}
      &(\text{the $(I, J)$ element of the $(r,j)$ block in $\hbar_p^{-1} T(\partial_{(a)}X^A) [T(X^A), T(\varphi)]$}) \nonumber\\
      &=\hbar_{p}^{-1} \sum_{r'} \sum_{l,K} T^{(r, r')} \left(\sqrt{\frac{d_{r'}}{d_r}} C^{v, r'^*; r}_{b(a), (l); (j)} \omega_{cb} \partial^c X^A\right)_{IK} \left[T(X^A \bm{1}), T^{(r', 1)}\left(\varphi^{\langle r' \rangle}_{(l)}\right)\right]_{KJ} \nonumber\\
      &= -i \sum_{r'} \sum_{l, K} T^{(r, r')} \left(\sqrt{\frac{d_{r'}}{d_r}} C^{v, r'^*; r}_{b(a), (l); (j)} \omega_{cb} \partial^c X^A\right)_{IK} T^{(r', 1)} \left(\left\{X^A, \varphi^{\langle r' \rangle}_{(l)}\right\}\right)_{KJ} + O\left(\frac{1}{p}\right) \nonumber\\
      &= -i \sum_{r'} \sum_{l} T^{(r, 1)} \left(\sqrt{\frac{d_{r'}}{d_r}} C^{v, r'^*; r}_{b(a), k(l); (j)} \omega_{cb} (\partial^c X^A) \left\{X^A, \varphi^{\langle r' \rangle}_{k(l)}\right\}\right)_{IJ} + O\left(\frac{1}{p}\right) \nonumber\\
      &= -i \sum_{r'} \sum_{l} T^{(r, 1)} \left(\sqrt{\frac{d_{r'}}{d_r}} C^{v, r'^*; r}_{b(a), k(l); (j)} \omega_{cb} (\partial^c X^A) W^{de} (\partial_d X^A)\left(\nabla_e \varphi^{\langle r' \rangle}_{k(l)}\right)\right)_{IJ} + O\left(\frac{1}{p}\right) \nonumber\\
      &= T^{(r, 1)} \left(-i \sum_{r'} \sum_{l} \sqrt{\frac{d_{r'}}{d_r}} C^{v, r'^*; r}_{b(a), k(l); (j)} \left(\nabla_b \varphi^{\langle r' \rangle}_{k(l)}\right)\right)_{IJ} + O\left(\frac{1}{p}\right) \nonumber\\
      &= (\text{the $(I,J)$ element of $(r,j)$ block in $T(-i \nabla_{(a)} \varphi)$}) + O\left(\frac{1}{p}\right) \ .
      \label{eq3.15}
    \end{align}
$[\ , \ ]$ in the second line is the commutator-like operation \eqref{eq2.36}, namely
\begin{align}
    \left[T(X^A \bm{1}), T^{(r', 1)}\left(\varphi^{\langle r' \rangle}_{(l)}\right)\right] = T^{(r', r')} (X^A \bm{1}_{E_{r'}}) T^{(r', 1)}\left(\varphi^{\langle r' \rangle}_{(l)}\right) - T^{(r', 1)}\left(\varphi^{\langle r' \rangle}_{(l)}\right) T^{(1, 1)} (X^A)\ .
\end{align}
In the second and third equalities, 
we have used the asymptotic properties \eqref{eq2.12} and \eqref{eq2.11} of the Toeplitz operator, respectively.
Note that in the third equality that the sum is taken over the index $k$.
In the fifth equality, we have used that 
$(\partial^c X^A)(\partial_d X^A)=\delta^c_d$ and 
$\omega_{cb}W^{ce}=\delta_b^e$. In the last equality, we have used 
\eqref{eq3.11}.

The second term in the RHS of \eqref{eq3.1} is shown to be $O(1/p)$ as follows: 
    \begin{align}
      &(\text{the $(I, J)$ element of the $(r,j)$ block in $\hbar_p^{-1} [T(\partial_{(a)}X^A), T(X^A)] T(\varphi)$}) \nonumber\\
      &= -\hbar_p^{-1} \sum_{r'} \sum_{l, K} \left[T(X^A \bm{1}), \ T^{(r, r')} \left(\sqrt{\frac{d_{r'}}{d_r}} C^{v, r'^*; r}_{b(a), (l); (j)} \omega_{cb} \partial^c X^A\right)\right]_{IK} T^{(r', 1)} \left(\varphi^{\langle r' \rangle}_{(l)}\right)_{KJ} \nonumber\\
      &= \sum_{r'} \sum_{l} T^{(r, 1)}\left(i W^{de} (\partial_d X^A) \nabla_e\left(\sqrt{\frac{d_{r'}}{d_r}} C^{v, r'^*; r}_{b(a), k(l); (j)} \omega_{cb} \partial^c X^A\right) \varphi^{\langle r' \rangle}_{k(l)}\right)_{IJ} + O\left(\frac{1}{p}\right) \nonumber\\
      &= T^{(r, 1)}\left(i \sum_{r'} \sum_{l} W^{de} (\partial_d X^A) \sqrt{\frac{d_{r'}}{d_r}} C^{v, r'^*; r}_{b(a), k(l); (j)} \omega_{cb} (\nabla_e \partial^c X^A) \varphi^{\langle r' \rangle}_{k(l)}\right)_{IJ} + O\left(\frac{1}{p}\right) \nonumber\\
      &= O\left(\frac{1}{p}\right). 
      \label{eq3.16}
    \end{align}
In the second equality, we have used the asymptotic properties of the Toeplitz operator 
\eqref{eq2.11} and  \eqref{eq2.12}. 
Note that the sum over $k$ is taken there.
In the third equality, we have used a relation $\nabla_e C^{v, r'^*; r}_{b(a), k(l); i(j)} = 0$, which follows from
the fact that $C^{v, r'^*; r}_{b(a), k(l); i(j)}$ is an invariant tensor, and an equality $\nabla_e \omega_{cb}=0$, which is
a general property of the K\"{a}hler structure.
In the last equality, we have used 
$(\partial_d X^A)(\nabla_e \partial^c X^A)=0$.
This equality can be shown as
    \begin{align}
      (\partial_d X^A)(\nabla_e \partial^c X^A)
      &= \nabla_e((\partial_d X^A)(\partial^c X^A)) - (\nabla_e \partial_d X^A)(\partial^c X^A) \nonumber\\
      &= - (\nabla_d \partial_e X^A)(\partial^c X^A) \nonumber\\
      &= (\partial_e X^A)(\nabla^c \partial_d X^A) \nonumber\\
      &= -(\nabla_e \partial^c X^A)(\partial_d X^A) \nonumber\\
      &= 0 \ .
      \label{eq3.17}
    \end{align}
In the second, third and fourth equalities of (\eqref{eq3.17}), 
$(\partial_d X^A)(\partial^c X^A)=\tensor{\delta}{_d^c}$ and $\nabla_e \partial_d X^A = \nabla_d \partial_e X^A$
have been used. 
From \eqref{eq3.15} and \eqref{eq3.16}, we showed \eqref{pa}.

Next, we show that
$\mathcal{P}_{(a)}$ is Hermitian for finite $p$.
The second term in the RHS of \eqref{eq3.1},
which was shown to be 
$O(1/p)$, is introduced such that this holds.
First, we define an inner product between the Toeplitz operators 
$T(\varphi)$ and $T(\varphi')$ for $\varphi, \ \varphi' \in \Gamma(E_{\text{reg}})$
in terms of the Frobenius inner product as
    \begin{equation}
      (T(\varphi), T(\varphi'))
      = \Tr (T(\varphi)^\dagger T(\varphi')) \ ,
      \label{eq3.19}
    \end{equation}
where $T(\varphi)^\dagger$ stands for the matrix Hermitian conjugate of 
$T(\varphi)$.
\eqref{eq3.19} is expressed as
    \begin{equation}
      \Tr (T(\varphi)^\dagger T(\varphi'))
      = \sum_{r} \sum_{j} \sum_{I,J} T^{(r, 1)} \left(\varphi^{\langle r \rangle}_{(j)}\right)_{IJ}^* T^{(r, 1)} \left(\varphi^{\prime \langle r \rangle}_{(j)}\right)_{IJ} \ .
    \end{equation}
By using this inner product, we define the Hermitian conjugate $\mathcal{P}_{(a)}^\dagger$ of 
$\mathcal{P}_{(a)}$ as
    \begin{equation}
      (\mathcal{P}_{(a)}^\dagger T(\varphi), T(\varphi'))
      = (T(\varphi), \mathcal{P}_{(a)} T(\varphi')) \ .
    \end{equation}
Then, we can show that $\mathcal{P}_{(a)}$ is Hermitian as follows:
    \begin{align}
      &(\mathcal{P}_{(a)}^\dagger T(\varphi), T(\varphi')) \nonumber\\
      &= (T(\varphi), \mathcal{P}_{(a)} T(\varphi')) \nonumber\\
      &= \Tr (T(\varphi)^\dagger \left(\hbar_p^{-1} T(\partial_{(a)}X^A) [T(X^A), T(\varphi')] - \frac{1}{2} \hbar_p^{-1} [T(\partial_{(a)}X^A), T(X^A)] T(\varphi')\right)) \nonumber\\
      &= \hbar_p^{-1} \mathrm{Tr} \biggl(T(\varphi)^\dagger T(\partial_{(a)}X^A) T(X^A) T(\varphi') - T(\varphi)^\dagger T(\partial_{(a)}X^A) T(\varphi') T'(X^A) \nonumber\\ 
      &\left. \qquad \qquad - \frac{1}{2} T(\varphi)^\dagger [T(\partial_{(a)}X^A), T(X^A)] T(\varphi')\right) \nonumber\\
      &= \hbar_p^{-1} \mathrm{Tr} \biggl(T(\varphi)^\dagger T(X^A) T(\partial_{(a)}X^A) T(\varphi') - T'(X^A) T(\varphi)^\dagger T(\partial_{(a)}X^A) T(\varphi')  \nonumber\\
      &\left. \qquad \qquad + \frac{1}{2} T(\varphi)^\dagger [T(\partial_{(a)}X^A), T(X^A)] T(\varphi')\right) \nonumber\\
      &= \hbar_p^{-1} \Tr ([T(\varphi)^\dagger, T(X^A)] T(\partial_{(a)}X^A) T(\varphi') + \frac{1}{2} T(\varphi)^\dagger [T(\partial_{(a)}X^A), T(X^A)] T(\varphi')) \nonumber\\
      &= \hbar_p^{-1} \Tr (\Bigl(T(\partial_{(a)}X^A) [T(X^A), T(\varphi)]\Bigr)^\dagger T(\varphi') - \frac{1}{2} \Bigl([T(\partial_{(a)}X^A), T(X^A)] T(\varphi)\Bigr)^\dagger T(\varphi')) \nonumber\\
      &= \Tr (\Bigl(\hbar_p^{-1} T(\partial_{(a)}X^A) [T(X^A), T(\varphi)] - \frac{1}{2} \hbar_p^{-1} [T(\partial_{(a)}X^A), T(X^A)] T(\varphi)\Bigr)^\dagger T(\varphi')) \nonumber\\
      &= (\mathcal{P}_{(a)} T(\varphi), T(\varphi'))  \ . 
      \label{eq3.23}
    \end{align}
In the fourth equality, we have used $T(\partial_{(a)}X^A)T(X^A) = T(X^A)T(\partial_{(a)}X^A) + [T(\partial_{(a)}X^A), T(X^A)]$ in
the first term, and the cyclic property of the trace in the second term.
In the fifth equality,
we note $[T(\varphi)^\dagger, T(X^A)] = T(\varphi)^\dagger T(X^A) - T'(X^A) T(\varphi)^\dagger$.
In the sixth equality, we have used the fact that 
$T(X^A)$, $T'(X^A)$, $T(\partial_{(a)}X^A)$ are Hermitian.
$T(\partial_{(a)}X^A)$ is shown to be Hermitian as 
    \begin{align}
      &(\text{the $(I,J)$ element of the $((r, j), (r', l))$ block in $T(\partial_{(a)} X^A)^\dagger$}) \nonumber\\
      &= (\text{the $(I,J)$ element of the $((r', l), (r, j))$ 
      block in $T(\partial_{(a)} X^A){}^*$}) \nonumber\\
      &= T^{(r', r)} \left(\sqrt{\frac{d_{r}}{d_{r'}}} C^{v, r^*; r'}_{b(a), (j); (l)} \omega_{cb} \partial^c X^A\right)_{JI}^* \nonumber\\
      &= \left(\int_M \mu \ \left(\psi^{\langle r' \rangle}_{J;k}\right)^\dagger \sqrt{\frac{d_{r}}{d_{r'}}} C^{v, r^*; r'}_{b(a), i(j); k(l)} \omega_{cb} (\partial^c X^A) \psi^{\langle r \rangle}_{I;i}\right)^* \nonumber\\
      &= \int_M \mu \ \left(\psi^{\langle r \rangle}_{I;i}\right)^\dagger \sqrt{\frac{d_{r}}{d_{r'}}} \left(C^{v, r^*; r'}_{b(a), i(j); k(l)}\right)^* \omega_{cb} (\partial^c X^A) \psi^{\langle r' \rangle}_{J;k} \ . 
      \label{eq3.24}
    \end{align}
Here, we show 
$\sqrt{\frac{d_{r}}{d_{r'}}} \left(C^{v, r^*; r'}_{b(a), i(j); k(l)}\right)^* = \sqrt{\frac{d_{r'}}{d_r}} C^{v, r'^*; r}_{b(a), k(l); 
i(j)}$. From \eqref{eq3.9}, we have
    \begin{equation}
      R^{\langle v \rangle}_{b(a)}(g) R^{\langle r'^* \rangle}_{k(l)}(g)
      = \sum_{r''} C^{v, r'^*; r''}_{b(a), k(l); i'(j')} R^{\langle r''^* \rangle}_{i'(j')}(g) \ .
    \end{equation}
Multiplying the both sides by $R^{\langle r \rangle}_{i(j)}(g)$, integrating over $g$ and
using the orthonormality of the representation matrices
$\int dg \ R^{\langle r''^* \rangle}_{i'(j')}(g) R^{\langle r \rangle}_{i(j)}(g) 
= \frac{1}{d_r} \delta^{\langle r'' \rangle \langle r \rangle} \delta_{i'i} \delta_{j'j}$,  leads to
    \begin{equation}
      C^{v, r'^*; r}_{b(a), k(l); i(j)}
      = d_r \int dg \ R^{\langle v \rangle}_{b(a)}(g) R^{\langle r'^* \rangle}_{k(l)}(g) R^{\langle r \rangle}_{i(j)}(g) \ .
      \label{eq3.26}
    \end{equation}
This implies that
    \begin{align}
      \sqrt{\frac{d_{r}}{d_{r'}}} \left(C^{v, r^*; r'}_{b(a), i(j); k(l)}\right)^*
      &= \sqrt{\frac{d_{r}}{d_{r'}}} \left(d_{r'} \int dg \ R^{\langle v \rangle}_{b(a)}(g) R^{\langle r^* \rangle}_{i(j)}(g) R^{\langle r' \rangle}_{k(l)}(g)\right)^* \nonumber\\
      &= \sqrt{d_r d_{r'}} \int dg \ R^{\langle v \rangle}_{b(a)}(g) R^{\langle r \rangle}_{i(j)}(g) R^{\langle r'^* \rangle}_{k(l)}(g) \nonumber\\
      &= \sqrt{\frac{d_{r'}}{d_{r}}} C^{v, r'^*; r}_{b(a), k(l); i(j)} \ .
    \end{align}
By using this relation, we rewrite \eqref{eq3.24} as
    \begin{align}
      &(\text{the $(I, J)$ element of the $((r, j), (r', l))$ 
      block in $T(\partial_{(a)} X^A)^\dagger$}) \nonumber\\
      &= \int_M \mu \ \left(\psi^{\langle r \rangle}_{I;i}\right)^\dagger \sqrt{\frac{d_{r'}}{d_{r}}} C^{v, r'^*; r}_{b(a), k(l); i(j)} \omega_{cb} (\partial^c X^A) \psi^{\langle r' \rangle}_{J;k} \nonumber\\
      &= (\text{the $(I,J)$ element of the $((r, j), (r', l))$
      block in $T(\partial_{(a)} X^A)$}). 
    \end{align}
Thus, $T(\partial_{(a)}X^A)$ are Hermitian, which implies that
the sixth equality in \eqref{eq3.23} holds. Namely,
we have shown that $\mathcal{P}_{(a)}$ are Hermitian for finite $p$.

Therefore, we conclude that $\mathcal{P}_{(a)}$ defined in \eqref{eq3.1} serves as the finite-matrix regularization of the covariant derivatives $-i \nabla_{(a)}$.

Finally, 
we calculate $[\mathcal{P}_{(a)}, \mathcal{P}_{(b)}]$ in the large-$p$ limit.
In appendix \ref{seca}, 
$[\mathcal{P}_{(a)}, \mathcal{P}_{(b)}] T(\varphi)$ is calculated as follows:
    \begin{align}
      &[\mathcal{P}_{(a)}, \mathcal{P}_{(b)}] T(\varphi) \nonumber\\
      &= T(\partial_{(a)} X^A) T(\partial_{(b)} X^B) T \left(-\frac{1}{4} (G^{\alpha \beta} G^{\gamma \delta} - G^{\delta \beta} G^{\gamma \alpha}) (\partial_{\alpha} X^A) R^{E_r}_{\beta \gamma}(\partial_{\delta} X^B)\right) T(\varphi) + O\left(\frac{1}{p}\right) \ .
      \label{eq3.31}
    \end{align}
Here, $T (- \frac{1}{4} (G^{\alpha \beta} G^{\gamma \delta} - G^{\delta \beta} G^{\gamma \alpha}) (\partial_{\alpha} X^A) R^{E_r}_{\beta \gamma}(\partial_{\delta} X^B))$ are given by
    \begin{align}
      & (\text{the $(I,J)$ element of the $((r, j), (r', l))$
      block in $T (- \tfrac{1}{4} (G^{\alpha \beta} G^{\gamma \delta} - G^{\delta \beta} G^{\gamma \alpha}) (\partial_{\alpha} X^A) R^{E_r}_{\beta \gamma}(\partial_{\delta} X^B))$}) \nonumber\\
      & = \delta^{\langle r \rangle \langle r' \rangle} \delta_{jl} T^{(r, r)} \left(-\frac{1}{4} (G^{\alpha \beta} G^{\gamma \delta} - G^{\delta \beta} G^{\gamma \alpha}) (\partial_{\alpha} X^A) R^{E_r}_{\beta \gamma}(\partial_{\delta} X^B) \bm{1}_{E_r}\right)_{IJ} \ . \label{eq3.32}
    \end{align}
As in section \ref{sec2.2}, $G^{\alpha \beta} = g^{\alpha \beta} + iW^{\alpha \beta}$, 
$R^{E_r}_{\beta \gamma}$ are components of  the curvature $R^{E_r} = \mathrm{d}A^{E_r} + A^{E_r} \wedge A^{E_r}$
of $E_r$. 

\subsection{Case of square Toeplitz operator for $\varphi$}\label{sec3.22}
 In this subsection, we consider the case of square Toeplitz operator for $\varphi$. In section \ref{sec3.1}, we construct 
finite-size matrices $\mathcal{P}_{(a)}$ 
that serve as a finite-matrix regularization
of $-i\nabla_{(a)}$ such that they
act on the rectangular $T(\varphi) \ (\varphi \in \Gamma(E_{\text{reg}}))$.
In this subsection, we construct finite-size matrices $\mathcal{P}_{(a)}$ that act on
a square Toeplitz operator
for $\varphi \in \Gamma(E_{\text{reg}})$.

First, we explain the outline of
how we construct a square Toeplitz operator for 
$\varphi \in \Gamma(E_{\text{reg}})$. 
Here, we note that the following isomorphism holds:
    \begin{equation}
      \Gamma(E_{\text{reg}} \otimes E_{\text{reg}}) \cong \Gamma(E_{\text{reg}}) \oplus \Gamma(E_{\text{reg}}) \oplus \cdots \ .\label{eq3.66}
    \end{equation}
By taking the same element $\varphi$ 
in each $\Gamma(E_{\text{reg}})$ appearing in 
the RHS of \eqref{eq3.66}, we 
obtain an element of $\Gamma(E_{\text{reg}} \otimes E_{\text{reg}})$
from $\varphi \in \Gamma(E_{\text{reg}})$. 
We construct a square Toeplitz operator for the element of 
$\Gamma(E_{\text{reg}} \otimes E_{\text{reg}})$ in a similar way as
section \ref{sec3.1}. Thus, we obtain 
a square Toeplitz operator for 
$\varphi \in \Gamma(E_{\text{reg}})$.

Next, we explain our construction in detail.
Let us consider
$\phi(x, g_1, g_2) \in \Gamma(E_{\text{reg}} \otimes E_{\text{reg}})$ ($x \in M, \  g_1, g_2 \in G$). 
Regarding $\phi$ as a function of 
$(x, g_1, g_1^{-1}g_2)$, we denote it by $\phi'$: 
$\phi'(x, g_1, g_1^{-1}g_2) = \phi(x, g_1, g_2)$. 
For 
$h \in G$,
$\phi'(x, g_1, g_1^{-1}g_2)$ is transformed as follows:
    \begin{equation}
      (\hat{h}\phi')(x, g_1, g_1^{-1}g_2)
      = \phi'(x, h^{-1}g_1, (h^{-1}g_1)^{-1}(h^{-1}g_2))
      = \phi'(x, h^{-1}g_1, g_1^{-1}g_2) \ . \label{eq3.3333}
    \end{equation}
We see that $g_1^{-1}g_2$ is invariant under action of
  $h \in G$. 
  
By using the Peter-Weyl theorem, $\phi'(x, g_1, g_1^{-1}g_2)$
is expanded as
    \begin{equation}
      \phi'(x, g_1, g_1^{-1}g_2)
      = \sum_{r'} \phi^{\prime \langle r'^* \rangle}_{(k)(l)}(x, g_1) \sqrt{d_{r'}} R^{\langle r' \rangle}_{(k)(l)}(g_1^{-1}g_2) \ . \label{eq3.34}
    \end{equation}
We see from
\eqref{eq3.3333} that under $G$
not only the index $(l)$ but also the index $(k)$ 
in $\phi^{\prime \langle r'^* \rangle}_{(k)(l)}(x, g_1)$
are invariant.
Thus, $r'^*$, $(k)$ and  $(l)$ just label
copies of $\Gamma(E_{\text{reg}})$ in the RHS of \eqref{eq3.66}. 
Then, for an arbitrary irreducible representation $r'^*$
of $G$ and arbitrary
$k, l = 1, \ldots, d_{r'}$, we put  
    \begin{equation}
      \phi^{\prime \langle r'^* \rangle}_{(k)(l)}(x, g_1)
      = \varphi(x, g_1) \ , \label{eq3.35}
    \end{equation}
where
$\varphi(x, g_1) \in \Gamma(E_{\text{reg}})$ ($x \in M, \  g_1 \in G$). 
In this way, we can construct an element of 
$\Gamma(E_{\text{reg}} \otimes E_{\text{reg}})$ from 
$\varphi \in \Gamma(E_{\text{reg}})$.

We represent an element 
of $\Gamma(E_{\text{reg}} \otimes E_{\text{reg}})$
as a square matrix with finite size in the following way.
First, by using the Peter-Weyl theorem, we expand
$\phi(x, g_1, g_2) \in \Gamma(E_{\text{reg}} \otimes E_{\text{reg}})$ ($x \in M, \  g_1, g_2 \in G$) as 
    \begin{equation}
      \phi(x, g_1, g_2)
      = \sum_{r, r'} \phi^{\langle r, r'^* \rangle}_{i(j), k(l)}(x) \sqrt{d_{r} d_{r'}} R^{\langle r^* \rangle}_{i(j)}(g_1) R^{\langle r' \rangle}_{k(l)}(g_2) \ . \label{eq3.4242}
    \end{equation}
    
The indices $l$ and $k$
in $\phi^{\langle r, r'^* \rangle}_{i(j), k(l)}(x)$
transform
under $r$ and $r'$ representations of
$G$, respectively,
while the indices $(j)$ and $(l)$ not.
For fixed $r$, $r'^*$, $(j)$ and $(l)$, 
$\phi^{\langle r, r'^* \rangle}_{i(j), k(l)}(x)$
is a field of the tensor product representation of 
 $r$ and $r'^*$ representations.
Namely,
$\phi^{\langle r, r'^* \rangle}_{i(j), k(l)}(x) \in \Gamma(\mathrm{Hom}(E_{r'}, E_r))$ with
$E_r$ and $E_{r'}$ being fiber bundles whose fibers
are the $r$ and $r'$ representation spaces of $G$, respectively.

As in section \ref{sec2.2}, we construct the Toeplitz operators
for $\phi^{\langle r, r'^* \rangle}_{i(j), k(l)}(x)$
by using the BT quantization.
We denote the Toeplitz operators for 
$\phi^{\langle r, r'^* \rangle}_{i(j), k(l)}(x)$
by $T^{(r, r')} \left(\phi^{\langle r, r'^* \rangle}_{(j), 
(l)}\right)$.
Namely, we define
    \begin{equation}
      T^{(r, r')} \left(\phi^{\langle r, r'^* \rangle}_{(j), (l)}\right)_{IJ}
      = \int_M \mu \  \left(\psi^{\langle r \rangle}_{I; i}(x) \right)^{\dagger} \phi^{\langle r, r'^* \rangle}_{i(j), k(l)}(x) \ \psi^{\langle r' \rangle}_{J; k}(x) \ ,
    \end{equation}
where $I=1, \ldots, \mathrm{dimKer}D^{(E_r)}$ and $J=1, \ldots, \mathrm{dimKer}D^{(E_{r'})}$.

The Toeplitz operator $T(\phi)$ for 
$\phi \in \Gamma(E_{\text{reg}} \otimes E_{\text{reg}})$
is given in terms of 
$T^{(r, r')} \left(\phi^{\langle r, r'^* \rangle}_{(j), (l)}\right)$ as 
    \begin{equation}
      (\text{the $(I,J)$ element of the $((r, j), (r', l))$
      block in $T(\phi)$})
      = T^{(r, r')} \left(\phi^{\langle r, r'^* \rangle}_{(j), (l)}\right)_{IJ} \ . \label{eq3.3939}
    \end{equation}
We introduce a cutoff $\Xi$ and restrict irreducible
representations $r$ and $r'$ 
such that
the Casimir of the tensor product representations $r \otimes 
r'^*$ are less than or equal to $\Xi$\footnote{In two dimensions,
we restrict charges $s$ and $s'$ such that
$|s|, |s'| \leq \Xi$. }. 
Then, we take the $p \rightarrow \infty$ limit 
and the $\Xi \rightarrow \infty$ limit with 
$\Xi \ll p$ being kept. 
Thus, 
$T(\phi)$ in 
\eqref{eq3.3939} is defined as a finite-size square matrix
that serves as a regularization of $\phi \in \Gamma(E_{\text{reg}}\otimes E_{\text{reg}})$.

We are now considering $\phi \in \Gamma(E_{\text{reg}} \otimes E_{\text{reg}})$ which is constructed from
$\varphi \in \Gamma(E_{\text{reg}})$ in \eqref{eq3.35}. 
We see how the expansion coefficients $\phi^{\langle r, r'^* \rangle}_{i(j), k(l)}(x)$ of $\phi$
depend on $\varphi$.
We start with \eqref{eq3.35}. The RHS of \eqref{eq3.35} can be expanded through the Peter-Weyl theorem as
    \begin{equation}
      \phi^{\prime \langle r'^* \rangle}_{(k)(l)}(x, g_1)
      = \sum_{r} \phi^{\prime \langle r, r'^* \rangle}_{i(j), (k)(l)}(x) \sqrt{d_r} R^{\langle r^* \rangle}_{i(j)}(g_1) \ . \label{eq3.36}
    \end{equation}
The index $i$ in $\phi^{\prime \langle r, r'^* \rangle}_{i(j), (k)(l)}(x)$
   transforms under $r$ representation of $G$, 
  while the indices $(j)$, $(k)$ and $(l)$ not.
On the other hand, the RHS of \eqref{eq3.35}
can be expanded 
through the Peter-Weyl theorem as in \eqref{eq3.2}:
    \begin{equation}
      \varphi(x, g_1)
      =\sum_{r} \varphi^{\langle r \rangle}_{i(j)}(x) \sqrt{d_r} R^{\langle r^* \rangle}_{i(j)}(g_1) \ . \label{eq3.4141}
    \end{equation}
\eqref{eq3.35} is equivalent to the following condition.
For an arbitrary representation $r'^*$ of $G$ and arbitrary 
$k, l = 1, \ldots, d_{r'}$,  
    \begin{equation}
      \phi^{\prime \langle r, r'^* \rangle}_{i(j), (k)(l)}(x)
      = \varphi^{\langle r \rangle}_{i(j)}(x) \ . \label{eq3.38}
    \end{equation}

We express the expansion coefficients $\phi^{\langle r, r'^* \rangle}_{i(j), k(l)}(x)$  
of $\phi  \in \Gamma(E_{\text{reg}} \otimes E_{\text{reg}})$ in \eqref{eq3.4242}
in terms of the expansion coefficients $\varphi^{\langle r \rangle}_{i(j)}(x)$ of
$\varphi \in \Gamma(E_{\text{reg}})$ in \eqref{eq3.4141}. 
We see from \eqref{eq3.38} that
this is achieved by expressing the expansion coefficients $\phi^{\langle r, r'^* \rangle}_{i(j), k(l)}(x)$
of $\phi$ in \eqref{eq3.4242} in terms of the expansion coefficients 
$\phi^{\prime \langle r, r'^* \rangle}_{i(j), (k)(l)}(x)$ of 
$\phi'$ in \eqref{eq3.36}. 
Substituting \eqref{eq3.36} into \eqref{eq3.34} gives rise to  
    \begin{align}
      \phi'(x, g_1, g_1^{-1}g_2) 
      &= \sum_{r, r'} \phi^{\prime \langle r, r'^* \rangle}_{i(j), (k)(l)}(x) \sqrt{d_r d_{r'}} R^{\langle r^* \rangle}_{i(j)}(g_1) R^{\langle r' \rangle}_{(k)(l)}(g_1^{-1}g_2) \nonumber\\
      &= \sum_{r, r'} \phi^{\prime\langle r, r'^* \rangle}_{i(j), (k)(l)}(x) \sqrt{d_r d_{r'}} R^{\langle r^* \rangle}_{i(j)}(g_1) R^{\langle r' \rangle}_{(k)k'}(g_1^{-1}) R^{\langle r' \rangle}_{k'(l)}(g_2) \nonumber\\
      &= \sum_{r, r'} \phi^{\prime \langle r, r'^* \rangle}_{i(j), (k)(l)}(x) \sqrt{d_r d_{r'}} R^{\langle r^* \rangle}_{i(j)}(g_1) R^{\langle r'^* \rangle}_{k'(k)}(g_1) R^{\langle r' \rangle}_{k'(l)}(g_2) \nonumber\\
      &= \sum_{r, r'} \phi^{\prime \langle r, r'^* \rangle}_{i(j), (k)(l)}(x) \sqrt{d_r d_{r'}} \sum_{r''} C^{r^*, r'^*; r''}_{i(j), k'(k); m(n)} R^{\langle r''^* \rangle}_{m(n)}(g_1) R^{\langle r' \rangle}_{k'(l)}(g_2) \nonumber\\
      &= \sum_{r'', r'} \left(\sum_{r} \sqrt{\frac{d_r}{d_{r''}}} C^{r^*, r'^*; r''}_{i(j), k'(k); m(n)} \phi^{\prime\langle r, r'^* \rangle}_{i(j), (k)(l)}(x)\right) \sqrt{d_{r''} d_{r'}} R^{\langle r''^* \rangle}_{m(n)}(g_1) R^{\langle r' \rangle}_{k'(l)}(g_2) \nonumber\\
      &= \sum_{r, r'} \left(\sum_{r''} \sqrt{\frac{d_{r''}}{d_{r}}} C^{r''^*, r'^*; r}_{m(n), k(k'); i(j)} \phi^{\prime \langle r'', r'^* \rangle}_{m(n), (k')(l)}(x)\right) \sqrt{d_{r} d_{r'}} R^{\langle r^* \rangle}_{i(j)}(g_1) R^{\langle r' \rangle}_{k(l)}(g_2) \ . \label{eq3.4545}
    \end{align}
By noticing that $\phi(x, g_1, g_2) = \phi'(x, g_1, g_1^{-1}g_2)$ and 
comparing \eqref{eq3.4242} and \eqref{eq3.4545}, we see that
$\phi^{\langle r, r'^* \rangle}_{i(j), k(l)}(x)$ are expressed in terms of
$\phi^{\prime \langle r, r'^* \rangle}_{i(j), (k)(l)}(x)$ as
    \begin{equation}
      \phi^{\langle r, r'^* \rangle}_{i(j), k(l)}(x)
      = \sum_{r''} \sqrt{\frac{d_{r''}}{d_{r}}} C^{r''^*, r'^*; r}_{m(n), k(k'); i(j)} \phi^{\prime\langle r'', r'^* \rangle}_{m(n), (k')(l)}(x) \ . 
    \end{equation}
Then, from \eqref{eq3.38}, we finally find  
    \begin{equation}
      \phi^{\langle r, r'^* \rangle}_{i(j), k(l)}(x)
      = \sum_{r''} \sum_{k'=1}^{d_{r'}} \sqrt{\frac{d_{r''}}{d_{r}}} C^{r''^*, r'^*; r}_{m(n), k(k'); i(j)} \varphi^{\langle r'' \rangle}_{m(n)}(x) \ . 
    \end{equation}

To summarize, the Toeplitz operator $\bar{T}(\varphi)$
for $\varphi \in \Gamma(E_{\text{reg}})$ that is 
a square matrix is given by
    \begin{align}
      &(\text{the $(I,J)$ element of the $((r, j), (r', l))$ block in $\bar{T}(\varphi)$}) \nonumber\\
      &= T^{(r, r')} \left(\sum_{r''} \sum_{k'=1}^{d_{r'}} \sqrt{\frac{d_{r''}}{d_{r}}} C^{r''^*, r'^*; r}_{m(n), (k'); (j)} \varphi^{\langle r'' \rangle}_{m(n)}\right)_{IJ} \ , \label{eq3.46}
    \end{align}
where $\varphi^{\langle r'' \rangle}_{m(n)}$
appear in the expansion of 
$\varphi$ through the Peter-Weyl theorem in \eqref{eq3.4141}.
Note that the RHS of \eqref{eq3.46} 
does not depend on $l$.

Next, we construct $\bar{T}(-i\nabla_{(a)} \varphi)$. \eqref{eq3.3333} indicates that
$\nabla_{(a)}$ act only on $g_1$
in the expansion coefficients 
$\phi^{\prime \langle r'^* \rangle}_{(k)(l)}(x, g_1)$
when they act on $\phi'(x, g_1, g_1^{-1}g_2)$ 
expanded in \eqref{eq3.34}. 
Namely, they do not act on the indices $(k)$
and $(l)$.
The constraint \eqref{eq3.35} is therefore 
maintained after the action of
$\nabla_{(a)}$. This implies that we can construct $\bar{T}(-i\nabla_{(a)} \varphi)$ in a similar way to $\bar{T}(\varphi)$. From \eqref{eq3.10} and \eqref{eq3.46}, we see
that the Toeplitz operators $\bar{T}(-i\nabla_{(a)} \varphi)$, which are square matrices, 
are given by
    \begin{align}
      &(\text{the $(I,J)$ element of the $((r, j), (r', l))$ block in $\bar{T}(-i\nabla_{(a)} \varphi)$}) \nonumber\\
      &= T^{(r, r')} \left(\sum_{r''} \sum_{k'=1}^{d_{r'}} \sqrt{\frac{d_{r''}}{d_{r}}} C^{r''^*, r'^*; r}_{m(n), (k'); (j)} \left(-i \sum_{r'''} \sqrt{\frac{d_{r'''}}{d_{r''}}} C^{v, r'''^*; r''}_{b(a), s(t); m(n)} \left(\nabla_b \varphi^{\langle r''' \rangle}_{s(t)}\right)\right)\right)_{IJ} \nonumber\\
      &= T^{(r, r')} \left(-i \sum_{r'', r'''} \sum_{k'=1}^{d_{r'}} \sqrt{\frac{d_{r'''}}{d_{r}}} C^{r''^*, r'^*; r}_{m(n), (k'); (j)} C^{v, r'''^*; r''}_{b(a), s(t); m(n)} \left(\nabla_b \varphi^{\langle r''' \rangle}_{s(t)}\right)\right)_{IJ} \ . \label{eq3.48}
    \end{align}

Now we can construct finite-size matrices $\bar{\mathcal{P}}_{(a)}$
for $-i\nabla_{(a)}$ acting on the square matrix  
$\bar{T}(\varphi)$. $\bar{\mathcal{P}}_{(a)}$ are defined by
    \begin{equation}
      \bar{\mathcal{P}}_{(a)} \bar{T}(\varphi)
      = \hbar_p^{-1} T(\partial_{(a)}X^A) [T(X^A), \bar{T}(\varphi)] - \frac{1}{2} \hbar_p^{-1} [T(\partial_{(a)}X^A), T(X^A)] \bar{T}(\varphi) \ . \label{eq3.47}
    \end{equation}
The only difference between \eqref{eq3.47}
and 
\eqref{eq3.1} is that $[T(X^A), \bar{T}(\varphi)]$
are ordinary commutators because
$\bar{T}(\varphi)$ is a square matrix.
In the present case, $T'(X^A)$ does not appear.
$T(X^A)$ and $T(\partial_{(a)} X^A)$ in \eqref{eq3.47} agree with \eqref{eq3.13} and \eqref{eq3.14}, respectively.

As in section \ref{sec3.1}, we can show
\begin{align}
    \bar{\mathcal{P}}_{(a)} T(\varphi)=\bar{T}(-i \nabla_{(a)} \varphi) + O(1/p)\ ,
\end{align}
and that $\bar{\mathcal{P}}_{(a)}$ with finite $p$
are Hermitian. Thus, we conclude that $\bar{\mathcal{P}}_{(a)}$ serve as the finite-matrix regularization of $-i \nabla_{(a)}$.
Similarly in appendix \ref{seca},
we can calculate
$[\bar{\mathcal{P}}_{(a)}, \bar{\mathcal{P}}_{(b)}] \bar{T}(\varphi)$ as 
    \begin{align}
      &[\bar{\mathcal{P}}_{(a)}, \bar{\mathcal{P}}_{(b)}] \bar{T}(\varphi) \nonumber\\
      &= T(\partial_{(a)} X^A) T(\partial_{(b)} X^B) \left[T \left(-\frac{1}{4} (G^{\alpha \beta} G^{\gamma \delta} - G^{\delta \beta} G^{\gamma \alpha}) (\partial_{\alpha} X^A) R^{E_r}_{\beta \gamma}(\partial_{\delta} X^B)\right), \bar{T}(\varphi)\right] + O\left(\frac{1}{p}\right) \, \label{eq3.49}
    \end{align}
where $T (- \frac{1}{4} (G^{\alpha \beta} G^{\gamma \delta} - G^{\delta \beta} G^{\gamma \alpha}) (\partial_{\alpha} X^A) R^{E_r}_{\beta \gamma}(\partial_{\delta} X^B))$ are given in
\eqref{eq3.32}.

\subsection{$n=1$ case} \label{sec3.2}
In this subsection, we
study the $n=1$ case in sections \ref{sec3.1}
and \ref{sec3.22}. 
We see that
$\bar{\mathcal{P}}_{(a)}$ acting on $\bar{T}(\varphi)$
agree with the regularization of $\nabla_{(a)}$ obtained in \cite{Hattori:2024btt}. 

For $n=1$, $M$ (a closed connected K\"{a}hler 2-dimensional manifold) possesses a spin structure, so that
we consider not 
$\mathrm{Spin}^c(2)$ but 
$\mathrm{Spin}(2)$.
The irreducible representations of
$\mathrm{Spin}(2) \cong \mathrm{U}(1)$
are specified by integer or half-integer charges.
The representation matrix ($1\times 1$ matrix) for the charge $s$ 
is given by 
    \begin{equation}
      R^{\langle s \rangle}(\theta)
      = e^{2is \theta} \quad (\theta \in [0, 2\pi)) \ . 
    \end{equation}

First, let us consider the Toeplitz
operator $T(\varphi)$ for $\varphi \in \Gamma(E_{\text{reg}})$, 
which is 
a rectangular matrix.
Through the Peter-Weyl theorem,
$\varphi(x, \theta) \ (x \in M, \ \theta \in \mathrm{Spin}(2))$
is expanded as follows\footnote{This is nothing but a
Fourier expansion.}: 
    \begin{equation}
      \varphi(x, \theta)
      = \sum_{s} \varphi^{\langle s \rangle}(x) e^{-2is 
      \theta} \ . \label{eq3.51}
    \end{equation}
where the sum is taken over all integers and half-integers.
For $\theta' \in \mathrm{Spin}(2)$,
$\varphi^{\langle s \rangle}(x)$
is transformed as 
$e^{2is \theta'} \varphi^{\langle s \rangle}(x)$. 
$\varphi^{\langle s \rangle}(x)$ is
a field with charge $s$ of
$\mathrm{Spin}(2)$. 
For instance, 
the elements of a contravariant vector,
$V^{\pm} = V^1 \pm i V^2$,
have charge $\pm 1$, respectively,
and those of a covariant vector,
$V_{\pm} = \frac{1}{2} (V^1 \mp iV^2)$,
have charge $\mp1$, respectively.
The metric $\delta_{ab}$
is given by 
$\delta_{++} = \delta_{--} = 0$
and $\delta_{+-} = \delta_{-+} = \frac{1}{2}$. 

Let $E_s$ be a fiber bundle whose fiber
is the space of charge $s$ of 
$\mathrm{Spin}(2)$ and $E_0$ be
a fiber bundle whose fiber is the space
of charge $0$ of $\mathrm{Spin}(2)$.
Then, 
$\varphi^{\langle s \rangle}(x)$
is viewed as a section of the 
Hom-bundle $\mathrm{Hom}(E_{s}, E_0)$: 
$\varphi^{\langle s \rangle}(x) \in \Gamma(\mathrm{Hom}
(E_s, E_0))$. 
We can therefore apply the BT quantization to
$\varphi^{\langle s \rangle}(x)$.
Its Toeplitz operator 
$T^{(s, 0)} \left(\varphi^{\langle s \rangle}\right)$
is given by
    \begin{equation}
      T^{(s, 0)} \left(\varphi^{\langle s \rangle}\right)_{IJ}
      = \int_M \mu \  \left(\psi^{\langle s \rangle}_{I}(x) \right)^{\dagger} \varphi^{\langle s \rangle}(x) \ \psi^{\langle 0 \rangle}_{J}(x) \ ,
    \end{equation}
where $\left\{\psi^{\langle s \rangle}_{I}(x)\right\}_{I=1, \ldots, \mathrm{dimKer}D^{(E_{s})}}$
is an orthonormal basis in the space of 
normalizable zero modes of the Dirac operator 
$D^{(E_s)}$, $\mathrm{Ker} D^{(E_{s})}$, while
$\left\{\psi^{\langle 0 \rangle}_{J}(x)\right\}_{J=1, \ldots, \mathrm{dimKer}D^{(E_{0})}}$ is an orthonormal basis
in the space in $\mathrm{Ker} D^{(E_{0})}$.

The Toeplitz operator $T(\varphi)$ for 
$\varphi(x, \theta)$ is given by
    \begin{equation}
      (\text{the $(I,J)$ element of the $s$ block in
      $T(\varphi)$})
      = T^{(s, 0)} \left(\varphi^{\langle s \rangle}\right)_{IJ} \  .
    \end{equation}
Namely,
    \begin{align}
      T(\varphi)=
      \begin{pmatrix}
        \vdots                                                                  \\
        T^{(s-\frac{1}{2},0)}(\varphi^{\langle s-\frac{1}{2} \rangle})  \\ 
        T^{(s,0)}(\varphi^{\langle s \rangle})                          \\
        T^{(s+\frac{1}{2},0)}(\varphi^{\langle s+\frac{1}{2} \rangle})  \\
        \vdots                                                                  \\
      \end{pmatrix}
     \  .
    \end{align}
We introduce a cutoff $\Xi$ 
for charge $s$ and restrict charge such that
$|s| \leq \Xi$. Then, 
$T(\varphi)$ becomes a finite-size rectangular matrix. 
The size of $T(\varphi)$
is $\left(\sum_{|s| \leq \Xi} \mathrm{dimKer}D^{(E_{s})}\right) 
\times \mathrm{dimKer}D^{(E_{0})}$. 
We take the limits $\Xi \rightarrow \infty$ and 
$p \rightarrow \infty$ with  $\Xi \ll p$ kept.
  
We construct
the Toeplitz operator  $T(-i \nabla_{(\pm)} \varphi)$
for $-i \nabla_{(\pm)} \varphi$. 
Noting that the lower indices $\pm$
have charge $\mp 1$,
we have
    \begin{align}
      \nabla_{(\pm)} \varphi(x, \theta) 
      &= R^{\langle \mp \rangle}(-\theta) \nabla_{\pm} \varphi(x, \theta) \nonumber\\
      &= R^{\langle \pm \rangle}(\theta) \nabla_{\pm} \varphi(x, \theta) \ .
    \end{align}
As in the same way as we
derived \eqref{eq3.10} from \eqref{eq3.5},
we obtain
    \begin{equation}
      \nabla_{(\pm)} \varphi(x, \theta)
      = \sum_{s} \left(\sum_{s'} C^{\pm, s'^*; s} \left(\nabla_{\pm} \varphi^{\langle s' \rangle}(x)\right)\right) R^{\langle s^* \rangle}(\theta) \ ,\label{eq3.55}
    \end{equation}
where $C^{\pm, s'^*; s}$ are defined by 
    \begin{equation}
      R^{\langle \pm \rangle}(\theta) R^{\langle s'^* \rangle}(\theta)
      = \sum_{s} C^{\pm, s'^*; s} R^{\langle s^* \rangle}(\theta)
      \ , 
    \end{equation}
which implies
    \begin{align}
      C^{\pm, s'^*; s}
      &= \int_{0}^{2\pi} \frac{d\theta}{2\pi} \ R^{\langle \pm \rangle}(\theta) R^{\langle s'^* \rangle}(\theta) R^{\langle s \rangle}(\theta) \nonumber\\
      &= \int_{0}^{2\pi} \frac{d\theta}{2\pi} \ e^{\pm 2i \theta} e^{-2is' \theta} e^{2is \theta} \nonumber\\
      &= \delta_{s', s \pm 1} \ . 
      \label{eq3.33}
    \end{align}
\eqref{eq3.55} and \eqref{eq3.33} give 
$T(-i \nabla_{(\pm)} \varphi)$ as
    \begin{align}
      (\text{the $(I,J)$ element of the $s$ block in 
      $T(-i \nabla_{(\pm)} \varphi)$})
      &= T^{(s, 0)} \left(-i \sum_{s'} \delta_{s', s \pm 1} \left(\nabla_{\pm} \varphi^{\langle s' \rangle}\right)\right)_{IJ} \nonumber\\
      &= T^{(s, 0)} \left(-i \nabla_{\pm} \varphi^{\langle s \pm 1 \rangle}\right)_{IJ} \ .
    \end{align}
  
The Toeplitz operators  $T(X^A)$ and 
$T'(X^A)$ for the isometric embedding 
functions $X^A$ are given by
    \begin{align}
      & (\text{the $(I,J)$ element of the $(s,s')$ block in $T(X^A)$})
      = \delta^{\langle s \rangle \langle s' \rangle} T^{(s, s)} (X^A)_{IJ}, \label{eq3.39}\\
      & (\text{the $(I,J)$ element in $T'(X^A)$})
      = T^{(0, 0)} (X^A)_{IJ}. \label{eq3.40}
    \end{align}
$T(X^A)$ are square matrices whose off-diagonal blocks are zero
and whose diagonal blocks are the Toeplitz operators
for $X^A$ with charge $s$ representation\footnote{$T(X^A)$ is also used in \cite{Ishii:2008ib}.}: 
    \begin{align}
      T(X^A)=
      \begin{pmatrix}
        \ddots \\
                & T^{(s-\frac{1}{2},s-\frac{1}{2})}(X^A)  &                 & \text{\huge{0}} \\ 
                &                                         & T^{(s,s)}(X^A) \\
                & \text{\huge{0}}                         &                 & T^{(s+\frac{1}{2},s+\frac{1}{2})}(X^A) \\
                &                                         &                 &                                         & \ddots \\
      \end{pmatrix}
    \ .
    \end{align}
$T'(X^A)$ are the Toeplitz operators for 
$X^A$ with charge $0$ representation.

The Toeplitz operators $T(\partial_{(a)} X^A)$
are given by 
    \begin{equation}
      (\text{the $(I,J)$ element of the $(s,s')$ block in $T(\partial_{(\pm)} X^A)$})
      = T^{(s, s')} \left(C^{\pm, s'^*; s} \omega_{c \pm} \partial^c X^A\right)_{IJ} \ . 
    \end{equation}
Here, since the symplectic form $\omega$
is expressed in terms of the vielbeins
as $\omega = e^1 \wedge e^2$, we have
$\omega_{cb} = e_c^{\mu} e_b^{\nu} \omega_{\mu \nu} = e_c^{\mu} e_b^{\nu} (e^1_{\mu} e^2_{\nu} - e^1_{\nu} e^2_{\mu}) = \delta_c^1 \delta_b^2 - \delta_c^2 \delta_b^1$. This gives
$\omega_{+ -} = \frac{i}{2}$. Using this equality,
$\partial^{\pm} X^A = 2\partial_{\mp} X^A $
and \eqref{eq3.33}, we obtain
    \begin{equation}
      (\text{the $(I,J)$ element of the $(s,s')$ block in $T(\partial_{(\pm)} X^A)$})
      = \delta_{s', s \pm 1} T^{(s, s \pm 1)} \left(\mp i\partial_{\pm} X^A\right)_{IJ} \ .
      \label{eq3.41}
    \end{equation}
In $T(\partial_{(\pm)} X^A)$,
only the $(s, s \pm 1)$ blocks are nonzero.
For instance, $T(\partial_{(+)} X^A)$ take the form
  \small
    \begin{align}
      T(\partial_{(+)} X^A) =
      \begin{pmatrix}
        \ddots  & \ddots & \ddots \\
                & \ddots & \ddots  & T^{(s-\frac{3}{2},s-\frac{1}{2})}(-i \partial_+ X^A)  &                                 & \text{\huge{0}}                                      \\
                &        & \ddots  & 0                                                     & T^{(s-1,s)}(-i \partial_+ X^A)  \\
                &        &         & 0                                                     & 0                               & T^{(s-\frac{1}{2},s+\frac{1}{2})}(-i \partial_+ X^A) \\
                &        &         &                                                       & 0                               & 0                                                     & \ddots \\
                &        & \text{\huge{0}}        &                                        &                                 & 0                                                     & \ddots \\
                &        &         &                                                       &                                 &                                                       & \ddots \\
      \end{pmatrix}
      \ .
    \end{align}
  \normalsize

By using the above results, $\mathcal{P}_{(\pm)}$ are written as
    \begin{align}
      (\text{the $s$ block in 
      $\mathcal{P}_{(\pm)} T(\varphi)$})
      &= \mp i \hbar_{p}^{-1} T^{(s, s \pm 1)} \left(\partial_{\pm} X^A\right) \left[T(X^A \bm{1}), T^{(s \pm 1, 0)}\left(\varphi^{\langle s \pm 1 \rangle}\right)\right] \nonumber\\
      & \quad \pm \frac{i}{2} \hbar_p^{-1} \left[T^{(s, s \pm 1)} \left(\partial_{\pm} X^A\right), \ T(X^A \bm{1})\right] T^{(s \pm 1, 0)} \left(\varphi^{\langle s \pm 1 \rangle}\right)  \ .
      \label{eq3.42}
    \end{align}
This serves as a finite-matrix regularization for 
$-i\nabla_{(\pm)}$:
    \begin{equation}
      \mathcal{P}_{(\pm)} T(\varphi)
      = T(-i \nabla_{(\pm)} \varphi) + O\left(\frac{1}{p}\right)
      \ . 
      \label{eq3.44}
    \end{equation}

By putting $n=1$ in \eqref{eq3.31},
$[\mathcal{P}_{(+)}, \mathcal{P}_{(-)}] T(\varphi)$
is given by
    \begin{align}
      &(\text{the $s$ block in $[\mathcal{P}_{(+)}, \mathcal{P}_{(-)}] T(\varphi)$}) \nonumber\\
      &= T^{(s, s+1)} (\partial_+ X^A) T^{(s+1, s)} (\partial_- X^B) \nonumber\\
      &\quad \times T^{(s, s)} \left(-\frac{1}{4} (G^{\alpha \beta} G^{\gamma \delta} - G^{\delta \beta} G^{\gamma \alpha}) (\partial_{\alpha} X^A) R^{E_s}_{\beta \gamma} (\partial_{\delta} X^B)\right) T^{(s, 0)} \left(\varphi^{\langle s \rangle}\right) + O\left(\frac{1}{p}\right) \ .
    \end{align}
Further calculation yields
    \begin{align}
      &(\text{the $s$ block in 
      $[\mathcal{P}_{(+)}, \mathcal{P}_{(-)}] T(\varphi)$}) \nonumber\\
      &= T^{(s, s)} \left((\partial_+ X^A) (\partial_- X^B) (-1) R^{E_s}_{12} W^{ab} (\partial_a X^A) (\partial_b X^B)\right) T^{(s, 0)} \left(\varphi^{\langle s \rangle}\right) + O\left(\frac{1}{p}\right) \ ,
      \label{eq3.45}
    \end{align}
where 
$R^{E_s}_{12} = e_1^{\beta} e_2^{\gamma} R^{E_s}_{\beta \gamma} = 
e_1^{\beta} e_2^{\gamma} \left(\partial_{\beta} A^{E_s}_{\gamma} - 
\partial_{\gamma} A^{E_s}_{\beta}\right)$,
and the Poisson tensor
$W^{ab}$ is an anti-symmetric tensor with
$W^{12} = 1$.

Next, we construct the Toeplitz operator
$\bar{T}(\varphi)$ for
$\varphi \in \Gamma(E_{\text{reg}})$ that is a square matrix.
By expanding $\varphi$ as in \eqref{eq3.51},
we see from \eqref{eq3.46} that the Toeplitz operator is
given by 
    \begin{align}
      (\text{the $(I,J)$ element of $(s,s')$ block in
      $\bar{T}(\varphi)$})
      &= T^{(s, s')} \left(\sum_{s''} C^{s''^*, s'^*; s} \varphi^{\langle s'' \rangle}\right)_{IJ} \nonumber\\
      &= T^{(s, s')} \left(\varphi^{\langle s-s' \rangle}\right)_{IJ} \ . \label{eq3.64}
    \end{align}
where $I=1, \ldots, \mathrm{dimKer}D^{(E_{s})}$ and
$J=1, \ldots, \mathrm{dimKer}D^{(E_{s'})}$. 
In the second equality, we have used
$C^{s''^*, s'^*; s} = \delta_{s'', s-s'}$. 
\eqref{eq3.64} is expressed as
    \begin{align}
      \bar{T}(\varphi)=
      \begin{pmatrix}
        \ddots  & \ddots                                                                 & \ddots                                                                & \ddots \\
        \ddots  & T^{(s-\frac{1}{2},s-\frac{1}{2})}(\varphi^{\langle 0 \rangle}) & T^{(s-\frac{1}{2},s)}(\varphi^{\langle -\frac{1}{2} \rangle}) & T^{(s-\frac{1}{2},s+\frac{1}{2})}(\varphi^{\langle -1 \rangle})  & \ddots \\ 
        \ddots  & T^{(s,s-\frac{1}{2})}(\varphi^{\langle \frac{1}{2} \rangle})   & T^{(s,s)}(\varphi^{\langle 0 \rangle})                        & T^{(s,s+\frac{1}{2})}(\varphi^{\langle -\frac{1}{2} \rangle})    & \ddots \\
        \ddots  & T^{(s+\frac{1}{2},s-\frac{1}{2})}(\varphi^{\langle 1 \rangle}) & T^{(s+\frac{1}{2},s)}(\varphi^{\langle \frac{1}{2} \rangle})  & T^{(s+\frac{1}{2},s+\frac{1}{2})}(\varphi^{\langle 0 \rangle})   & \ddots \\
                & \ddots                                                                 & \ddots                                                                & \ddots                                                                   & \ddots \\
      \end{pmatrix}
    \end{align}
  \eqref{eq3.64} agrees with
the Toeplitz operator for 
$\varphi \in \Gamma(E_{\text{reg}})$
obtained in \cite{Hattori:2024btt}. 
Introducing a cutoff $\Xi$ 
and restricting $s$ and $s'$ such that 
$|s|, |s'| \leq \Xi$ make 
$\bar{T}(\varphi)$ a square matrix with a finite size. 
We take the limits $\Xi \rightarrow \infty$
and $p \rightarrow \infty$ with $\Xi \ll p$ kept. 
\eqref{eq3.47} gives a finite-size matrix 
$\bar{\mathcal{P}}_{(\pm)}$
for $-i\nabla_{(\pm)}$ as
    \begin{align}
      (\text{the $(s,s')$ block in 
      $\bar{\mathcal{P}}_{(\pm)} \bar{T}(\varphi)$})
      &= \mp i \hbar_{p}^{-1} T^{(s, s \pm 1)} \left(\partial_{\pm} X^A\right) \left[T(X^A \bm{1}), \bar{T}^{(s \pm 1, s')}\left(\varphi^{\langle s - s' \pm 1 \rangle}\right)\right] \nonumber\\
      & \quad \pm \frac{i}{2} \hbar_p^{-1} \left[T^{(s, s \pm 1)} \left(\partial_{\pm} X^A\right), \ T(X^A \bm{1})\right] \bar{T}^{(s \pm 1, s')} \left(\varphi^{\langle s - s' \pm 1 \rangle}\right) \ , \label{eq3.65}
    \end{align}
where $[T(X^A), \bar{T}(\varphi)]$
are ordinary commutators.
$T(X^A)$ and $T(\partial_{(a)} X^A)$ in \eqref{eq3.65}
are given in \eqref{eq3.39} and 
\eqref{eq3.41}, respectively. 
\eqref{eq3.65} gives a finite-matrix regularization with a square matrix
for $-i\nabla_{(\pm)}$: 
    \begin{equation}
      \bar{\mathcal{P}}_{(\pm)} \bar{T}(\varphi)
      = \bar{T}(-i \nabla_{(\pm)} \varphi) + O\left(\frac{1}{p}\right) \ ,
    \end{equation}
where \eqref{eq3.48} gives 
$\bar{T}(-i \nabla_{(\pm)} \varphi)$ as 
    \begin{align}
      &(\text{the $(I,J)$ element of $(s,s')$ block in
      $\bar{T}(-i \nabla_{(\pm)} \varphi)$}) \nonumber\\
      &= T^{(s, s')} \left(-i \sum_{s'', s'''} C^{s''^*, s'^*; s} C^{\pm, s'''^*; s''} \left(\nabla_{\pm} \varphi^{\langle s''' \rangle}\right)\right)_{IJ} \nonumber\\
      &= T^{(s, s')} \left(-i \nabla_{\pm} \varphi^{\langle s - s' \pm 1 \rangle}\right)_{IJ} 
      \ . 
    \end{align}
Here, in the second equality,
we have used 
$C^{s''^*, s'^*; s} = \delta_{s'', s-s'}$
and $C^{\pm, s'''^*; s''} = \delta_{s''', s'' \pm 1}$. $\bar{\mathcal{P}}_{(\pm)}$ in \eqref{eq3.65}
agree with the corresponding finite-size matrices ${\mathcal{P}}_{(\pm)}$ obtained in
\cite{Hattori:2024btt}\footnote{At first sight, the overall signs are
different. However, this is because
the finite-size matrices for $-i\nabla_{(\pm)}$
are obtained in this paper, while those for 
$i\nabla_{(\pm)}$ are obtained in \cite{Hattori:2024btt}. }. 

From \eqref{eq3.49}, 
$[\bar{\mathcal{P}}_{(+)}, \bar{\mathcal{P}}_{(-)}] \bar{T}
(\varphi)$ is calculated as
    \begin{align}
      &(\text{the $(s,s')$ block in 
      $[\bar{\mathcal{P}}_{(+)}, \bar{\mathcal{P}}_{(-)}] \bar{T}(\varphi)$}) \nonumber\\
      &= T^{(s, s+1)} (\partial_+ X^A) T^{(s+1, s)} (\partial_- X^B) \nonumber\\
      &\quad \times \left[ T^{(s, s)} \left(-\frac{1}{4} (G^{\alpha \beta} G^{\gamma \delta} - G^{\delta \beta} G^{\gamma \alpha}) (\partial_{\alpha} X^A) R^{E_s}_{\beta \gamma} (\partial_{\delta} X^B)\right) T^{(s, s')} \left(\varphi^{\langle s-s' \rangle}\right) \right. \nonumber\\
      &\quad \qquad \left. - T^{(s, s')} \left(\varphi^{\langle s-s' \rangle}\right) T^{(s', s')} \left(-\frac{1}{4} (G^{\alpha \beta} G^{\gamma \delta} - G^{\delta \beta} G^{\gamma \alpha}) (\partial_{\alpha} X^A) R^{E_{s'}}_{\beta \gamma} (\partial_{\delta} X^B)\right) \right] + O\left(\frac{1}{p}\right) \nonumber\\
      &= T^{(s, s)} \left((\partial_+ X^A) (\partial_- X^B) (-1) R^{E_s}_{12} W^{ab} (\partial_a X^A) (\partial_b X^B)\right) T^{(s, s')} \left(\varphi^{\langle s-s' \rangle}\right) \nonumber\\
      &\quad - T^{(s, s')} \left(\varphi^{\langle s-s' \rangle}\right) T^{(s', s')} \left((\partial_+ X^A) (\partial_- X^B) (-1) R^{E_{s'}}_{12} W^{ab} (\partial_a X^A) (\partial_b X^B)\right) + O\left(\frac{1}{p}\right) \ . 
    \end{align}

\section{Examples} \label{sec4}
In this section, as concrete examples, we construct finite-size matrices that serve as 
a finite-matrix regularization for $-i\nabla_{(a)}$ on
$T^{2n}$ and $S^2$. We consider the case where $T(\varphi) \ (\varphi \in \Gamma(E_{\text{reg}}))$ is a rectangular
matrix.

\subsection{$T^{2n}$}
In this subsection, we construct $\mathcal{P}_{(a)}$ for
$-i\nabla_{(a)}$ on $2n$-dimensional torus $T^{2n} \cong (S^1)^{2n}$
and see that $[\mathcal{P}_{(a)}, \mathcal{P}_{(b)}] = 0$ in the large-$p$ limit. 
See \cite{Adachi:2022mln} for the detailed derivation of the Dirac zero modes and the Toeplitz operators $T^{(1,1)}(X^A)$. 

We define an equivalence relation $\sim$ in $\mathbb{R}^{2n}$ with the flat metric
as follows. For all $\forall x = (x^1, x^2, \ldots, x^{2n}) \in \mathbb{R}^{2n}$ and 
$\forall x' = (x'^1, x'^2, \ldots, x'^{2n}) \in \mathbb{R}^{2n}$, 
    \begin{equation}
      x \sim x' \ 
      \Leftrightarrow \ \exists m_a \in \mathbb{Z}: \ x^a - x'^a = 2 \pi l_a m_a \quad (a=1, 2, \ldots, 2n) \ , 
    \end{equation}
where $l_a$ are certain constants. 
$T^{2n}$ is defined as a quotient space of $\mathbb{R}^{2n}$ divided by $\sim$:
    \begin{equation}
      T^{2n} = \mathbb{R}^{2n}/{\sim} \ . 
    \end{equation}
The flat metric on $T^{2n}$ and the associated K\"{a}hler form are given by
    \begin{equation}
      g = \sum_{a=1}^{2n} \mathrm{d}x^a \otimes \mathrm{d}x^a \ , 
      \quad 
      \omega = \sum_{m=1}^{n} \mathrm{d}x^{2m-1} \wedge \mathrm{d}x^{2m} 
      = i \mathrm{d}z^{\mu} \wedge \mathrm{d}\bar{z}^{\mu} \ .
    \end{equation}
Here, the real and complex coordinates are related as 
$z^{\mu} = \frac{1}{\sqrt{2}} (x^{2\mu -1} + ix^{2\mu}) \ (\mu = 1, 2, \ldots, n)$. 
We introduce functions $u_m$ and $v_m$ on $T^{2n}$ as
    \begin{equation}
      u_m = e^{ix^{2m-1}/l_{2m-1}}, \quad v_m = e^{ix^{2m}/l_{2m}} \quad (m=1, 2, \ldots, n) \ . 
    \end{equation}
The coordinate functions $X^A$ that embed $T^{2n}$ isometrically into  
$\mathbb{R}^{4n}$ are given by
    \begin{align}
      & X^{4m-3} = l_{2m-1}\cos(\frac{x^{2m-1}}{l_{2m-1}}) = \frac{l_{2m-1}}{2}(u_m + u_m^*) \ ,\label{eq4.5} \\
      & X^{4m-2} = l_{2m-1}\sin(\frac{x^{2m-1}}{l_{2m-1}}) = \frac{l_{2m-1}}{2i}(u_m - u_m^*) \ , \label{eq4.6} \\ 
      & X^{4m-1} = l_{2m}\cos(\frac{x^{2m}}{l_{2m}}) = \frac{l_{2m}}{2}(v_m + v_m^*) \ , \label{eq4.7} \\
      & X^{4m} = l_{2m}\sin(\frac{x^{2m}}{l_{2m}}) = \frac{l_{2m}}{2i} (v_m - v_m^*) \ , \label{eq4.8}
    \end{align}
where $m = 1, 2, \ldots, n$.

Because $T^{2n}$ is a spin manifold, we can use the spin bundle $S$.
The flatness of $T^{2n}$ implies that the spin connection of $S$ vanishes.
The 2-cycles of $T^{2n}$ are $T^{2}$, and 
the curvature $R^L = -ik\omega$ is nonzero on $T^2$ parametrized by $(x^{2m-1}, x^{2m})$
for $m=1, 2, \ldots, n$.
As stated in section 2.2, a manifold $M$ is quantizable when
$R^L = -ik\omega$ and for an appropriate choice of a constant $k$ there exists a complex line bundle such that
$\frac{i}{2 \pi} \int_{\Sigma} R^L \in \mathbb{Z}$ for any 2-cycle $\Sigma \subseteq M$.
As for $T^{2n}$, this condition is  satisfied for $k$ and $l_a$ that satisfy
    \begin{equation}
      \forall m \in \{1, 2, \ldots, n\}: \ 
      q_m 
      = \frac{i}{2 \pi} \int_{T^2} R^L 
      = 2 \pi k l_{2m-1} l_{2m} 
      \in \mathbb{N} \ .
    \end{equation}
This is equivalent to a condition that the ratio between the areas for arbitrary $m$ and $m'$
$\frac{l_{2m-1} l_{2m}}{l_{2m'-1} l_{2m'}}$ is a rational number.

We construct an orthonormal basis of the Dirac zero modes on $T^{2n}$. 
Let $E_r$ be a fiber bundle over $T^{2n}$ whose fiber is the representation space of an irreducible 
representation $r$ of $\mathrm{Spin}(2n)$. 
As stated in section
\ref{sec2.2}, the normalizable zero modes of the Dirac operator
$D^{(E_r)}$ on 
$\Gamma(S \otimes L^{\otimes p} \otimes E_r)$
are given by 
$\psi = f \ket{+}^{\otimes n}$, 
where $f \in \Gamma(L^{\otimes p} \otimes E_r)$
and $\ket{+}$ is a two-dimensional spinor $(1, 0)^{\mathsf{T}}$.
Noting that the spin connection of $E_r$ is
zero since $T^{2n}$ is flat, 
we see that $f = f^{(p)} \ket{r}$, where
$f^{(p)} \in \Gamma(L^{\otimes p})$ and 
$\ket{r}$ is an element of the fiber of $E_r$.
\eqref{eq2.48} reduces, therefore, the equation for 
the Dirac zero modes $D^{(E_r)} \psi = 0$
to an equation for $f^{(p)}$
    \begin{equation}
      \left(\partial_{\bar{\mu}} + p A^{L}_{\bar{\mu}}\right) f^{(p)} = 0 \ .
      \label{eq4.10}
    \end{equation}
  
We fix the connection 1-form $A^L$ 
of the complex line bundle $L$ to
    \begin{equation}
      A^L 
      = -i k \sum_{m=1}^{n} x^{2m-1} \mathrm{d}x^{2m}
      = -\frac{k}{2}(z^{\mu} + \bar{z}^{\mu}) (\mathrm{d}z^{\mu} - \mathrm{d}\bar{z}^{\mu}) \ .
    \end{equation}  
In fact, this satisfies $R^L = \mathrm{d}A^L = -ik\omega$.
The equation for the zero modes becomes
    \begin{equation}
      \left(\partial_{\bar{\mu}} + \frac{kp}{2} (z^{\mu} + \bar{z}^{\mu})\right) f^{(p)} = 0 \ .
    \end{equation}

$A^L(x)$ is invariant under a coordinate transformation,
so that $f^{(p)}(x)$ is also invariant. 
On the other hand, since the former is transformed as
$A^L(x) \mapsto A^L(x) - \mathrm{d}\lambda(x)$ under
a coordinate transformation
$x^{2m-1} \mapsto x^{2m-1} + 2 \pi l_{2m-1}$,
the latter should be transformed as
$f^{(p)}(x) \mapsto e^{p \lambda(x)} f^{(p)}(x)$,
where $\lambda(x) = i 2 \pi k l_{2m-1} x^{2m}$. 
This can be viewed as a boundary condition for
$f^{(p)}(x)$. 
Noting this boundary condition and the fact that
the differential equation   \eqref{eq4.10} is closed
on $T^2$ with the coordinates
$(x^{2m-1}, x^{2m})$, we see that general solutions are
given by 
    \begin{equation}
      f^{(p)}(x)
      = \prod_{m=1}^{n} \left(e^{-\frac{kp}{2}(x^{2m-1})^2} \phi_m(x^{2m-1} + ix^{2m})\right) \ .
    \end{equation}
The boundary condition for
$f^{(p)}$ are reinterpreted as the ones for
$\phi_m \ (m=1, 2, \ldots, n)$: 
    \begin{align}
      & \phi_m(x^{2m-1} + ix^{2m} + i2\pi l_{2m}) = \phi_m(x^{2m-1} + ix^{2m}) \ , \\
      & \phi_m(x^{2m-1} + ix^{2m} + 2\pi l_{2m-1}) = e^{-i \pi p q_m \tau_m} e^{p q_m (x^{2m-1} + ix^{2m})/l_{2m}} \phi_m(x^{2m-1} + ix^{2m}) \ ,
    \end{align}
where $\tau_m = i l_{2m-1}/l_{2m}$
is the moduli parameter of the $m$-th $T^2$. 
The number of  linearly independent 
$\phi_m(x^{2m-1} + ix^{2m})$ 
that satisfy the above two boundary conditions is $p q_m$,
and they are given by 
    \begin{equation}
      \phi_m(x^{2m-1} + ix^{2m}) 
      = c^{(p)}_{I_m} \sum_{l \in \mathbb{Z}} e^{i \pi \left(l + \frac{I_m}{p q_m}\right)^2 p q_m \tau_m} e^{\left(l + \frac{I_m}{p q_m}\right) \frac{p q_m}{l_{2m}}(x^{2m-1} + ix^{2m})} \ ,
    \end{equation}
where the independent solutions are labeled by
$I_m \in \{0, 1, \ldots, p q_m - 1\}$, and
$c^{(p)}_{I_m}$ is a certain complex constant.

In this way,
we find an orthonormal basis for 
the zero modes of the Dirac operator $D^{(E_r)}$ on
$\Gamma(S \otimes L^{\otimes p} \otimes E_r)$:
    \begin{equation}
      \psi_{(I, I')}(x) = f^{(p)}_I(x) \ket{+}^{\otimes n} \ket{r, I'} \ .
    \end{equation}
Here, 
$I = (I_1, I_2, \ldots, I_n)$, $I' = 1, 2, \ldots, d_r$, and 
$f^{(p)}_I(x)$ is given by
    \begin{equation}
      f^{(p)}_I(x)
      = \prod_{m=1}^{n}f^{(p)}_{I_m}(x^{2m-1}, x^{2m}) \ . 
    \end{equation}
For $m = 1, 2, \ldots, n$, $I_m \in \{0, 1, \ldots, p q_m - 1\}$, and $f^{(p)}_{I_m}(x^{2m-1}, x^{2m})$ is given by
    \begin{equation}
      f^{(p)}_{I_m}(x^{2m-1}, x^{2m})
      = \left(\frac{kp}{4{\pi}^3 l_{2m}^2}\right)^{\frac{1}{4}} e^{-\frac{kp}{2}(x^{2m-1})^2} \sum_{l \in \mathbb{Z}} e^{i \pi \left(l + \frac{I_m}{p q_m}\right)^2 p q_m \tau_m} e^{\left(l + \frac{I_m}{p q_m}\right) \frac{p q_m}{l_{2m}}(x^{2m-1} + ix^{2m})} \ . \label{eq4.19}
    \end{equation}
$\ket{r, I'} \ (I' = 1, 2, \ldots, d_r)$
are an orthonormal basis for the fiber of
$E_r$. The independent zero modes are 
labeled by $(I, I')$. 
  
We construct the Toeplitz operators for
the embedding functions $X^A$. 
From \eqref{eq3.13}, we obtain
    \begin{align}
      & (\text{the $((I, I'), (J, J'))$ element of the
      $((r, j), (r', l))$ block in 
      $T(X^A)$}) \nonumber\\
      &= \delta^{\langle r \rangle \langle r' \rangle} \delta_{jl} T^{(r, r)} (X^A \bm{1}_{E_r})_{(I, I'), (J, J')} \nonumber\\
      &= \delta^{\langle r \rangle \langle r' \rangle} \delta_{jl} \delta_{I'J'} \int_{T^{2n}} \mu \left(f^{(p)}_I\right)^* X^A f^{(p)}_J \nonumber\\
      &= \delta^{\langle r \rangle \langle r' \rangle} \delta_{jl} \delta_{I'J'} T^{(1, 1)} (X^A)_{IJ}
    \end{align}
In the last equality, 
we have put $T^{(1, 1)} (X^A)_{IJ} = \int_{T^{2n}} \mu \left(f^{(p)}_I\right)^* X^A f^{(p)}_J$.
From \eqref{eq3.13.2}, we obtain
    \begin{align}
      (\text{the $(I,J)$ element in $T'(X^A)$})
      = T^{(1, 1)} (X^A)_{IJ} \ .
    \end{align}

We can calculate $T^{(1, 1)} (X^A)_{IJ}$
by using the following formula:
    \begin{align}
      &\int_{0}^{2\pi l_{2m-1}} dx^{2m-1} \int_{0}^{2\pi l_{2m}} dx^{2m} \left(f^{(p)}_{I_m}\right)^* e^{i \frac{a x^{2m-1}}{l_{2m-1}}} e^{i \frac{b x^{2m}}{l_{2m}}} f^{(p)}_{J_m} \nonumber\\
      &= e^{-\frac{1}{4kp} \left(\frac{a^2}{l_{2m-1}^2} + i\frac{2ab}{l_{2m-1} l_{2m}} + \frac{b^2}{l_{2m}^2}\right)} e^{i \frac{2\pi a I_m}{pq_m}} \delta^{(\mathrm{mod} \ pq_m)}_{I_m - J_m - b, 0} \ ,
    \end{align}
where $a, b \in \mathbb{Z}$, 
$f^{(p)}_{I_m}$ are defined in \eqref{eq4.19}, and 
$\delta^{(\mathrm{mod} \ n)}_{a,b}$ are defined by
    \begin{equation}
      \delta^{(\mathrm{mod} \ n)}_{a, b} =
        \begin{dcases}
          1 & \text{if $a-b \in n\mathbb{Z}$}, \\
          0 & \text{otherwise} \ .       \end{dcases}
    \end{equation}

  $T^{(1, 1)} (X^A)_{IJ}$ are calculated as follows.
    \begin{align}
      & T^{(1, 1)} (X^{4m-3})_{IJ} 
      = \frac{l_{2m-1}}{2} \left(U^{(p)}_m + U^{(p)\dagger}_m \right)_{IJ} \ , 
      && T^{(1, 1)} (X^{4m-2})_{IJ} 
      = \frac{l_{2m-1}}{2i} \left(U^{(p)}_m - U^{(p)\dagger}_m \right)_{IJ} \ ,
      \nonumber \\
      & T^{(1, 1)} (X^{4m-1})_{IJ} 
      = \frac{l_{2m}}{2} \left(V^{(p)}_m + V^{(p)\dagger}_m \right)_{IJ} \ ,
      && T^{(1, 1)} (X^{4m})_{IJ} 
      = \frac{l_{2m}}{2i} \left(V^{(p)}_m - V^{(p)\dagger}_m \right)_{IJ} \ .
      \label{eq4.22}
    \end{align}
Here,
    \begin{align}
      & \left(U^{(p)}_m\right)_{IJ}
      = (U_{p q_m})_{I_m J_m} \prod_{1 \leq k \leq n, \ k \neq m} \delta_{I_k J_k} \ , \\
      & \left(V^{(p)}_m\right)_{IJ}
      = (V_{p q_m})_{I_m J_m} \prod_{1 \leq k \leq n, \ k \neq m} \delta_{I_k J_k} \ .
    \end{align}
Note that
$I = (I_1, I_2, \ldots, I_n)$ and $J = (J_1, J_2, \ldots, J_n)$
and that for $m=1, 2, \ldots, n$, 
$I_m, J_m \in \{0, 1, \ldots, p q_m-1\}$. 
$U_{p q_m}$ and $V_{p q_m}$ are the so-called clock-shift matrices
defined by
    \begin{align}
      &U_{p q_m} = e^{- \frac{1}{4kpl_{2m-1}^2}}
        \begin{pmatrix}
          1  & {} & {} & {} \\
          {} & e^{i\frac{2 \pi}{p q_m}} & {} & {} \\
          {} & {} & \ddots & {} \\
          {} & {} & {} & e^{i\frac{2(p q_m - 1)\pi}{p q_m}}
        \end{pmatrix} \ , \\
      & V_{p q_m} = e^{- \frac{1}{4kpl_{2m}^2}}
        \begin{pmatrix}
          {} & {} & {}     & {} & 1  \\
          1  & {} & {}     & {} & {} \\
          {} & 1  & {}     & {} & {} \\
          {} & {} & \ddots & {} & {} \\
          {} & {} & {}     & 1  & {} \\
        \end{pmatrix} \ .
    \end{align}
$U_{p q_m}$ and $V_{p q_m}$ satisfy
the 't Hooft-Weyl algebra
$U_{p q_m} V_{p q_m} = e^{i\frac{2 \pi}{p q_m}} V_{p q_m} U_{p 
q_m}$.

Next, we construct 
$T(\partial_{(a)} X^A)$. From \eqref{eq3.14}, we obtain
    \begin{align}
      & (\text{the $((I, I'), (J, J'))$ element
      of the $((r, j), (r', l))$ block in
      $T(\partial_{(a)} X^A)$}) \nonumber\\
      &= T^{(r, r')} \left(\sqrt{\frac{d_{r'}}{d_r}} C^{v, r'^*; r}_{b(a), (l); (j)} \omega_{cb} \partial^c X^A\right)_{(I, I'), (J, J')} \nonumber\\
      &= \int_{T^{2n}} \mu \sum_{I'', J''} \delta_{I', I''} \left(f^{(p)}_I\right)^* \sqrt{\frac{d_{r'}}{d_r}} C^{v, r'^*; r}_{b(a), J''(l); I''(j)} \omega_{cb} \partial^c X^A f^{(p)}_J \delta_{J', J''} \nonumber\\
      &= \sqrt{\frac{d_{r'}}{d_r}} C^{v, r'^*; r}_{b(a), J'(l); I'(j)} \omega_{cb} \int_{T^{2n}} \mu \left(f^{(p)}_I\right)^* \partial^c X^A f^{(p)}_J \nonumber\\
      &= \sqrt{\frac{d_{r'}}{d_r}} C^{v, r'^*; r}_{b(a), J'(l); I'(j)} \omega_{cb} T^{(1, 1)} (\partial^c X^A)_{IJ} \ .
    \end{align}
In the third equality, since $\mathrm{d}x^{2m-1} \wedge \mathrm{d}x^{2m}$, 
we take out 
$\omega_{cb} = \delta_c^{\mu} \delta_b^{\nu} \omega_{\mu \nu}$
outside the integral.
In the last equality, we put
$T^{(1, 1)} (\partial^c X^A)_{IJ} = \int_{T^{2n}} \mu 
\left(f^{(p)}_I\right)^* \partial^c X^A f^{(p)}_J$. 
By using \eqref{eq4.5}--\eqref{eq4.8}, 
for $c = 1, 2, \ldots, 2n$, $A = 1, 2, \ldots, 4n$,
$\partial^c X^A = \delta_c^{\mu} \partial_{\mu} X^A$
are calculated as 
    \begin{equation}
      \partial^c X^A = 
        \begin{dcases}
          -\frac{1}{l_c} X^{2c} & \text{if $A=2c-1$}, \\
          \frac{1}{l_c} X^{2c-1} & \text{if $A=2c$}, \\
          0 & \text{otherwise} \ . 
        \end{dcases}
    \end{equation}
Thus, we have
    \begin{align}
      & (\text{the $((I, I'), (J, J'))$ element of
      the $((r, j), (r', l))$ block in
      $T(\partial_{(a)} X^A)$}) \nonumber\\
      &= 
        \begin{dcases}
          \sqrt{\frac{d_{r'}}{d_r}} C^{v, r'^*; r}_{2m(a), J'(l); I'(j)} \left(-\frac{1}{l_{2m-1}}\right)T^{(1, 1)} (X^{4m-2})_{IJ} & \text{if $A=4m-3$}  \ , \\
          \sqrt{\frac{d_{r'}}{d_r}} C^{v, r'^*; r}_{2m(a), J'(l); I'(j)} \frac{1}{l_{2m-1}} T^{(1, 1)} (X^{4m-3})_{IJ} & \text{if $A=4m-2$} \ , \\
          \sqrt{\frac{d_{r'}}{d_r}} C^{v, r'^*; r}_{2m-1(a), J'(l); I'(j)} \frac{1}{l_{2m}} T^{(1, 1)} (X^{4m})_{IJ} & \text{if $A=4m-1$} \ , \\
          \sqrt{\frac{d_{r'}}{d_r}} C^{v, r'^*; r}_{2m-1(a), J'(l); I'(j)} \left(-\frac{1}{l_{2m}}\right)T^{(1, 1)} (X^{4m-1})_{IJ} & \text{if $A=4m$} \ , \\
        \end{dcases}
    \end{align}
where, $T^{(1, 1)} (X^A)_{IJ}$ are given in
\eqref{eq4.22}. 

$[U^{(p)}_m, U^{(p) \dagger}_m] = 0$
leads to $[T^{(1, 1)} (X^{4m-3}), T^{(1, 1)} (X^{4m-2})] = 0$,
while $[V^{(p)}_m, V^{(p) \dagger}_m] = 0$ to $[T^{(1, 1)} (X^{4m-1}), T^{(1, 1)} (X^{4m})] = 0$. Thus,
we have 
    \begin{equation}
      T(\partial_{(a)} X^A) T(X^A) = T(X^A) T(\partial_{(a)} X^A) = 0 \ .
    \end{equation}
Note that the sum over the repeated indices $A$ are taken.

We therefore find that the Toeplitz operators  $\mathcal{P}_{(a)}$
for $T^{2n}$ are given by 
    \begin{equation}
      \mathcal{P}_{(a)} T(\varphi) 
      = -\hbar_p^{-1} T(\partial_{(a)} X^A) T(\varphi) T'(X^A) \ . 
    \end{equation}
  
Since $T^{2n}$ is flat, the curvature of
$R^{E_r}_{\beta \gamma}$ of 
$E_r$ is zero. Thus, we see from \eqref{eq3.31}
that 
$[\mathcal{P}_{(a)}, \mathcal{P}_{(b)}]T(\varphi)$
behave as
    \begin{equation}
      [\mathcal{P}_{(a)}, \mathcal{P}_{(b)}]T(\varphi) = O\left(\frac{1}{p}\right) \ . 
    \end{equation}
Namely,  $\mathcal{P}_{(a)}$ and 
$\mathcal{P}_{(b)}$ commute in the $p \rightarrow \infty$:
    \begin{equation}
      [\mathcal{P}_{(a)}, \mathcal{P}_{(b)}] = 0 \ .
    \end{equation}
The flatness of $T^{2n}$ implies
$[\nabla_{(a)}, \nabla_{(b)}] = 0$, and 
the above result reproduces this.

\subsection{$S^2$} 
In this subsection, we construct $\mathcal{P}_{(a)}$
for a unit sphere $S^2$ and see
that in the large-$p$ limit 
$\mathcal{P}_{(\pm)}$ and 
$[\mathcal{P}_{(+)}, \mathcal{P}_{(-)}]$ forms the $\mathfrak{su}(2)$ 
Lie algebra \cite{Hattori:2024btt}.

Let $X^A \ (A=1, 2, 3)$ be isometric embedding functions of $S^2$
into $\mathbb{R}^3$ 
and 
$(z, \bar{z})$
be a stereographic projection of $S^2$
from the north pole on the $X^1-X^2$ plane. 
$X^A$ are related to $(z, \bar{z})$ as
    \begin{equation}
      X^1 = \frac{z + \bar{z}}{1 + |z|^2}, \quad X^2 = -i \frac{z - \bar{z}}{1 + |z|^2}, \quad X^3 = \frac{1 - |z|^2}{1 + |z|^2} \ .
    \end{equation}
As in section \ref{sec2.22}, 
the induced metric, the zweibein, the inverse of
the zweibein and the spin connection 
are given in \eqref{S^2 metric}, \eqref{S^2 zweibein}, \eqref{S^2 inverse of zweibein} and \eqref{S^2 spin connection},
respectively.
Since $M$ is a two-dimensional manifold,
the symplectic form $\omega$
given by
    \begin{equation}
      \omega = i\sqrt{\mathrm{det}g} \ \mathrm{d}z \wedge \mathrm{d}\bar{z} = i\frac{2}{(1+|z|^2)^2} \mathrm{d}z \wedge \mathrm{d}\bar{z} 
    \end{equation}
agrees with
the volume form $\mu$, 
and $k$ in \eqref{eq2.41} is calculated as 
$k = 2\pi / \int_{S^2} \omega = \frac{1}{2}$.

As a connection 1-form $A^L$ of the complex line bundle $L$,
we adopt the Wu-Yang monopole:
    \begin{equation}
      A^L = -\frac{1}{2}\frac{\bar{z}\mathrm{d}z - z\mathrm{d}\bar{z}}{1+|z|^2} \ . 
    \end{equation}
This satisfies 
$R^L = \mathrm{d}A^L = -ik\omega$.

Let $E_s$ be a fiber bundle whose fiber
is the space of charge s of $\mathrm{Spin}(2)$. 
The connection 1-form $A^{E_s}$ of $E_s$ is given by
    \begin{equation}
      A^{E_s} = -is\Omega^{12} \ .
    \end{equation}
As stated in section \ref{sec2.2}, 
the normalizable zero modes of the Dirac operator
$D^{(E_r)}$ in 
$\Gamma(S \otimes L^{\otimes p} \otimes E_s)$
are given by $\psi = f \ket{+}$, where 
$f \in \Gamma(L^{\otimes p} \otimes E_s)$
and $\ket{+}$ is a two-dimensional spinor 
$(1, 0)^{\mathsf{T}}$. 
When we consider not a spin c bundle but a spin bundle,
the differential equation for 
$f$ \eqref{eq2.48} is given by 
    \begin{equation}
      \left(\partial_{\bar{\mu}} + \frac{1}{2} \sum_{m=1}^{n} \Omega_{m \bar{m} \bar{\mu}} + p A^L_{\bar{\mu}} + A^E_{\bar{\mu}}\right) f
      = 0 \ .
      \label{eq4.43}
    \end{equation}
Here, note that the indices 
$m$ and $\bar{m} \ (m = 1, \ldots, n)$
are defined by 
$V^m = \frac{1}{\sqrt{2}}(V^{2m-1} + iV^{2m})$
and $V^{\bar{m}} = \frac{1}{\sqrt{2}}(V^{2m-1} - iV^{2m})$
for vector fields $V^a$
in the local Lorentz frame, respectively.
In the case of $S^2$, \eqref{eq4.43} becomes 
    \begin{equation}
      \left(\partial_{\bar{z}} + \frac{(p + 2s - 1)z}{2(1+|z|^2)}\right) f = 0 \ .
    \end{equation}
By solving this equation, we obtain
an orthonormal basis of
$\mathrm{Ker}D^{(E_s)}$ as 
    \begin{align}
      & \psi_I(z, \bar{z}) = f_I(z, \bar{z}) \ket{+}, \\
      & f_I(z, \bar{z}) = \sqrt{\frac{p + 2s}{4\pi}} \binom{p + 2s - 1}{I}^{\frac{1}{2}} \frac{z^I}{{(1 + |z|^2)}^{\frac{p + 2s - 1}{2}}} \ ,
    \end{align}
where $I = 0, 1, \ldots, p+2s-1$ 
is an index that labels the independent
zero modes.

We construct the Toeplitz operators for $X^A$.
\eqref{eq3.39} and \eqref{eq3.40}
give $T(X^A)$ and $T'(X^A)$: 
    \begin{align}
      &(\text{the $(I,J)$ element of the $(s,s')$ block in $T(X^A)$})
      = \delta^{\langle s \rangle \langle s' \rangle} T^{(s, s)} (X^A)_{IJ}, \nonumber\\
      &T'(X^A)_{IJ} 
      = T^{(0, 0)} (X^A)_{IJ} \ .
    \end{align}
By using the formulae
    \begin{equation}
      \frac{1}{2\pi} \int i dz d\bar{z}\frac{z^a \bar{z}^b}{(1 + |z|^2)^c}
      = \delta_{ab} \frac{\Gamma(a+1) \Gamma(c-a-1)}{\Gamma(c)} \ ,
    \end{equation}
where $\Gamma(a)$ is the gamma function,
we can obtain the Toeplitz operators 
$T^{(s, s)} (X^+)$, $T^{(s, s)} (X^-)$ and $T^{(s, s)} (X^3)$
for 
$X^+ = X^1 + iX^2$, $X^- = X^1 - iX^2$ and 
$X^3$, respectively.
Some algebra gives
    \begin{align}
      & T^{(s, s)}(X^+)_{IJ} = \delta_{I-1,J} \frac{2}{p+2s+1} \sqrt{I(p+2s-I)} \ , \\
      & T^{(s, s)}(X^-)_{IJ} = \delta_{I+1,J} \frac{2}{p+2s+1} \sqrt{(I+1)(p+2s-I-1)} \ , \\
      & T^{(s, s)}(X^3)_{IJ} = \delta_{I,J} \frac{p+2s-2I-1}{p+2s+1} \ ,
    \end{align}
where $I, J = 0, 1, \ldots, p+2s-1$.
These form the $\mathfrak{s}\mathfrak{u}(2)$ Lie algebra:
    \begin{equation}
      [T^{(s, s)}(X^A), T^{(s, s)}(X^B)] = - \frac{2}{p+2s+1} i\epsilon^{ABC} T^{(s, s)}(X^C) \ .
    \end{equation}

We next construct the Toeplitz operators for 
the derivatives of $X^A$. 
From \eqref{eq3.41}, $T(\partial_{(\pm)} X^A)$
are given by
    \begin{equation}
      (\text{the $(I,J)$ element of the $(s,s')$ block in $T(\partial_{(\pm)} X^A)$})
      = \mp i \delta_{s', s \pm 1} T^{(s, s \pm 1)} (\partial_{\pm} X^A)_{IJ} \ .
    \end{equation}
$T^{(s, s \pm 1)} (\partial_{\pm} X^A)_{IJ}$
are calculated as follows:
    \begin{align}
      & T^{(s-1,s)}(\partial_+ X^+)_{IJ} = \delta_{I, J} \sqrt{\frac{(p+2s-I-1)(p+2s-I-2)}{(p+2s)(p+2s-1)}} \ , \label{eq4.52}\\
      & T^{(s-1,s)}(\partial_+ X^-)_{IJ} = - \delta_{I+2,J} \sqrt{\frac{(I+1)(I+2)}{(p+2s)(p+2s-1)}} \ , \\
      & T^{(s-1,s)}(\partial_+ X^3)_{IJ} = - \delta_{I+1,J} \sqrt{\frac{(I+1)(p+2s-I-2)}{(p+2s)(p+2s-1)}} \ , \label{eq4.54}\\
      & T^{(s+1,s)}(\partial_- X^+)_{IJ} = - \delta_{I-2,J} \sqrt{\frac{I(I-1)}{(p+2s+2)(p+2s+1)}} \ , \label{eq4.55}\\
      & T^{(s+1,s)}(\partial_- X^-)_{IJ} = \delta_{I,J} \sqrt{\frac{(p+2s-I+1)(p+2s-I)}{(p+2s+2)(p+2s+1)}} \ , \\
      & T^{(s+1,s)}(\partial_- X^3)_{IJ} = - \delta_{I-1,J} \sqrt{\frac{I(p+2s-I+1)}{(p+2s+2)(p+2s+1)}} \ .
      \label{eq4.57}
    \end{align}
Here, $I = 0, \ldots, p+2s-3,\ J = 0, \ldots, p+2s-1$ in \eqref{eq4.52}--\eqref{eq4.54}, and
$I = 0, \ldots, p+2s+1,\ J = 0, \ldots, p+2s-1$ in 
\eqref{eq4.55}--\eqref{eq4.57}.

By substituting the above
$T(X^A)$, $T'(X^A)$ and $T(\partial_{(a)} X^A)$
into \eqref{eq3.42}, we obtain $\mathcal{P}_{(a)}$ for 
$-i\nabla_{(a)}$ on $S^2$.

We examine 
$[\mathcal{P}_{(+)}, \mathcal{P}_{(-)}] T(\varphi)$
in the large-$p$ limit.
We note that $R^{E_s}_{12} = -is$ since
$R^{E_s} = \mathrm{d} A^{E_s} = -is \mathrm{d} \Omega^{12} = 2s \mathrm{d} A^L = -is\omega$ on $S^2$.
From this and \eqref{eq3.45}, we see  
    \begin{align}
      &(\text{the $s$ block in $[\mathcal{P}_{(+)}, \mathcal{P}_{(-)}] T(\varphi)$}) \nonumber\\
      &= is T^{(s, s)} \left((\partial_+ X^A) (\partial_- X^B) W^{ab} (\partial_a X^A) (\partial_b X^B)\right) T^{(s, 0)} \left(\varphi^{\langle s \rangle}\right) + O\left(\frac{1}{p}\right) \ . 
    \end{align}
$(\partial_+ X^A) (\partial_- X^B) W^{ab} (\partial_a X^A) 
(\partial_b X^B)$ is calculated as
    \begin{align}
      (\partial_+ X^A) (\partial_- X^B) W^{ab} (\partial_a X^A) (\partial_b X^B)
      &= (\partial_+ X^A) (\partial_- X^B) W^{-+} (\partial_- X^A) (\partial_+ X^B) \nonumber\\
      &= W^{-+} \delta_{+-} \delta_{+-} \nonumber\\
      &= 2i \ \frac{1}{2} \ \frac{1}{2} \nonumber\\
      &= \frac{i}{2} \ . 
    \end{align}
Here, in the first equality, we have used
$(\partial_+ X^A)(\partial_+ X^A) = \delta_{++} = 0$.
Thus, 
we obtain
    \begin{equation}
       (\text{the $(I,J)$ element of the $s$ block in $[\mathcal{P}_{(+)}, \mathcal{P}_{(-)}] T(\varphi)$})
       = -\frac{s}{2} T^{(s, 0)} \left(\varphi^{\langle s \rangle}\right)_{IJ} + O\left(\frac{1}{p}\right) \ .
       \label{eq4.60}
    \end{equation}
$[\mathcal{P}_{(+)}, \mathcal{P}_{(-)}]$ is proportional to
the charge of the fields. This is expected from
\eqref{eq2.37} 
because the charges are eigenvalues of the Lorentz generator $\mathcal{O}_{+-}$. 

If we define  
$\mathcal{P}_3 = [\mathcal{P}_{(+)}, \mathcal{P}_{(-)}]$, we see from \eqref{eq3.44} and \eqref{eq4.60} that 
    \begin{equation}
      [\mathcal{P}_3, \mathcal{P}_{(\pm)}] T(\varphi)
      = \pm \frac{1}{2} \mathcal{P}_{(\pm)} T(\varphi) + O\left(\frac{1}{p}\right) \ . 
    \end{equation}
In the $p \rightarrow \infty$ limit, therefore,
$\mathcal{P}_{(\pm)}$ and 
$\mathcal{P}_3$ form the $\mathfrak{su}(2)$ Lie algebra:
    \begin{align}
      &[\mathcal{P}_{(+)}, \mathcal{P}_{(-)}] = \mathcal{P}_3 \ , \\
      &[\mathcal{P}_3, \mathcal{P}_{(\pm)}] = \pm \frac{1}{2} \mathcal{P}_{(\pm)} \ . 
    \end{align}
These reproduce \eqref{eq2.37} and \eqref{eq2.38}. 

\section{Conclusion and discussion} \label{sec5}
In this paper, by using the BT quantization,
we develop a finite-matrix regularization of the covariant derivatives $\nabla_{(a)}$
on closed connected K\"{a}hler 2$n$-dimensional manifolds in the covariant derivative
interpretation of matrix models.
Our method is a higher dimensional 
generalization of the one developed 
for closed Riemann surfaces in \cite{Hattori:2024btt}.
As concrete examples,
we considered regularizing
the covariant derivatives on 
$T^{2n}$ and $S^2$ as finite-size matrices,
and verified that the matrices satisfy the properties
observed in the covariant derivative interpretation in the limit
in which the size of the matrices becomes large.

We embedded the K\"{a}hler  manifolds into the Euclidean
spaces in regularizing the covariant derivatives as finite-size
matrices.
In general, there exist numerous ways of
such embedding.
Different ways of embeddings are considered to
correspond to different regularizations.
Since physical quantities must be 
independent of regularizations,
they are expected to be independent of the choice of
the embedding functions $X^A$.
It is a future problem to verify this. 

As we have regularized 
the covariant derivatives $\nabla_{(a)}$
as finite-size matrices, 
we would like to perform quantum mechanical calculations in the covariant derivative interpretation of the type IIB matrix model
such as calculation of the effective action.
Eventually, we hope to extract geometry from the results
of numerical simulations of the type IIB matrix model  by using the method that we developed
in this paper. 

\section*{Acknowledgements}
We would like to thank 
Yuki Mizuno for collaboration at the early stage of this work. We are also grateful to Hikaru Kawai for discussions.
A. T.  was supported in part by JSPS KAKENHI Grant 
Numbers JP21K03532 and JP25K07319.
T. S. was supported by JST SPRING, Japan Grant Number JPMJSP2167.

\appendix

\section{Calculation of $[\mathcal{P}_{(a)}, \mathcal{P}_{(b)}]$} \label{seca}
In this appendix,
we see how $[\mathcal{P}_{(a)}, \mathcal{P}_{(b)}] T(\varphi)$
behave in the large-$p$ limit.
We calculate $\mathcal{P}_{(a)}(\mathcal{P}_{(b)} T(\varphi))$
as follows:
\begin{align}
      \mathcal{P}_{(a)}(\mathcal{P}_{(b)} T(\varphi))
      &= \mathcal{P}_{(a)} (\hbar_p^{-1} T(\partial_{(b)} X^B)[T(X^B), T(\varphi)]) + O\left(\frac{1}{p}\right) \nonumber\\
      &= \hbar_p^{-2} T(\partial_{(a)} X^A) [T(X^A), T(\partial_{(b)} X^B)[T(X^B), T(\varphi)]] + O\left(\frac{1}{p}\right) \nonumber\\
      &= \hbar_p^{-2} T(\partial_{(a)} X^A) T(\partial_{(b)} X^B) [T(X^A), [T(X^B), T(\varphi)]] \nonumber\\
      &\quad + \hbar_p^{-2} T(\partial_{(a)} X^A) [T(X^A), T(\partial_{(b)} X^B)] [T(X^B), T(\varphi)] + O\left(\frac{1}{p}\right) \ .
    \end{align}
Here, in the first and second equalities,
we have used the fact that
the second term in the RHS of 
\eqref{eq3.1} is $O(1/p)$. 
In the third equality, 
we have used a relation
\begin{align}
    &[T(X^A), T(\partial_{(b)} X^B)[T(X^B), T(\varphi)]] \nonumber\\
    &= T(\partial_{(b)} X^B) [T(X^A), [T(X^B), T(\varphi)]] + [T(X^A), T(\partial_{(b)} X^B)] [T(X^B), T(\varphi)],
\end{align}
which holds despite the fact that 
$[\ , \ ]$ is different from the ordinary commutator.
Thus, we have
    \begin{align}
      [\mathcal{P}_{(a)}, \mathcal{P}_{(b)}] T(\varphi)
      &= \hbar_p^{-2} \left( T(\partial_{(a)} X^A) T(\partial_{(b)} X^B) [T(X^A), [T(X^B), T(\varphi)]]\right. \nonumber\\
      &\qquad \quad + T(\partial_{(a)} X^A) [T(X^A), T(\partial_{(b)} X^B)] [T(X^B), T(\varphi)] \nonumber\\
      &\qquad \quad - T(\partial_{(b)} X^B) T(\partial_{(a)} X^A) [T(X^B), [T(X^A), T(\varphi)]] \nonumber\\
      &\left. \qquad \quad - T(\partial_{(b)} X^B) [T(X^B), T(\partial_{(a)} X^A)] [T(X^A), T(\varphi)] \right) + O\left(\frac{1}{p}\right) \nonumber\\
      &= \hbar_p^{-2} \left( T(\partial_{(a)} X^A) T(\partial_{(b)} X^B) [T(X^A), [T(X^B), T(\varphi)]]\right. \nonumber\\
      &\qquad \quad + T(\partial_{(a)} X^A) [T(X^A), T(\partial_{(b)} X^B)] [T(X^B), T(\varphi)] \nonumber\\
      &\qquad \quad - T(\partial_{(a)} X^A) T(\partial_{(b)} X^B) [T(X^B), [T(X^A), T(\varphi)]] \nonumber\\
      &\qquad \quad + [T(\partial_{(a)} X^A), T(\partial_{(b)} X^B)] [T(X^B), [T(X^A), T(\varphi)]] \nonumber\\
      &\left. \qquad \quad - T(\partial_{(b)} X^B) [T(X^B), T(\partial_{(a)} X^A)] [T(X^A), T(\varphi)] \right) + O\left(\frac{1}{p}\right) \nonumber\\
      &= \hbar_p^{-2} \left( T(\partial_{(a)} X^A) T(\partial_{(b)} X^B) [[T(X^A), T(X^B)], T(\varphi)]\right. \nonumber\\
      &\qquad \quad + T(\partial_{(a)} X^A) [T(X^A), T(\partial_{(b)} X^B)] [T(X^B), T(\varphi)] \nonumber\\
      &\qquad \quad + [T(\partial_{(a)} X^A), T(\partial_{(b)} X^B)] [T(X^B), [T(X^A), T(\varphi)]] \nonumber\\
      &\left. \qquad \quad - T(\partial_{(b)} X^B) [T(X^B), T(\partial_{(a)} X^A)] [T(X^A), T(\varphi)] \right) + O\left(\frac{1}{p}\right) \ . 
      \label{eqa.2}
    \end{align}
In the second equality, we have applied a relation
$T(\partial_{(b)} X^B) T(\partial_{(a)} X^A) = T(\partial_{(a)} 
X^A) T(\partial_{(b)} X^B) - [T(\partial_{(a)} X^A), 
T(\partial_{(b)} X^B)]$ to the third term.
In the third equality, we have used
the Jacobi identity in the first and third terms.
Note that 
$[[T(X^A), T(X^B)], T(\varphi)] = (T(X^A) T(X^B) - T(X^B) T(X^A)) T(\varphi) - T(\varphi) (T'(X^A)T'(X^B) - T'(X^B)T'(X^A))$.

We will show that the third term in the last equality
of \eqref{eqa.2} is $O(1/p)$.
First, we see from 
\eqref{eq2.12} that
$\hbar_p^{-2} [T(X^B), [T(X^A), T(\varphi)]] = O(1)$.
Thus, if we show that
    \begin{equation}
      [T(\partial_{(a)} X^A), T(\partial_{(b)} X^B)] = O\left(\frac{1}{p}\right) \ ,
      \label{eqa.3}
    \end{equation}
we see that the third term in the last equality of 
\eqref{eqa.2} is $O(1/p)$.

We show \eqref{eqa.3}. 
We calculate 
$T(\partial_{(a)} X^A) T(\partial_{(b)} X^B)$ as follows:
    \begin{align}
      &(\text{the $(I,J)$ element of the 
      $((r, j), (r', l))$ block in 
      $T(\partial_{(a)} X^A) T(\partial_{(b)} X^B)$}) \nonumber\\
      &= \sum_{r''} \sum_{n, K} T^{(r, r'')} \left(\sqrt{\frac{d_{r''}}{d_r}} C^{v, r''^*; r}_{a'(a), (n); (j)} \omega_{ca'} \partial^c X^A\right)_{IK} T^{(r'', r')} \left(\sqrt{\frac{d_{r'}}{d_{r''}}} C^{v, r'^*; r''}_{b'(b), (l); (n)} \omega_{db'} \partial^d X^B\right)_{KJ} \nonumber\\
      &= T^{(r, r')} \left(\sqrt{\frac{d_{r'}}{d_r}} \sum_{r''} \sum_{m, n} C^{v, r''^*; r}_{a'(a), m(n); (j)} C^{v, r'^*; r''}_{b'(b), (l); m(n)} \omega_{ca'} \partial^c X^A \omega_{db'} \partial^d X^B \right)_{IJ} + O\left(\frac{1}{p}\right) \ .
    \end{align}
In the second equality, we have used
\eqref{eq2.11}. Note here that
the sum over repeated indices are taken.
Thus, we have  
    \begin{align}
      &(\text{the $(I,J)$ element of the 
      $((r, j), (r', l))$ block in $[T(\partial_{(a)} X^A), T(\partial_{(b)} X^B)]$}) \nonumber\\
      &= T^{(r, r')} \left(\sqrt{\frac{d_{r'}}{d_r}} \sum_{r''} \sum_{m, n} C^{v, r''^*; r}_{a'(a), m(n); (j)} C^{v, r'^*; r''}_{b'(b), (l); m(n)} \omega_{ca'} \partial^c X^A \omega_{db'} \partial^d X^B \right)_{IJ} \nonumber\\
      & \quad - T^{(r, r')} \left(\sqrt{\frac{d_{r'}}{d_r}} \sum_{r''} \sum_{m, n} C^{v, r''^*; r}_{a'(b), m(n); (j)} C^{v, r'^*; r''}_{b'(a), (l); m(n)} \omega_{ca'} \partial^c X^B \omega_{db'} \partial^d X^A \right)_{IJ} + O\left(\frac{1}{p}\right) \nonumber\\
      &= T^{(r, r')} \left(\sqrt{\frac{d_{r'}}{d_r}} \sum_{r''} \sum_{m, n} C^{v, r''^*; r}_{a'(a), m(n); (j)} C^{v, r'^*; r''}_{b'(b), (l); m(n)} \omega_{ca'} \partial^c X^A \omega_{db'} \partial^d X^B \right)_{IJ} \nonumber\\
      & \quad - T^{(r, r')} \left(\sqrt{\frac{d_{r'}}{d_r}} \sum_{r''} \sum_{m, n} C^{v, r''^*; r}_{b'(b), m(n); (j)} C^{v, r'^*; r''}_{a'(a), (l); m(n)} \omega_{db'} \partial^d X^B \omega_{ca'} \partial^c X^A \right)_{IJ} + O\left(\frac{1}{p}\right) \ .
      \label{eqa.5}
    \end{align}
In the second equality,
we have exchanged $a'$ and $b'$ as well as 
$c$ and $d$. Here, we show the following equality 
    \begin{equation}
      \sum_{r''} \sum_{m, n} C^{v, r''^*; r}_{a'(a), m(n); i(j)} C^{v, r'^*; r''}_{b'(b), k(l); m(n)} 
      = \sum_{r''} \sum_{m, n} C^{v, r''^*; r}_{b'(b), m(n); i(j)} C^{v, r'^*; r''}_{a'(a), k(l); m(n)}
      \label{eqa.6}
    \end{equation}
Using \eqref{eq3.26} in the LHS leads to 
    \begin{align}
      & \sum_{r''} \sum_{m, n} C^{v, r''^*; r}_{a'(a), m(n); i(j)} C^{v, r'^*; r''}_{b'(b), k(l); m(n)} \nonumber\\
      &= \sum_{r''} \sum_{m, n} d_r \int dg \ R^{\langle v \rangle}_{a'(a)}(g) R^{\langle r''^* \rangle}_{m(n)}(g) R^{\langle r \rangle}_{i(j)}(g) \ d_{r''} \int dh \ R^{\langle v \rangle}_{b'(b)}(h) R^{\langle r'^* \rangle}_{k(l)}(h) R^{\langle r'' \rangle}_{m(n)}(h) \nonumber\\
      &= d_r \int dg \ R^{\langle v \rangle}_{a'(a)}(g) R^{\langle r \rangle}_{i(j)}(g) \int dh \ R^{\langle v \rangle}_{b'(b)}(h) R^{\langle r'^* \rangle}_{k(l)}(h) \sum_{r''} d_{r''} \sum_{n} R^{\langle r'' \rangle}_{n(n)}(g^{-1} h) \nonumber\\
      &= d_r \int dg \ R^{\langle v \rangle}_{a'(a)}(g) R^{\langle r \rangle}_{i(j)}(g) \int dh' \ R^{\langle v \rangle}_{b'(b)}(gh') R^{\langle r'^* \rangle}_{k(l)}(gh') \sum_{r''} d_{r''} \sum_{n} R^{\langle r'' \rangle}_{n(n)}(h') \ .
      \label{eqa.7}
    \end{align}
In the second equality, it has been used that
$R$ is unitary, namely
$R^{\langle r''^* \rangle}_{m(n)}(g) = R^{\langle r'' 
\rangle}_{n(m)}(g^{-1})$. 
In the third equality,
we have put $h' = g^{-1}h$ and used 
$dh = dh'$.
By using 
    \begin{equation}
      \int dh' \ f(h') \sum_{r''} d_{r''} \sum_{n} R^{\langle r'' \rangle}_{n(n)}(h')
      = f(1)
      \label{eqa.8}
    \end{equation}
which holds for arbitrary smooth functions 
$f: \mathrm{Spin}^c(2n) \rightarrow \mathbb{C}$\footnote{\eqref{eqa.8} can be shown by
expanding $f(h') = \sum_{r} f^{\langle r \rangle}_{i(j)} \sqrt{d_r} R^{\langle r^* \rangle}_{i(j)}(h')$ and using
the orthonormality of the representation matrices
$\int dg \ R^{\langle r'^* \rangle}_{i'(j')}(g) R^{\langle r \rangle}_{i(j)}(g) = \frac{1}{d_r} \delta^{\langle r' \rangle \langle r \rangle} \delta_{i'i} \delta_{j'j}$. },
we find that \eqref{eqa.7} becomes
    \begin{equation}
      \sum_{r''} \sum_{m, n} C^{v, r''^*; r}_{a'(a), m(n); i(j)} C^{v, r'^*; r''}_{b'(b), k(l); m(n)}
      = d_r \int dg \ R^{\langle v \rangle}_{a'(a)}(g) R^{\langle r \rangle}_{i(j)}(g) R^{\langle v \rangle}_{b'(b)}(g) R^{\langle r'^* \rangle}_{k(l)}(g) \ .
      \label{eqa.9}
    \end{equation}
Since \eqref{eqa.9} is invariant when 
$a'$ is exchanged with $b'$ and $(a)$ with $(b)$, 
we see that \eqref{eqa.6} holds.
Thus, in the second equality of \eqref{eqa.5},
the first and second terms are canceled out each other.
In this way, we have shown that
$[T(\partial_{(a)} X^A), T(\partial_{(b)} X^B)] = O(1/p)$. 
This implies that \eqref{eqa.2} becomes 
    \begin{align}
      [\mathcal{P}_{(a)}, \mathcal{P}_{(b)}] T(\varphi)
      &= \hbar_p^{-2} \left( T(\partial_{(a)} X^A) T(\partial_{(b)} X^B) [[T(X^A), T(X^B)], T(\varphi)]\right. \nonumber\\
      &\qquad \quad + T(\partial_{(a)} X^A) [T(X^A), T(\partial_{(b)} X^B)] [T(X^B), T(\varphi)] \nonumber\\
      &\left. \qquad \quad - T(\partial_{(b)} X^B) [T(X^B), T(\partial_{(a)} X^A)] [T(X^A), T(\varphi)] \right) + O\left(\frac{1}{p}\right). 
      \label{eqa.10}
    \end{align}

Next, we will show that the second term in
the RHS of \eqref{eqa.10} is $O(1/p)$. 
Calculation similar to \eqref{eq3.15} gives
    \begin{align}
      &(\text{the $(I,J)$ element of the
      $((r, j), (r', l))$ block in $\hbar_p^{-1} T(\partial_{(a)}X^A) [T(X^A), T(\partial_{(b)}X^B)]$}) \nonumber\\
      &= T^{(r, r')} \left(-i \sum_{r''} \sqrt{\frac{d_{r''}}{d_r}} C^{v, r''^*; r}_{a'(a), m(n); (j)} \left(\nabla_{a'} C^{v, r'^*; r''}_{b'(b), (l); m(n)} \omega_{cb'} \partial^c X^B\right)\right)_{IJ} + O\left(\frac{1}{p}\right) \nonumber\\ 
      &= T^{(r, r')} \left(-i \sum_{r''} \sqrt{\frac{d_{r''}}{d_r}} C^{v, r''^*; r}_{a'(a), m(n); (j)} C^{v, r'^*; r''}_{b'(b), (l); m(n)} \omega_{cb'} (\nabla_{a'} \partial^c X^B)\right)_{IJ} + O\left(\frac{1}{p}\right). 
      \label{eqa.11}
    \end{align}
Note that the sums over $m$ and $n$ are taken similarly as
the other repeated indices.
In the second equality, we have used 
$\nabla_{a'} C^{v, r'^*; r''}_{b'(b), k(l); m(n)} = 0$
and $\nabla_{a'} \omega_{cb'} = 0$,
which is a property of K\"{a}hler manifolds. 
From \eqref{eq2.12}, we obtain 
    \begin{align}
      &(\text{the $(I,J)$ element of the $(r, j)$ block in $\hbar_p^{-1} [T(X^B), T(\varphi)]$}) \nonumber\\
      &= -i T^{\langle r, 1 \rangle} \left(W^{de} (\partial_d X^B)\left(\nabla_e \varphi^{\langle r \rangle}_{(j)}\right)\right)_{IJ} + O\left(\frac{1}{p}\right) \ .
      \label{eqa.12}
    \end{align}
Using \eqref{eqa.11} and \eqref{eqa.12} yields 
    \begin{align}
      &(\text{the $(I,J)$ element of the $(r,j)$ block in $\hbar_p^{-2} T(\partial_{(a)}X^A) [T(X^A), T(\partial_{(b)}X^B)] [T(X^B), T(\varphi)]$}) \nonumber\\
      &= - \sum_{r'} \sum_{l, K} T^{(r, r')} \left(\sum_{r''} \sqrt{\frac{d_{r''}}{d_r}} C^{v, r''^*; r}_{a'(a), m(n); (j)} C^{v, r'^*; r''}_{b'(b), (l); m(n)} \omega_{cb'} (\nabla_{a'} \partial^c X^B)\right)_{IK} \nonumber\\
      & \qquad \qquad \quad \times T^{(r', 1)} \left(W^{de} (\partial_d X^B)\left(\nabla_e \varphi^{\langle r' \rangle}_{(l)}\right)\right)_{KJ} + O\left(\frac{1}{p}\right) \nonumber\\
      &= - T^{(r, 1)} \left(\sum_{r'} \sum_{k, l} \sum_{r''} \sqrt{\frac{d_{r''}}{d_r}} C^{v, r''^*; r}_{a'(a), m(n); (j)} C^{v, r'^*; r''}_{b'(b), k(l); m(n)} \omega_{cb'} (\nabla_{a'} \partial^c X^B)\right. \nonumber\\
      & \qquad \qquad \qquad \qquad \qquad \times W^{de} (\partial_d X^B)\left(\nabla_e \varphi^{\langle r' \rangle}_{k(l)}\right) \Biggr) + O\left(\frac{1}{p}\right)
      \ .
    \end{align}
Calculation similar to that in 
\eqref{eq3.17} gives
    \begin{equation}
      (\nabla_{a'} \partial^c X^B) (\partial_d X^B) = 0 \ .
    \end{equation}
Thus, we have
    \begin{equation}
      \hbar_p^{-2} T(\partial_{(a)}X^A) [T(X^A), T(\partial_{(b)}X^B)] [T(X^B), T(\varphi)] 
      = O\left(\frac{1}{p}\right)  \ .
    \end{equation}
Similarly, we can show that third term of RHS of \eqref{eqa.10} is also $O(1/p)$.
In this way, \eqref{eqa.10} becomes
    \begin{equation}
      [\mathcal{P}_{(a)}, \mathcal{P}_{(b)}] T(\varphi)
      = \hbar_p^{-2} T(\partial_{(a)} X^A) T(\partial_{(b)} X^B) [[T(X^A), T(X^B)], T(\varphi)] + O\left(\frac{1}{p}\right)  \ . 
      \label{eqa.17}
    \end{equation}
Here, 
\begin{align}
    [[T(X^A), T(X^B)], T(\varphi)]=[T(X^A), T(X^B)]T(\varphi)-T(\varphi)[T'(X^A), T'(X^B)]\ .
\end{align}
We further calculate \eqref{eqa.17}.
We expand $[T(X^A), T(X^B)]$ up to the second order in $\hbar_p$: 
    \begin{align}
      &(\text{the $(I,J)$ element of the $((r, j), (r', l))$ block in $[T(X^A), T(X^B)]$}) \nonumber\\
      &= \delta^{\langle r \rangle \langle r' \rangle} \delta_{jl} [T^{(r, r)} (X^A \bm{1}_{E_r}), T^{(r, r)} (X^B \bm{1}_{E_r})]_{IJ} \nonumber\\
      &= \delta^{\langle r \rangle \langle r' \rangle} \delta_{jl} \biggl(-i \hbar_p T^{(r, r)} \left(\{X^A, X^B\} \bm{1}_{E_r}\right)_{IJ} \nonumber\\
      & \left. \qquad \qquad \qquad + \hbar_p^2 T^{(r, r)} \left((C_2(X^A, X^B) - C_2(X^B, X^A))\bm{1}_{E_r}\right)_{IJ} + O\left(\frac{1}{p^3}\right)\right) \ .
      \label{eqa.18}
    \end{align}
In the second equality, we have used \eqref{eq2.9} and \eqref{eq2.10}. Here, we define
$\tilde{C}_2(X^A, X^B)$ by 
    \begin{equation}
      \tilde{C}_2(X^A, X^B)
      = C_2(X^A, X^B) - C_2(X^B, X^A) + \frac{1}{4} (G^{\alpha \beta} G^{\gamma \delta} - G^{\delta \beta} G^{\gamma \alpha}) (\partial_{\alpha} X^A) R^{E_r}_{\beta \gamma}(\partial_{\delta} X^B) \ . 
    \end{equation}
Note that while $C_2$ is dependent on the representation $r$
because it includes
$R^{E_r}_{\beta \gamma}$, 
$\tilde{C}_2$ is independent of representation $r$
because the term including 
$R^{E_r}_{\beta \gamma}$ is subtracted.
By using $\tilde{C}_2(X^A, X^B)$, 
\eqref{eqa.18} is expressed as 
    \begin{align}
      &(\text{the $(I,J)$ element of 
      the $((r, j), (r', l))$ block in 
      $[T(X^A), T(X^B)]$}) \nonumber\\
      &= \delta^{\langle r \rangle \langle r' \rangle} \delta_{jl} \biggl(-i \hbar_p T^{(r, r)} \left(\{X^A, X^B\} \bm{1}_{E_r}\right)_{IJ} \nonumber\\
      & \qquad \qquad \qquad + \hbar_p^2 T^{(r, r)} \left(\tilde{C}_2(X^A, X^B) \bm{1}_{E_r}\right)_{IJ} \nonumber\\
      & \left. \qquad \qquad \qquad + \hbar_p^2 T^{(r, r)} \left(-\frac{1}{4} (G^{\alpha \beta} G^{\gamma \delta} - G^{\delta \beta} G^{\gamma \alpha}) (\partial_{\alpha} X^A) R^{E_r}_{\beta \gamma}(\partial_{\delta} X^B) \bm{1}_{E_r}\right)_{IJ} + O\left(\frac{1}{p^3}\right) \right) \ .
      \label{eqa.20}
    \end{align}
Since $R^{E_{\mathrm{trivial}}}_{\beta \gamma} = 0$,
$[T'(X^A), T'(X^B)]$ is calculated similarly as
    \begin{align}
      &(\text{the $(I,J)$ element in
      $[T'(X^A), T'(X^B)]$}) \nonumber\\
      &= -i \hbar_p T^{(1, 1)} \left(\{X^A, X^B\}\right)_{IJ} + \hbar_p^2 T^{(1, 1)} \left(\tilde{C}_2(X^A, X^B) \right)_{IJ} + O\left(\frac{1}{p^3}\right) \ .
      \label{eqa.21}
    \end{align}

We examine how the first terms in the RHS of 
\eqref{eqa.20} and \eqref{eqa.21}, 
which are the first order in $\hbar_p$,
contribute to
\eqref{eqa.17}.
We calculate
$\hbar_p^{-2} T(\partial_{(a)} X^A) T(\partial_{(b)} X^B) [-i \hbar_p T(\{X^A, X^B\}), T(\varphi)]$ as
    \begin{align}
      &(\text{the $(I,J)$ element of
      the $(r,j)$ block in $\hbar_p^{-2} T(\partial_{(a)} X^A) T(\partial_{(b)} X^B) [-i \hbar_p T(\{X^A, X^B\}), T(\varphi)]$}) \nonumber\\
      &= - T^{(r, 1)} \left(\sum_{r'} \sum_{k, l} \sqrt{\frac{d_{r'}}{d_r}} C^{v, r'^*; r}_{a'(a), k(l); (j)} \omega_{c'a'} (\partial^{c'} X^A) \sum_{r''} \sum_{m, n} \sqrt{\frac{d_{r''}}{d_{r'}}} C^{v, r''^*; r'}_{b'(b), m(n); k(l)} \omega_{cb'} (\partial^{c} X^B) \right. \nonumber\\
      & \qquad \qquad \qquad  \times W^{fg} \left(\partial_f W^{de} (\partial_d X^A) (\partial_e X^B)\right) \left(\nabla_g \varphi^{\langle r'' \rangle}_{m(n)}\right)\Biggr)_{IJ} + O\left(\frac{1}{p}\right) \nonumber\\
      &= - T^{(r, 1)} \left(\sum_{r'} \sum_{k, l} \sqrt{\frac{d_{r'}}{d_r}} C^{v, r'^*; r}_{a'(a), k(l); (j)} \omega_{c'a'} (\partial^{c'} X^A) \sum_{r''} \sum_{m, n} \sqrt{\frac{d_{r''}}{d_{r'}}} C^{v, r''^*; r'}_{b'(b), m(n); k(l)} \omega_{cb'} (\partial^{c} X^B) \right. \nonumber\\
      & \qquad \qquad \qquad  \times W^{fg} W^{de} \left(\nabla_f (\partial_d X^A) (\partial_e X^B)\right) \left(\nabla_g \varphi^{\langle r'' \rangle}_{m(n)}\right)\Biggr)_{IJ} + O\left(\frac{1}{p}\right) \nonumber\\
      &= O\left(\frac{1}{p}\right) \ .
    \end{align}
In the second equality, we have used $\nabla_f W^{de} = 0$, 
which is derived from a general property of the
K\"{a}hler structure.
In the third equality, we have used 
    \begin{align}
      (\partial^{c'} X^A) (\partial^{c} X^B) \left(\nabla_f (\partial_d X^A) (\partial_e X^B)\right)
      &= (\partial^{c'} X^A) (\partial^{c} X^B) (\nabla_f \partial_d X^A) (\partial_e X^B) \nonumber\\
      &\quad + (\partial^{c'} X^A) (\partial^{c} X^B) (\partial_d X^A) (\nabla_f \partial_e X^B) \nonumber\\
      &= 0 \ .
    \end{align} 
In the second equality,
we have used $(\partial^{c'} X^A) (\nabla_f \partial_d X^A) = 0$ and $(\partial^{c} X^B) (\nabla_f \partial_e X^B) = 0$, 
as in \eqref{eq3.14}. 
We see, therefore, that the first terms in the RHS of 
\eqref{eqa.20} and \eqref{eqa.21}
contribute to \eqref{eqa.17} as $O(1/p)$. 

Next, we examine how the second terms in the RHS of \eqref{eqa.20}
and \eqref{eqa.21} contribute to 
$[\mathcal{P}_{(a)}, \mathcal{P}_{(b)}] T(\varphi)$
in \eqref{eqa.17}.
Noting that $\tilde{C}_2(X^A, X^B)$
is independent of the representation $r$,
we see from \eqref{eq2.11} and
\eqref{eq2.12} that their contribution is $O(1/p)$: 
    \begin{equation}
      \hbar_p^{-2} T(\partial_{(a)} X^A) T(\partial_{(b)} X^B) [\hbar_p^2 T(\tilde{C}_2(X^A, X^B)), T(\varphi)]
      = O\left(\frac{1}{p}\right) \ . 
    \end{equation}

In this way, we see that it is only the third term in \eqref{eqa.20} that contribute to
$[\mathcal{P}_{(a)}, \mathcal{P}_{(b)}] T(\varphi)$
as $O(1)$. 
Thus, we finally find that
$[\mathcal{P}_{(a)}, \mathcal{P}_{(b)}] T(\varphi)$ become
    \begin{align}
      &[\mathcal{P}_{(a)}, \mathcal{P}_{(b)}] T(\varphi) \nonumber\\
      &= T(\partial_{(a)} X^A) T(\partial_{(b)} X^B) T \left(-\frac{1}{4} (G^{\alpha \beta} G^{\gamma \delta} - G^{\delta \beta} G^{\gamma \alpha}) (\partial_{\alpha} X^A) R^{E_r}_{\beta \gamma}(\partial_{\delta} X^B)\right) T(\varphi) + O\left(\frac{1}{p}\right) \ , 
    \end{align}
where 
$T (- \frac{1}{4} (G^{\alpha \beta} G^{\gamma \delta} - G^{\delta \beta} G^{\gamma \alpha}) (\partial_{\alpha} X^A) R^{E_r}_{\beta \gamma}(\partial_{\delta} X^B))$ is given by
    \begin{align}
      & (\text{the $(I,J)$ element of the
      $((r, j), (r', l))$ block in 
      $T (- \tfrac{1}{4} (G^{\alpha \beta} G^{\gamma \delta} - G^{\delta \beta} G^{\gamma \alpha}) (\partial_{\alpha} X^A) R^{E_r}_{\beta \gamma}(\partial_{\delta} X^B))$}) \nonumber\\
      & = \delta^{\langle r \rangle \langle r' \rangle} \delta_{jl} T^{(r, r)} \left(-\frac{1}{4} (G^{\alpha \beta} G^{\gamma \delta} - G^{\delta \beta} G^{\gamma \alpha}) (\partial_{\alpha} X^A) R^{E_r}_{\beta \gamma}(\partial_{\delta} X^B) \bm{1}_{E_r}\right)_{IJ} \ .
      \label{eqa.26}
    \end{align}

\bibliography{reference}
\bibliographystyle{utphys}

\end{document}